\documentclass[letter]{emulateapj}
\usepackage{color}
\usepackage{xspace}

\def\MIT{1}
\def\KICPChicago{2}
\def\EFIChicago{3}
\def\Cardiff{4}
\def\UChicago{5}
\def\Illinois{6}
\def\PhysicsUChicago{7}
\def\CfA{8}
\def\AAUChicago{9}
\def\McGill{10}
\def\PSU{11}
\def\Berkeley{12}
\def\Colorado{13}
\def\Harvard{14}
\def\NASA{15}
\def\Davis{16}
\def\LBNL{17}
\def\Michigan{18}
\def\Munich{19}
\def\ExcellenceCluster{20}
\def\MPE{21}
\def\CaseWestern{22}
\def\Yale{23}
\def\Caltech{24}
\def\Russia{25}

\slugcomment{Submitted to ApJ}

\begin{document}  

\title{X-ray Properties of the First SZE-selected Galaxy Cluster Sample from the South Pole Telescope}

\author{
K.~Andersson,\altaffilmark{\MIT}
B.~A.~Benson,\altaffilmark{\KICPChicago,\EFIChicago}
P.~A.~R.~Ade,\altaffilmark{\Cardiff}
K.~A.~Aird,\altaffilmark{\UChicago}
B.~Armstrong,\altaffilmark{\Illinois}
M.~Bautz,\altaffilmark{\MIT}
L.~E.~Bleem,\altaffilmark{\KICPChicago,\PhysicsUChicago}
M.~Brodwin,\altaffilmark{\CfA}
J.~E.~Carlstrom,\altaffilmark{\KICPChicago,\AAUChicago,\EFIChicago,\PhysicsUChicago} 
C.~L.~Chang,\altaffilmark{\KICPChicago,\EFIChicago}
T.~M.~Crawford,\altaffilmark{\KICPChicago,\AAUChicago}
A.~T.~Crites,\altaffilmark{\KICPChicago,\AAUChicago}
T.~de~Haan,\altaffilmark{\McGill}
S.~Desai,\altaffilmark{\Illinois}
M.~A.~Dobbs,\altaffilmark{\McGill}
J.~P.~Dudley,\altaffilmark{\McGill}
R.~J.~Foley,\altaffilmark{\CfA} 
W.~R.~Forman,\altaffilmark{\CfA} 
G.~Garmire,\altaffilmark{\PSU} 
E.~M.~George,\altaffilmark{\Berkeley}
M.~D.~Gladders,\altaffilmark{\KICPChicago,\AAUChicago}
N.~W.~Halverson,\altaffilmark{\Colorado}
F.~W.~High,\altaffilmark{\Harvard} 
G.~P.~Holder,\altaffilmark{\McGill}
W.~L.~Holzapfel,\altaffilmark{\Berkeley}
J.~D.~Hrubes,\altaffilmark{\UChicago}
C.~Jones,\altaffilmark{\CfA} 
M.~Joy,\altaffilmark{\NASA}
R.~Keisler,\altaffilmark{\KICPChicago,\PhysicsUChicago}
L.~Knox,\altaffilmark{\Davis}
A.~T.~Lee,\altaffilmark{\Berkeley,\LBNL}
E.~M.~Leitch,\altaffilmark{\KICPChicago,\AAUChicago}
M.~Lueker,\altaffilmark{\Berkeley}
D.~P.~Marrone,\altaffilmark{\KICPChicago,\UChicago}
J.~J.~McMahon,\altaffilmark{\KICPChicago,\EFIChicago,\Michigan}
J.~Mehl,\altaffilmark{\KICPChicago,\AAUChicago}
S.~S.~Meyer,\altaffilmark{\KICPChicago,\EFIChicago,\PhysicsUChicago,\AAUChicago}
J.~J.~Mohr,\altaffilmark{\Munich,\ExcellenceCluster,\MPE}
T.~E.~Montroy,\altaffilmark{\CaseWestern}
S.~S.~Murray,\altaffilmark{\CfA} 
S.~Padin,\altaffilmark{\KICPChicago,\AAUChicago}
T.~Plagge,\altaffilmark{\Berkeley,\AAUChicago}
C.~Pryke,\altaffilmark{\KICPChicago,\AAUChicago,\EFIChicago} 
C.~L.~Reichardt,\altaffilmark{\Berkeley}
A.~Rest,\altaffilmark{\Harvard}
J.~Ruel,\altaffilmark{\Harvard}
J.~E.~Ruhl,\altaffilmark{\CaseWestern} 
K.~K.~Schaffer,\altaffilmark{\KICPChicago,\EFIChicago} 
L.~Shaw,\altaffilmark{\McGill,\Yale}
E.~Shirokoff,\altaffilmark{\Berkeley} 
J.~Song,\altaffilmark{\Illinois}
H.~G.~Spieler,\altaffilmark{\LBNL}
B.~Stalder,\altaffilmark{\CfA}
Z.~Staniszewski,\altaffilmark{\CaseWestern}
A.~A.~Stark,\altaffilmark{\CfA} 
C.~W.~Stubbs,\altaffilmark{\Harvard,\CfA} 
K.~Vanderlinde,\altaffilmark{\McGill}
J.~D.~Vieira,\altaffilmark{\KICPChicago,\PhysicsUChicago,\Caltech}
A.~Vikhlinin,\altaffilmark{\CfA, \Russia} 
R.~Williamson,\altaffilmark{\KICPChicago,\AAUChicago} 
Y.~Yang,\altaffilmark{\Illinois}
O.~Zahn,\altaffilmark{\Berkeley}
and
A.~Zenteno\altaffilmark{\Munich,\ExcellenceCluster}
}

\altaffiltext{\MIT}{MIT Kavli Institute for Astrophysics and Space Research, Massachusetts Institute of Technology, 77 Massachusetts Avenue, Cambridge, MA 02139}

\altaffiltext{\KICPChicago}{Kavli Institute for Cosmological Physics, University of Chicago, 5640 South Ellis Avenue, Chicago, IL 60637}
\altaffiltext{\EFIChicago}{Enrico Fermi Institute, University of Chicago, 5640 South Ellis Avenue, Chicago, IL 60637}
\altaffiltext{\Cardiff}{Department of Physics and Astronomy, Cardiff University, CF24 3YB, UK}
\altaffiltext{\UChicago}{University of Chicago, 5640 South Ellis Avenue, Chicago, IL 60637}
\altaffiltext{\Illinois}{Department of Astronomy, University of Illinois, 1002 West Green Street, Urbana, IL 61801}
\altaffiltext{\PhysicsUChicago}{Department of Physics, University of Chicago, 5640 South Ellis Avenue, Chicago, IL 60637}
\altaffiltext{\CfA}{Harvard-Smithsonian Center for Astrophysics, 60 Garden Street, Cambridge, MA 02138}
\altaffiltext{\AAUChicago}{Department of Astronomy and Astrophysics, University of Chicago, 5640 South Ellis Avenue, Chicago, IL 60637}
\altaffiltext{\McGill}{Department of Physics, McGill University, 3600 Rue University, Montreal, Quebec H3A 2T8, Canada}
\altaffiltext{\PSU}{Department of Astronomy and Astrophysics, Pennsylvania State University, 525 Davey Lab, University Park, PA 16802}
\altaffiltext{\Berkeley}{Department of Physics, University of California, Berkeley, CA 94720}
\altaffiltext{\Colorado}{Department of Astrophysical and Planetary Sciences and Department of Physics, University of Colorado, Boulder, CO 80309}
\altaffiltext{\Harvard}{Department of Physics, Harvard University, 17 Oxford Street, Cambridge, MA 02138}
\altaffiltext{\NASA}{Department of Space Science, VP62, NASA Marshall Space Flight Center, Huntsville, AL 35812}
\altaffiltext{\Davis}{Department of Physics, University of California, One Shields Avenue, Davis, CA 95616}
\altaffiltext{\LBNL}{Physics Division, Lawrence Berkeley National Laboratory, Berkeley, CA 94720}
\altaffiltext{\Michigan}{Department of Physics, University of Michigan, 450 Church Street, Ann Arbor, MI, 48109}
\altaffiltext{\Munich}{Department of Physics, Ludwig-Maximilians-Universit\"{a}t, Scheinerstr.\ 1, 81679 M\"{u}nchen, Germany}
\altaffiltext{\ExcellenceCluster}{Excellence Cluster Universe, Boltzmannstr.\ 2, 85748 Garching, Germany}
\altaffiltext{\MPE}{Max-Planck-Institut f\"{u}r extraterrestrische Physik, Giessenbachstr.\ 85748 Garching, Germany}
\altaffiltext{\CaseWestern}{Physics Department and CERCA, Case Western Reserve University, 10900 Euclid Ave., Cleveland, OH 44106}
\altaffiltext{\Yale}{Department of Physics, Yale University, P.O. Box 208210, New Haven, CT 06520-8120}
\altaffiltext{\Caltech}{California Institute of Technology, Pasadena, CA 91125}
\altaffiltext{\Russia}{Space Research Institute (IKI), Profsoyuznaya 84/32, Moscow, Russia}

\email{kanderss@space.mit.edu}

\begin{abstract}
We present results of X-ray observations of a sample of 15 clusters 
selected via their imprint on the cosmic microwave background (CMB)
from the thermal Sunyaev-Zel'dovich (SZ) effect.  
These clusters are a 
subset of the first SZ-selected cluster catalog, obtained from observations of 
178 deg$^2$ of sky surveyed by the South Pole Telescope. 
Using X-ray observations with {\sl Chandra} and {\sl XMM-Newton}, we estimate 
the temperature, $T_X$, and mass, $M_g$, of the intracluster medium (ICM) 
within $r_{500}$ for each cluster.  
From these, we calculate $Y_X=M_g~T_X$ and estimate the total cluster mass 
using a $M_{500}-Y_X$ scaling relation measured from previous X-ray studies. 
The integrated Comptonization, $Y_{SZ}$, is derived from the SZ measurements, 
using additional information from the X-ray measured gas density profiles 
and a universal temperature profile.  
We calculate scaling relations between the X-ray and SZ observables, and find results 
generally consistent with other measurements and the expectations from simple 
self-similar behavior. 
Specifically, we fit a $Y_{SZ}-Y_{X}$ relation and find a normalization
of $0.82 \pm 0.07$, marginally consistent with the predicted ratio of 
$Y_{SZ}/Y_{X}=0.91 \pm 0.01$ that would be expected from the density and 
temperature models used in
this work.
Using the 
$Y_X$ derived mass estimates, we fit a $Y_{SZ}-M_{500}$ relation 
and find a slope consistent with the self-similar expectation of $Y_{SZ} \propto M^{5/3}$ 
with a normalization consistent with predictions from other X-ray studies.  
We compare the X-ray mass estimates to previously published SZ mass 
estimates derived from cosmological simulations of the SPT survey. 
We find that the SZ mass estimates are lower by a factor of $0.89\pm0.06$, 
which is within the $\sim$15\% systematic uncertainty quoted for the 
simulation-based SZ masses.  
Overall, the X-ray measurements confirm that the SZ-selected sample consists of 
very massive systems which exhibit general 
properties consistent with X-ray selected samples, even allowing for the broad 
redshift range ($0.29<z<1.08$) of the sample.

\end{abstract}

\clearpage 
\section{Introduction}
Large area cluster surveys extending to high redshift can be used to study the evolution 
of the abundance of galaxy clusters, thereby delivering precise constraints on the amount and nature 
of the dark energy \citep{wang98, haiman01}.  
The accuracy with which the observed mass proxy can be linked to the true cluster mass 
fundamentally limits the cosmological constraints from the survey.  In particular, a 
redshift dependent bias on a survey's cluster mass estimates could mimic a time-evolving 
dark energy, so this systematic must be understood and constrained.   
Clusters are known to evolve --- through mergers, 
galaxy and star formation, and variable contributions from active galactic
nuclei (AGN) --- and these effects will influence the evolution of any
cluster observable at some level.

Most of the baryonic mass in clusters is in the form of an intra-cluster gas that can be 
heated to several keV as it virializes in a massive cluster's gravitational potential well.  
This gas is visible in the X-ray band via thermal bremsstrahlung and also from its 
distortion of the Cosmic Microwave Background (CMB) from 
inverse Compton scattering, otherwise known as the Sunyaev-Zel'dovich (SZ) effect \citep{sun72}.
The lowest scatter cluster observables that scale with cluster mass, $M$, 
are likely those most closely related to the gas pressure \citep[e.g.,][]{kra06}, and 
hence related to the total energy of the gas. 
The SZ intensity is proportional to the Comptonization, the line-of-sight integral 
of the gas pressure. Hence, the SZ signal integrated over the cluster's extent, $Y_{SZ}$, 
measures the total pressure in the cluster.  An X-ray analogue, $Y_X$, can be 
constructed from the product of the cluster's total gas mass and the X-ray 
spectroscopic temperature \citep{kra06}.  

The intrinsic scatter in the $M-Y_X$ relation has been found to be smaller than  
the measurement error in X-ray studies conducted to date \citep{vikh06, vikh09, sun09}, and 
has been used to place interesting constraints on the dark energy equation of state 
with only a relatively small sample of clusters \citep{vikh09b}.  
The slope and scatter of the 
relationship between $Y_{SZ}$ and the cluster mass, $M$,
has been studied extensively in simulations and 
is expected to be relatively insensitive to non-gravitational physics \citep{nag06},  
the dynamical state of clusters \citep{jeltema08}, and the 
presence of cool cores \citep{motl05}.   
The expected close correlation between cluster mass and $Y_{SZ}$, as well 
as the redshift independence of the SZ brightness, strongly motivates 
using SZ cluster surveys for cosmological studies \citep{carlstrom02}.

Recently, there has been significant progress in measuring the SZ-signal from clusters. 
High resolution imaging has been obtained for single objects \citep[e.g.,][]{nord09,mason10} and 
ICM profiles have been measured for moderately sized samples \citep[e.g.,][]{mroc09,plagge10}. 
The first clusters discovered with a blind SZ survey were reported in \citet{staniszewski09}, 
and showed the capability of the SZ-signal as a cluster finder.  
From larger samples, scaling relations have been measured between the SZ-signal 
and mass estimates from both X-ray and weak lensing measurements  
\citep[e.g.,][]{bonamente08, mar09, melin10}.  Notably, \citet{melin10} measured a 
$Y_{SZ}$-$M$ relation from binned WMAP fluxes at the location of known X-ray selected 
clusters combined with X-ray luminosity based mass estimates.  They found a $Y_{SZ}$-$M$ 
relation with a normalization and slope that matched the X-ray prediction, however 
similar analyses have found conflicting results \citep{kom10}.   
More detailed observations comparing 
SZ and X-ray measurements will improve our understanding of the gas pressure in clusters.

The first SZ-selected catalog of clusters was  
presented in \citet{van10} (hereafter V10), and included the first meaningful cosmological 
constraints from an SZ cluster survey.  
In V10, the sample of 21 clusters 
had a median redshift of $z = 0.74$ and was predicted to be 100$\%$ complete above a mass 
threshold of $M_{500}  \approx 3 \times 10^{14}  h^{-1} M_{\sun}$ at $z > 0.6$. 
The cosmological constraints from V10 were limited by uncertainties in the 
cluster mass calibration.  This calibration relied on the dark matter simulations of \citet{shaw09}, 
with gas physics based on the models in \citet{bode07}, to link the SZ-significance to cluster mass.
This introduced
a $\sim 15\%$ systematic uncertainty in the mass calibration due to uncertainties 
in the pressure normalization of the simulations.  
Therefore, an important first step to improve the cosmological constraints 
of V10 is to tie the SZ observables to observationally calibrated X-ray scaling relations, 
such as those in \citet{vikh09}.  The V10 cluster sample is also unique because of 
its high median redshift and SZ-selection.  X-ray observations of these clusters 
could allow additional constraints on the redshift evolution 
of X-ray scaling relations, which typically have been studied from 
X-ray selected samples concentrated at $z < 0.6$ \citep[e.g.,][]{maughan07, 
vikh09, mantz09}.

In this work, we present results from X-ray observations of a subset of 
15 clusters with the highest SZ-significance from V10.  We report on the 
X-ray observables of this sample, and use X-ray scaling relations from 
\citet{vikh09} to estimate each cluster's mass.  For each cluster, we also note 
details of the cluster's X-ray morphology and the identification of X-ray sources 
with objects in other catalogs.  We use the X-ray measured 
gas density profiles to improve the SZ estimates of integrated Comptonization, $Y_{SZ}$.   
Finally we construct X-ray and SZ scaling relations, specifically the $Y_{SZ}$-$Y_{X}$ and 
$Y_{SZ}$-$M_{500}$ relations, and compare these relations to expectations and 
other results.  

The paper is organized as follows: \S 2 describes the cluster sample and 
the SZ, optical and X-ray observations. \S 3 discusses the X-ray data 
analysis, the estimation of gas mass and temperature while \S 4 describes 
the estimate of gravitational mass. In \S 5 we discuss the analysis of 
the SPT data and deprojection of the SZ measurements. In \S 6 the X-ray and 
SZ scaling relations are investigated and we conclude with a discussion in \S 7.
The properties of individual clusters are discussed in the Appendix.

In all calculations we have assumed a {\sl WMAP}7+BAO+{\sl $H_0$} $\Lambda$CDM cosmology 
\citep{kom10} with $\Omega_M = 0.272$, $\Omega_\Lambda = 0.728$ and 
$H_0 = 70.2~$km$~$s$^{-1}~$Mpc$^{-1}$ with distance measurements from Baryon Acoustic 
Oscillations (BAO) in the distribution of galaxies \citep{per10} and the 
Hubble constant ($H_0$) measurement from \citet{rie09}.
Everywhere, we define $M_{500}$ as the mass inside $r_{500}$, within which the matter density is 
500 times the critical density at the cluster redshift, $\rho_{crit}(z) = 3H^2(z)/8\pi G$, where 
$H(z)$ is the Hubble parameter.

\section{Cluster Sample and Observations}
The clusters used in this work were selected from a SZ detection significance-limited 
catalog from the South Pole Telescope (SPT) cluster survey described in V10.  For all clusters, we 
performed follow-up optical imaging to identify galaxy cluster counterparts and measure redshifts 
\citep{high10}.  
Sixteen cluster candidates with SZ detection significance above 5.4 were selected for 
X-ray follow-up with {\sl Chandra} and {\sl XMM-Newton}.  One candidate was later discovered 
to be a spurious detection, and is discussed in more detail in \S \ref{2343nondet}.  This section briefly 
describes the SZ, optical, and X-ray data sets.  

\subsection{SPT SZ Observations and the Cluster Sample}
\label{sec:spt}
The 10-m diameter South Pole Telescope is a millimeter wavelength telescope
located at the South Pole.  Its primary science goal is to conduct a $\sim$2000 deg$^2$ survey 
to find clusters of galaxies via measurements of the SZ effect.   
The receiver consists of a 960 element bolometer array that is sensitive in three frequency bands centered
at 95, 150, and 220 GHz. In this work, we only use observations at 150 GHz, the SPT frequency 
band with the most SZ sensitivity. Details of the telescope and 
receiver can be found in \citet{padin08} and \citet{carlstrom09}.  
 
In 2008, the SPT surveyed two $\sim$100 deg$^2$ regions.  These two fields are approximately square on the sky, 
and centered at right ascension (R.A.) 5$^h$30$^m$, declination (decl.) -55$^{\circ}$ and R.A. 23$^h$30$^m$, 
decl. -55$^{\circ}$.   The data from the first of 
these fields was used to report the first SZ-discovered clusters \citep{staniszewski09}, to measure 
source counts of extragalactic mm-wavelength emitting objects \citep{vieira09}, and to measure small scale temperature anisotropies
due to the SZ effect from unresolved clusters and emission from point sources \citep{lueker09, hall09}.  In V10, the data from 
both fields was used to report the first SZ detection significance-limited catalog of 22 cluster candidates.  

The SZ observations, data processing, and mapmaking used in this work are described in detail in V10, 
and only a brief overview is provided here.  Each field was scanned in azimuth at a 
constant velocity, with the scans stepped in elevation.  
With the velocity and elevation step used, it takes $\sim$2 hours to cover an entire 100 deg$^2$ field. 
This scan is then repeated several hundred times to decrease the noise in the co-added map.  Each detector timestream is 
filtered to remove both long 
timescale drifts and sky signal that is spatially correlated across the focal plane.  
The filtering effectively acts as a high pass filter; the filter used for the SZ 
cluster analysis removes signal on spatial scales larger than $\sim0.5^{\circ}$.  
The data is combined to make a map by reconstructing the pointing
for each detector and then averaging the
data from all the detectors using inverse-variance weighting.
The 150 GHz maps were calibrated to an accuracy of 
3.6$\%$ by direct comparison to the WMAP 5-year maps \citep{lueker09}.  The final 
map of each field has a sensitivity limit of 18 $\mu$K-arcmin. 

Cluster candidates were identified in the SPT maps by using a matched spatial filter technique
 \citep{haehnelt96, melin06}.  
Here we summarize the method and results as used in V10.  The SPT maps consist of several sources of signal, 
including primary CMB anisotropy, unresolved point sources, and SZ signal from clusters.  To 
identify cluster candidates, the SPT maps are filtered in Fourier space to give more weighting 
to signals matching the expected spatial scales of clusters.  Twelve different spatial filters were 
constructed using spherical $\beta$-models with $\beta$ fixed to 1 and 
core radii evenly spaced between $0.25'$ and $3.0'$.  With these, the SPT maps were filtered and cluster 
candidates were identified as decrements in the filtered map.  The significance of a candidate was 
quantified by their signal relative to the standard deviation in the filtered map, or 
signal-to-noise.   For a given cluster candidate, the highest signal-to-noise across all filter scales was defined 
as $\xi$.   To avoid spurious identifications from ringing around spatially filtered bright point sources, 
a $4\arcmin$ radius region around all 5$\sigma$ positive sources was masked before spatial filtering.  The total 
sky area used in both fields after masking was 178 deg$^2$.  
Further details of the SZ analysis relevant to the results presented here are given in 
\S \ref{sec:sze}.

The sixteen cluster candidates with the largest $\xi$ were selected for 
X-ray follow-up with the {\sl Chandra} and {\sl XMM-Newton} X-ray telescopes.  The original candidate list was based on a similar, 
but earlier version of the list that appeared in V10. This change re-ordered the list somewhat, such that 
sixteen of the seventeen most significant candidates from V10 had X-ray follow-up.  In Table \ref{obstable}, we give the position 
and $\xi$ from V10 for the fifteen confirmed clusters with X-ray observations.  One of the candidates 
with X-ray follow-up, SPT-CL J2343-5521, is not listed because it is very likely a false detection.  
Its X-ray observation is discussed further in the Appendix.

\subsection{Optical Imaging and Spectroscopy}
\label{sec:opt}

Optical counterparts for the clusters selected by SZ detection significance were identified and 
photometric redshifts were measured via a combination of 
imaging from the Blanco Cosmology Survey (BCS) and targeted observations using the Magellan telescopes.    
For a subset of clusters, spectroscopic redshifts were also obtained.  Further details of the optical 
data and analysis can be found in \citet{high10}, and are briefly described below.  

The BCS is an optical survey of two $\sim$50 deg$^2$ fields that lie inside the two 2008 SPT fields described 
in \S \ref{sec:spt}.  BCS used the Mosaic-II wide field imager on the Blanco 4-m telescope at the 
Cerro Tololo Inter-American Observatory in Chile.  
in Chile. The BCS obtained contiguous deep optical imaging in the
griz bands across their survey fields, and these data have been
processed using a development version of the DES data management system
\citep{ngeow06,mohr08} and then used in the study of the galaxy
population and in the redshift estimation of the first SPT survey fields
\citep{ngeow09,staniszewski09,high10,brodwin10,zenteno10}.  These data
are publicly available through the NOAO Survey Program and have been
used by other groups to study clusters in the SPT survey region
\citep{menanteau09, menanteau09a,men10, mcinnes09,suhada10}.  
For the nine clusters outside the BCS fields, 
optical imaging in the $griz$ bands was obtained with the Inamori Magellan Areal Camera 
and Spectrograph \citep[IMACS;][]{dressler06} on the 
Magellan Baade 6.5-m telescope in Las Campanas, Chile. 
Five cluster candidates in the BCS fields were also re-observed with Magellan.  
In contrast to BCS, the Magellan observations were performed adaptively, 
where the candidates were observed in $\sim$100 sec increments until the galaxy cluster was detected.  

Optical counterparts 
of each SPT candidate were identified by searching for red sequence objects within a 
$2\arcmin$  radius of the SPT candidate location.  A cluster was identified through an excess of red 
sequence objects relative to the background, and the photometric redshift was estimated by 
fitting a red sequence model.  
The redshift uncertainty varies over the sample, however they are
typically $\Delta z \sim 0.03$ and can be as large as $\Delta z = 0.10$
for clusters at $z \sim 1$ \citep[see][for details]{high10}.

For 8 of 16 cluster candidates, spectroscopic measurements were obtained using the Low Dispersion Survey Spectrograph (LDSS3) 
on the Magellan Clay 6.5-m telescope.  These observations are described in \citet{high10}. 
For one candidate, SPT-CL J0546-5345, multislit spectroscopy was done with IMACS on  
the Magellan Baade 6.5-m telescope \citep{brodwin10}.  
The spectroscopic targets 
were chosen to span the redshift range of the sample to help calibrate 
the photometric redshifts.  
One cluster, SPT-CL J0516-5430, has a previous spectroscopic redshift from the REFLEX cluster 
survey \citep{bohr04}. 
In Table \ref{obstable}, we give the photometric redshifts for each 
of the 15 clusters included in this work, and the spectroscopic redshift where available.

\subsection{Chandra \& XMM-Newton observations}
As described in \S \ref{sec:spt}, the sixteen cluster candidates with the highest 
detection significance, $\xi$, were selected for an X-ray follow-up program.  This program was 
split between several observing cycles and proposals between the {\sl Chandra} and {\sl XMM-Newton} 
X-ray satellites.  

In the original planning of X-ray observations, the observing time required 
for each cluster was estimated from preliminary predictions of the SZ significance-to-mass 
relation and a mass-luminosity scaling. The uncertainty in the relations does not allow an accurate
estimate of the exposure needed for a required number of photons.
The estimates were informed by ROSAT fluxes where 
available. 
For clusters scheduled for multiple observations with {\sl Chandra}, the integration time was 
modified accordingly once a flux had been measured. 

To date, the {\sl Chandra} program consists of two GTO programs in AO-9 (295 ks, total), 
two GTO programs in AO-11 (340 ks, total) and a GO program in AO-11 (310 ks). 
When completed, the {\sl Chandra} program will have collected at least 1500 cluster photons 
within $0.5 r_{500}$ and in the $0.5$-$7.0$ keV energy band for each of 12 clusters in the sample. 
This limit was chosen to enable measuring of the ICM temperature, $T_{X}$, to 15\% accuracy. 
For one very faint object, SPT-CL J0553-5005, the data will not be sufficient to measure 
the temperature to this accuracy in the current program.  The {\sl Chandra} ACIS-I count-rate 
in the $0.5$-$7.0$ keV band for this object is only $0.005~$ct$~$s$^{-1}$. 

Four of the candidates have been observed with {\sl XMM-Newton}, one of which is an 
archival observation (SPT-CL J0516-5430, also RXCJ0516.7-5430); the other three are 
from a 2008 program (SPT-CL J0559-5249, SPT-CL J2332-5358 and SPT-CL J2337-5942). 
All of these observations have more than the required $1500$ counts.
For two of the clusters, there exists both {\sl Chandra} and {\sl XMM-Newton} data 
(SPT-CL J0516-5430 and SPT-CL J2337-5942) and these are all analyzed in this work. 
We chose to re-observe these two clusters with {\sl Chandra} to better identify and remove 
X-ray point-sources in the analysis. The low redshift and high mass of the clusters made 
these observations possible with a comparably small amount of observing time. 

While we find consistent results from the analysis of data from the two different 
satellites, we choose to use the {\sl Chandra} data for the two clusters with 
data from both instruments since both the high spatial resolution and stable background 
of {\sl Chandra} are desirable.
For the two clusters where only {\sl XMM-Newton} data were available, we derive 
our results from these data using the methods as described below.  

X-ray measurements of SPT-CL J2332-5358 and SPT-CL J2342-5411 were also
reported by \citet{suhada10} from the XMM-BCS survey.  These measurements
were not as deep as the observations discussed in our work, however we
find X-ray observables consistent with their measurements.

Currently, there are eight out of the fifteen clusters for which the required 1500 source counts have 
not yet been collected with either instrument. However, we include these in the 
analysis since the existing data provide useful constraints.

\section{X-ray data analysis}
\subsection{Data reduction}

\begin{deluxetable*}{lrrrrrrrrrc}
\tabletypesize{\scriptsize}
\tablecaption{Chandra and XMM-Newton observations \label{obstable}}
\tablewidth{0pt}
\tablehead{
\colhead{Name} &
\colhead{R.A\footnotemark[a]} &
\colhead{Dec.\footnotemark[a]} &
\colhead{$\xi$\footnotemark[b]} & 
\colhead{photo-\footnotemark[c]} &
\colhead{spec-\footnotemark[c]} &
\colhead{exposure} &
\colhead{Source \footnotemark[d]} &
\colhead{$n_H$ \footnotemark[e]} &
\colhead{$d_{X-SZ}$\footnotemark[f] }& 
\colhead{Merger?} \\
\colhead{} &
\colhead{} &
\colhead{} &
\colhead{} & 
\colhead{$z$} & 
\colhead{$z$} & 
\colhead{[ks]} &
\colhead{counts} &
\colhead{$10^{20}~$cm$^{-2}$} &
\colhead{$\arcmin$} &
\colhead{} 
}
\footnotetext[X]{{\sl XMM-Newton} observation. The listed exposure times and source counts are for 
MOS1, MOS2 and PN combined.}
\footnotetext[a]{R.A. and Dec. determined from the SPT detection.}
\footnotetext[b]{Signal-to-noise measured in 150 GHz SPT maps as described in V10.}
\footnotetext[c]{Listed photometric and spectroscopic redshifts are based on analysis in \citet{high10}.}
\footnotetext[d]{Based on Chandra/XMM-Newton count-rate in the 0.5-7.0 keV band, within $0.5~r_{500}$}
\footnotetext[e]{Hydrogen column density from the Leiden-Argentine-Bonn survey \citep{kab05}.}
\footnotetext[f]{Distance between SPT detection and X-ray centroid.}
\footnotetext[g]{Spectroscopic redshift from the REFLEX cluster survey \citep{bohr04}.}
\footnotetext[h]{Spectroscopic redshift from \citep{brodwin10}.}
\footnotetext[i]{The position and $\xi$ of this cluster are modified from V10 after correcting 
for the effects of a point source coincident with the cluster, as described in \S \ref{sec:point_source}.}

\startdata
SPT-CL J0000-5748 & 0.25 & -57.807 & 5.48 & 0.74 & - & 28.1 & 1451 & 1.37  & 0.16 & \quad \\
SPT-CL J0509-5342 & 77.336 & -53.705 & 6.61 & 0.47 & 0.4626 & 27.3 & 2441 & 1.46 & 0.13 & \checkmark  \\
SPT-CL J0516-5430 & 79.148 & -54.506 & 9.42 & 0.25 & ~0.2952\footnotemark[g] & 8.4 & 2136 & 2.05 & 0.15  & \checkmark  \\
SPT-CL J0528-5300 & 82.017 & -53.0 & 5.45 & 0.75 & 0.7648 & 36.5 & 356 & 3.23 & 0.20 & \quad \\
SPT-CL J0533-5005 & 83.398 & -50.092 & 5.59 & 0.83 & 0.8810 & 41.5 & 201 & 2.95 & 0.53 & \checkmark  \\
SPT-CL J0546-5345 & 86.654 & -53.761 & 7.69 & 1.16 & ~1.0665\footnotemark[h] & 55.6 & 1304 & 6.78 & 0.13 & \checkmark  \\
SPT-CL J0551-5709 & 87.902 & -57.156 & 6.13 & 0.41 & 0.4230 & 19.8 & 876 & 6.27 & 0.53 & \checkmark \\
SPT-CL J0559-5249\footnotemark[X] & 89.925 & -52.826 & 9.28 & 0.66 & 0.6112 & 42.0 & 2006 & 5.08 & 0.42 & \checkmark \\
SPT-CL J2331-5051 & 352.958 & -50.864 & 8.04 & 0.55 & 0.5707 & 34.2 & 2428 & 1.12 & 0.23 &  \checkmark \\
SPT-CL J2332-5358\footnotemark[X,i] & 353.109 & -53.976 & 13.05 & 0.32 & - & 18.9 & 4826 & 1.28 & 0.36 & \quad \\
SPT-CL J2337-5942 & 354.354 & -59.705 & 14.94 & 0.77 & 0.7814 & 19.8 & 1488 & 1.45 & 0.12  & \checkmark \\
SPT-CL J2341-5119 & 355.299 & -51.333 & 9.65 & 1.03 & 0.9983 & 79.0 & 2090 & 1.21 & 0.26 & \quad \\
SPT-CL J2342-5411 & 355.69 & -54.189 & 6.18 & 1.08 & - & 133.7 & 1193 & 1.49 & 0.23 & \quad \\
SPT-CL J2355-5056 & 358.955 & -50.937 & 5.89 & 0.35 & - & 20.6 & 1798 & 1.27 & 0.61 & \quad \\
SPT-CL J2359-5009 & 359.921 & -50.16 & 6.35 & 0.76 & - & 57.9 & 713 & 1.33 & 0.78 & \checkmark \\
\enddata
\end{deluxetable*}

The exposure times and resulting source counts from the {\sl Chandra} and {\sl XMM-Newton} 
observations are listed in Table \ref{obstable}. We also list the cluster coordinates and 
SZ signal-to-noise as described in V10 as well as the optical redshifts and the galactic 
absorbing hydrogen column from the Leiden-Argentine-Bonn survey \citep{kab05}. 
Source counts are quoted within $0.5 r_{500}$ which is estimated from the $Y_{X}$ parameter 
and the $M_{500}$-$Y_X$ relation (see \S \ref{sec:yxresults}). 

The {\sl Chandra} data were reduced using CIAO 4.1 and CALDB 4.1.3. All data were taken 
with the ACIS-I nominal aimpoint in VFAINT telemetry mode and additional screening to reject 
particle background was applied. 
To remove periods of flaring background we extracted point-source subtracted lightcurves in the 
0.3-12 keV band and filtered these using a $3 \sigma$ threshold. 
{\sl XMM-Newton} data were reduced using SAS 9.0 and reprocessed. Source-free light curves 
were generated in hard (MOS:10-12 keV, pn:12-14 keV) and soft (MOS:0.3-10 keV, pn:0.3-10 keV) energy bands 
separately and a $3 \sigma$ cut was applied to remove periods of high background. 

\subsection{Data analysis methods}
For the X-ray observables, $T_X$, $M_g$ and $Y_X$ we use scaling relations from 
\citet{vikh09} iteratively to determine the $r_{500}$ radius where the observable is measured. 

The $r_{500}$ radii used for the measurement of $T_X$ were estimated using 
the $M$-$T_X$ relation in \citet{vikh09} from samples of local clusters 
with deep exposures for which hydrostatic masses could be determined:
\begin{eqnarray}
\nonumber M_{500} & = & (3.02 \pm 0.11) ~ 10^{14} ~ h^{-1} ~ M_{\odot} \\
& \times & \left( \frac{kT}{5~\mathrm{keV}} \right)^{(1.53 \pm 0.08)} ~ E(z)^{-1} 
\label{eq500}
\end{eqnarray}
where
\begin{eqnarray}
\nonumber E(z) & = & H(z)/H_0 \\
& = & \sqrt{(1+z)^2  (1+\Omega_M z) - z(2+z)\Omega_\Lambda}.
\end{eqnarray}
The radius is then defined as 
\begin{equation}
r_{500} =  \left( \frac{3 M_{500}}{4 \pi 500 \rho_{crit}(z)}\right)^{1/3}.
\end{equation}

Similarly, in the estimation of the gas mass, $M_{g}$, described below, 
we use the gas-fraction relation 
\begin{eqnarray}
\nonumber f_{g,500} & = & (0.0764\pm0.004) ~ h^{-1.5} \\ 
& + & (0.0225\pm0.004) ~ h^{-1.5} ~ \log M_{15} 
\label{eqfgas}
\end{eqnarray}
where $f_{g,500}$ is the gas mass fraction within $r_{500}$, $f_{g,500}=M_{g}/M_{500}$, and 
$M_{15}$ is the total mass, $M_{500}$, in units of 
$10^{15}~h^{-1}~M_{\sun}$\footnote{This equation contains a typo in \citet{vikh09}, Table 3.}. 
With this relation we explicitly take 
into account the observed trend of $f_{g,500}$ with cluster mass \citep[e.g.,][]{vikh09}. 
The gas mass estimation is much more dependent on the aperture radius than the estimate of 
temperature and $M_{g}$ must be estimated iteratively to obtain a self-consistent 
result. Given a gas mass, we use equation \ref{eqfgas} to determine the total mass and, hence, $r_{500}$. 

For the estimation of the $Y_{X}=M_{g} \times T_{X}$ parameter we analogously estimate the aperture 
radius through the determination of  
\begin{eqnarray}
\nonumber M_{500} & = & (5.77\pm0.20) ~ 10^{14} ~ h^{1/2} M_{\odot} \\
& \times & \left( \frac{Y_X}{3 \times 10^{14} ~ M_{\odot} \mathrm{keV}} \right)^{(0.57\pm0.03)} ~ E(z)^{-2/5}, 
\label{eqYgas}
\end{eqnarray}
also from \citet{vikh09}, and iteratively determine $Y_X$ and the total mass within 
$r_{500}$ (See \S \ref{sec:yxresults} below). 

For each cluster, the center of the X-ray emission is determined using a centroid calculation. 
X-ray point sources were excluded in this process. The distance between the X-ray centroid and 
the SPT position is listed in Table \ref{obstable}.
The SZ and X-ray positions given are both centroid measurements. However, due to the different weighting of their
signals on the ICM density and temperature, the centroids are not expected to agree for a cluster which is not
azimuthally symmetric.
For {\sl Chandra} data, point sources were identified using the CIAO tool {\sl wavdetect} 
and removed from subsequent analysis. In the {\sl XMM-Newton} data, point sources were 
detected with the SAS task {\sl edetect chain}. 
Additionally, extended secondary maxima were identified and removed 
following the method prescribed in \citet{vikh09}. Extended substructures were included in the 
estimation of the total cluster luminosity. 

\subsection{Spectral analysis}
\label{specsec}
Spectra and response files were generated from the {\sl Chandra} data for the spectral extraction 
regions using {\sl specextract}. We use two independent methods of background subtraction in this 
work and compare the results. 
Due to the limited spatial extent of most of the SPT candidates it is possible to use in-field 
background subtraction. 
Background was extracted from regions near the source but outside of the $r_{500}$ radius 
where the source flux is a small fraction of the background flux. Since the nominal pointing was chosen to be offset 
from the cluster position it 
is possible to extract the background at a detector coordinate where the mirror effective area is similar to that of 
the source extraction region and where the particle-induced background is similar. 
This limits complications related to under-subtracted X-ray background which can be a problem when using in-field 
background.
As a second background subtraction method 
we also use the latest blank field backgrounds collected by M. Markevitch\footnote{http://cxc.harvard.edu/contrib/maxim/acisbg}, 
re-project these on the sky to match our cluster datasets and normalize the exposure times to match the  
count rates in the $9.5$-$12$ keV band. The background spectra can then be subtracted from 
the same detector region as the source.  
We compare the temperatures and flux estimates from these two techniques and find good agreement. 
Since the blank-field background subtraction provides better signal-to-noise, we 
use this method throughout as this allows for temperature estimates within 
the $r_{500}$ aperture for most objects.

For the {\sl XMM-Newton} observations we use the spectral analysis as described 
in \citet{andersson04} and \citet{andersson09}. For this work we limit the analysis 
to use in-field background since we are only interested in obtaining one spectrum 
for each cluster. 
Spectra for both {\sl Chandra} and {\sl XMM-Newton} datasets were extracted inside $r_{500}$, excluding 
the central emission within $0.15 r_{500}$ 
to avoid the effects of cool cores, known to cause additional scatter in X-ray scaling 
relations. We model the data using a MEKAL model for the 
thermal plasma with emission lines and a WABS absorption model. 
For our sample, the data are not deep enough to measure metal abundances reliably and we fix the 
abundance to 0.3 solar. The hydrogen equivalent absorbing column, $n_H$, was 
fixed at the weighted average from the Leiden-Argentine-Bonn survey \citep{kab05}. These are 
listed in Table \ref{obstable}.
The best fit temperatures are shown in Table \ref{proxytable}.

\subsection{X-ray imaging}
\label{imagingsec}
The X-ray surface brightness is extracted in $n$ concentric annuli defined by 
\begin{equation}
\label{eq:binning}
r_i = \left(i  \times \frac{{r_{max}}^{1/c}}{n} \right)^c \quad \quad i=1\cdots n, 
\end{equation}
where we have used $r_{max}=1.5~r_{500}$, $c=1.2$ and $n=20$.
The limiting radius $r_{max}$ is chosen to be large enough so that confusion between 
background and cluster flux is negligible and so that the integrated gas density 
is not overestimated due to projection effects. The values of $c$ and $n$ are chosen 
to balance high central resolution with achieving a similar number of X-ray 
counts per radial bin.

For every annular region the average exposure is calculated, taking into account 
bad pixels and chip gaps. Similarly, the average effective mirror area is calculated 
for each region taking into account the mirror vignetting. 
The radial model for the X-ray counts includes a component for the cluster X-ray surface 
brightness, a spatially flat unvignetted component for the particle induced X-ray 
background and soft X-ray background plus an additional spatially flat vignetted 
component taking into account under-subtraction of the soft X-ray background flux due to 
mirror vignetting. This last component is always small. 

The cluster surface brightness is taken to be proportional to the integrated 
emission measure, $EM= \int n_e n_p dV$, which is a good approximation in the 
$0.5 - 2$ keV band for gas temperatures present in our sample. 
The inferred density distribution has been shown previously to have little 
dependence on the temperature when determined from the surface brightness in 
this energy band \citep{mohr99,forman07}.
The radial gas density distribution is modeled using a modified $\beta$-model, 
\begin{equation}
\label{n2model}
n_p n_e = n_0^2 \frac{(r/r_c)^{-\alpha}}{(1+r^2/r_c^2)^{3\beta-\alpha/2}} \quad \frac{1}{(1+r^\gamma/r_s^\gamma)^{\epsilon/\gamma}},
\end{equation}
\citep{vikh06}. The model accounts for the cuspy centers of clusters as well as the steepening of 
the profile seen at larger radii. We fix $\gamma = 3$ in the above expression \citep[see ][]{vikh06}.

For the {\sl XMM-Newton} data, the projected model is convolved with a model of the 
{\sl XMM} point-spread function (PSF), as described in  \citet{ghizzardi01}, prior to fitting. 
The model is normalized by integrating equation \ref{n2model} over a cylindrical volume 
with a radius of $r_{500}$, truncated at $3 r_{500}$ along the line-of-sight.  
This is compared to the spectral normalization of a MEKAL model with parameters derived 
in \S \ref{specsec} within $r_{500}$. 
This way the model normalization, $n_0$, can be calculated using the angular distance $D_A$ 
determined from the redshift listed in Table \ref{obstable}.
The gas mass $M_{g,500}$ is calculated by setting $n_e = Z n_p$ and $\rho_{g} = m_p n_e A/Z$ 
where $Z=1.199$ and $A=1.397$ are the average nuclear charge and mass respectively for a 
plasma with $0.3$ solar abundances (implying a mean molecular weight per free electron $\mu_e = 1.165$) 
assuming abundances by \citet{anders89}. 
The gas density $\rho_g$ is integrated over a sphere of radius $r_{500}$ determined iteratively 
for $M_{g}$ and $Y_{X}$ estimates using $M_{500}-M_{g}$ and $M_{500}-Y_{X}$ relations, 
respectively.

The parameter space of the model is explored via Markov chain Monte-Carlo iteration 
and the uncertainties on the parameters are estimated using the Markov chain posterior. 

\subsubsection{Cluster images}
X-ray images in the $0.5$ - $2.0$ keV band are generated using the filtered event files and binned 
into $4\arcsec \times 4\arcsec$ pixels. These are shown in Figures \ref{0000app}-\ref{2359app} 
alongside optical images of the clusters from \citet{high10}. Images are smoothed with a $8 \arcsec$ 
Gaussian filter. 
The images are shown with SZ S/N contours from V10 and the white crosses show the location of 
the brightest cluster galaxy (BCG). 
In all cases, the SPT detection is within $1 \arcmin$ of the X-ray centroid and in 
a majority of cases (9 of 15) it is within $20 \arcsec$. 
Additionally, Sydney University Molonglo Sky Survey \citep[SUMSS,][]{sumss} radio sources are marked 
in the images with small circles ($15\arcsec$ radius). 
Extended X-ray structures are marked with larger circles ($30\arcsec$ radius) or arrows and are 
discussed for each individual cluster in the Appendix.

\subsection{ICM temperature, gas mass, luminosity and $Y_X$}
\label{sec:yxresults}
The average spectral temperature of the intra-cluster medium is measured within the [0.15-1]$r_{500}$ aperture  
as described in \S \ref{specsec}. A new value of $r_{500}$ is then estimated using this temperature with 
equation \ref{eq500} and a new spectrum is extracted using the new radius. The process is repeated 
until the value of $T_X$ has converged. 
For 5 of the clusters, the number of source photons within [0.15-1]$r_{500}$ is less than 1000 and the 
signal-to-noise is less than 20. This leads to large uncertainties in the spectral fits and could 
potentially cause systematic biases in the temperature. For these objects, we instead use the estimate 
of $T_{X}$ within [0.15-0.5]$r_{500}$, where the signal-to-noise is higher and extrapolate using the fitted 
relation between the two temperatures from \citet{vikh09}:
\begin{eqnarray}
\nonumber T_{X,(0.15-1) r_{500}}/ T_{X, (0.15-0.5) r_{500}}  = \\
0.9075 + 0.0625~T_{X, (0.15-0.5) r_{500}}. 
\label{eqTxTx2}
\end{eqnarray}
We also add a 10\% uncertainty to these temperature estimates to account for the uncertainty 
in this relation. The clusters for which this procedure was performed are marked in Table \ref{proxytable}. 
For SPT-CL J0551-5709, a cluster which is spatially coincident with the local cluster 
Abell S0552, we also perform this procedure, but with a different motivation. 
The surface brightness to the south and north-west of this cluster suggests that there is 
a significant amount of flux from cooler gas associated with the Abell cluster. 
This is also seen in the temperature estimates. $T_{X}$ drops from $4.4\pm0.7$ keV to $3.3\pm0.4$ keV 
when comparing the inner and outer apertures. 
We include SPT-CL J0551-5709 in the scaling relation fits using the corrected temperature  
and mark is with a red square in the scaling relation plots.

Similarly, the gas mass within $r_{500}$, $M_{g,500}$, is estimated as described in \S  
\ref{imagingsec} using an initial estimate of $r_{500}$ from equation \ref{eq500} and the 
spectrally derived temperature, $T_X$. The estimate of $r_{500}$ is then revised using 
the implied gas mass fraction and equation \ref{eqfgas}, and the process 
is repeated until the gas mass converges. 

In this work, we adopt redshift evolution of $f_{g,500}$, in the same way as it was  
applied in the work of \citet{vikh09}. 
The redshift dependence of the gas-mass fraction is not well constrained observationally
to high-$z$ and independent measurements of the mass are needed to study
this further.
The $f_{g,500}$-$M$ relation is used here only to determine the radius $r_{500}$ within which 
to estimate $M_{g}$ and has no impact on the $Y_{X}$-based mass estimates.
We do not quote any total masses based on the $f_{g,500}$-$M$ relation in this work.

SPT-CL J0516-5430 has an unusually extended morphology compared to other objects in the 
sample and the assumption of spherical symmetry is very approximate. 
The line-of-sight dimension for this cluster and 
the inferred gas mass are likely to be overestimated for this reason \citep[see, e.g.][]{nag07}. 
To keep the analysis analogous to \citet{vikh09} we do not attempt to correct for 
this here. SPT-CL J0516-5430 is marked with a blue triangle in the scaling relation plots. 

When estimating $Y_X$, we use equation \ref{eqYgas} to determine $r_{500}$ and re-calculate both 
$T_X$ and $M_g$ in a similar way.  Cluster simulations show that $Y_X$ exhibits less scatter 
with mass compared to $M_{g}$ and $T_X$ individually due to the typical anti-correlation of 
deviations from the mean of these two observables \citep{kra06}. 
This makes the $Y_{X}$ mass 
estimator less dependent on the dynamical state of clusters, as far as it can be estimated from simulations. 
The X-ray observables are listed in Table \ref{proxytable}. 

For some of the clusters in the sample, the X-ray data are not deep enough to detect the X-ray emission 
with high significance out to $r_{500}$. 
For five of the clusters, the signal-to-noise in our radial bins (equation \ref{eq:binning}) drops below 3 
at a radius of about $0.6~r_{500}$. These are the same five objects with low signal-to-noise spectra 
described above.
The gas density profile is then constrained primarily by the central surface brightness profile 
and extrapolated out to $r_{500}$. This leads to systematic uncertainties that are difficult to account for 
as they depend on cluster morphology. 
In our sample we find a typical density slope $\rho_g \propto r^{\alpha}$ of $\alpha=-1.90\pm0.30$ 
at $0.6~r_{500}$, where we quote the mean and standard deviation. 
Varying the slope outside of this radius by $\pm0.30$ typically changes 
the mass outside $0.6~r_{500}$ by 8\%.  This is also an over-estimate of the uncertainty since 
density profiles generally become steeper with increasing radius out to $r_{500}$. To be conservative, we add 
an uncertainty of 10\% on the total gas mass for clusters with poor signal-to-noise outside $0.6~r_{500}$.

\begin{deluxetable*}{llrrrr}
\tabletypesize{\scriptsize}
\tablecaption{X-ray observables $T_{X,500}$, $M_{g,500}$, $Y_{X,500}$ and $L_{X,500}(0.5-2.0~\mathrm{keV})$}
\tablewidth{0pt}
\tablehead{
\colhead{Name} &
\colhead{$z$} &
\colhead{$T_{X,500}$} & 
\colhead{$M_{g,500}$} &
\colhead{$Y_{X,500}$} &
\colhead{$L_{X,500}$} \\
\colhead{\quad} &
\colhead{\quad} &
\colhead{keV} &
\colhead{$10^{13}$ $M_\sun$} &
\colhead{$10^{14}$ $M_\sun$ keV} &
\colhead{$10^{44}$ erg s$^{-1}$} \\
}
\startdata
J0000-5748\footnotemark[a] & 0.74 & $8.6^{+3.8}_{-2.3}$ & $4.7^{+0.5}_{-0.6}$ & $4.7 \pm 1.8$ &  $5.7 \pm 0.4 $ \\
J0509-5342 & 0.4626 & $7.0^{+1.4}_{-1.1}$ & $5.6^{+0.2}_{-0.2}$ & $4.3 \pm 0.8$ &  $3.2 \pm 0.1 $ \\
J0516-5430 & 0.2952 & $9.8^{+1.7}_{-1.2}$ & $17.0^{+0.4}_{-0.4}$ & $15.9 \pm 2.4$ &  $4.7 \pm 0.1 $ \\
J0528-5300\footnotemark[a] & 0.7648 & $5.2^{+3.5}_{-1.7}$ & $2.9^{+0.4}_{-0.4}$ & $1.7 \pm 0.9$ &  $1.9 \pm 0.2 $ \\
J0533-5005\footnotemark[a] & 0.8810 & $4.0^{+1.9}_{-1.2}$ & $2.1^{+0.5}_{-0.4}$ & $0.9 \pm 0.4$ &  $1.2 \pm 0.3 $ \\
J0546-5345 & 1.0665 & $7.5^{+1.7}_{-1.1}$ & $7.3^{+0.4}_{-0.3}$ & $5.3 \pm 1.0$ &  $6.4 \pm 0.4 $ \\
J0551-5709\footnotemark[b] & 0.4230 & $4.1^{+0.9}_{-0.7}$ & $5.1^{+0.3}_{-0.4}$ & $2.0 \pm 0.4$ &  $1.9 \pm 0.2 $ \\
J0559-5249 & 0.6112 & $7.7^{+1.1}_{-0.8}$ & $8.3^{+0.3}_{-0.2}$ & $6.1 \pm 0.8$ &  $3.3 \pm 0.2 $ \\
J2331-5051 & 0.5707 & $5.9^{+1.3}_{-0.8}$ & $5.7^{+0.2}_{-0.2}$ & $3.5 \pm 0.6$ &  $4.4 \pm 0.2 $ \\
J2332-5358 & 0.32 & $7.4^{+1.2}_{-0.7}$ & $5.6^{+0.2}_{-0.2}$ & $4.4 \pm 0.6$ &  $3.0 \pm 0.1 $ \\
J2337-5942 & 0.7814 & $8.9^{+2.0}_{-1.4}$ & $9.5^{+0.4}_{-0.6}$ & $8.5 \pm 1.7$ &  $8.9 \pm 0.5 $ \\
J2341-5119 & 0.9983 & $8.0^{+1.9}_{-1.6}$ & $5.6^{+0.2}_{-0.2}$ & $4.7 \pm 1.0$ &  $5.6 \pm 0.3 $ \\
J2342-5411\footnotemark[a] & 1.08 & $5.0^{+0.9}_{-0.8}$ & $2.6^{+0.3}_{-0.3}$ & $1.4 \pm 0.3$ &  $2.9 \pm 0.3 $ \\
J2355-5056 & 0.35 & $5.5^{+1.0}_{-0.8}$ & $4.4^{+0.2}_{-0.1}$ & $2.6 \pm 0.4$ &  $2.1 \pm 0.1 $ \\
J2359-5009\footnotemark[a] & 0.76 & $6.4^{+2.3}_{-1.8}$ & $2.8^{+0.3}_{-0.3}$ & $2.2 \pm 0.7$ &  $1.6 \pm 0.2 $ \\

\enddata
\footnotetext[a]{Low signal-to-noise within the [$0.15$-$1$]$r_{500}$ aperture. $T_{X,500}$ is estimated using temperature within [$0.15$-$0.5$]$r_{500}$ and scaled using equation \ref{eqTxTx2}.}

\footnotetext[b]{SPT-CL J0551-5709 is coincident with Abell S0552. The gas temperature is estimated through equation \ref{eqTxTx2}.} 
 
\label{proxytable}
\end{deluxetable*}

\section{Total mass estimates}
\label{sec:grav}
Using the mass proxies $T_X$ and $Y_X$, we estimate the gravitational 
mass, $M_{500}$ for the clusters using equations \ref{eq500} and \ref{eqYgas}. 
The mass estimates are listed in Table \ref{masstab}.
Clusters are identified as mergers based on their morphology. Clusters with secondary 
maxima, filamentary structure or significant isophotal centroid shifts are classified 
as unrelaxed \citep[e.g.,][]{mohr93}. As discussed in \citet{kra06}, the $T_X$ based masses should be 
multiplied by a factor $1.17$ for clusters identified as mergers. 
Following \citet{vikh09} we correct our masses upward with this factor and add 
an uncertainty of $5$\% on the masses (Table \ref{masstab}). 
Our merger classification is listed in Table \ref{obstable}.

In Table \ref{masstab}, we also give the SZ-derived mass estimates from V10.  In V10, 
the SPT cluster survey was used to constrain cosmological parameters while 
simultaneously fitting a cluster detection significance-mass relation.  The significance-mass 
relation had priors imposed on its normalization, slope, and redshift evolution that 
were motivated by simulated thermal SZ maps of the sky.  These maps were generated from 
large dark matter simulations \citep{shaw09} that used a semi-analytic gas model of \citet{bode07} 
which was calibrated such that the simulated clusters matched observed X-ray scaling 
relations for low redshift ($z < 0.25$) clusters.  
The maximum likelihood significance-mass relation was then used to generate mass estimates 
for each cluster, including both statistical 
and systematic uncertainties. 
These systematic uncertainties were equivalent to a $\sim$15\% uncertainty on the mass 
estimate for each cluster.  

Comparing the X-ray and SZ-based mass estimates, we find that the SZ-derived 
masses are lower by a factor of $0.83 \pm 0.06$, 
where we use only the statistical uncertainties to quote the uncertainty on the average ratio 
from a fit to a scaling relation between the two with the slope fixed to 1.  
Two clusters are 2$\sigma$ inconsistent with unity and both have larger X-ray mass estimates; 
SPT-CL J0516-5430 and SPT-CL J0546-5345.  V10 notes that clusters at $z \lesssim 0.3$ will have mass estimates 
biased low because the power-law scaling that is assumed for the significance-mass 
relation does not fully capture the effects of CMB-confusion on the SZ signal.  This effect 
could possibly explain the relatively low SZ-inferred mass found for SPT-CL J0516-5430, which 
is the only cluster in this sample at $z < 0.3$.   Leaving this cluster out, the average ratio of the SZ- to 
the X-ray-mass is $0.89 \pm 0.06$.  

Overall, we consider the agreement between the X-ray and SZ mass estimates reasonable
given the $\sim15$\% systematic uncertainty on the mass estimates quoted in V10. However, there is 
some evidence that the SZ masses in V10 are biased low by $\sim$10\%.  

This difference could have several explanations. For example, the
semi-analytic gas model used to calibrate
the simulations in V10 could have a redshift evolution that differs from self-similar
evolution. 
There could also be differences in
the gas profiles at large radii that cause systematic differences between
the measured $Y_X$-based mass and the derived mass from the $M$-$\xi$
scaling in V10 that were not included in the simulations.  It should also
be noted that the $M_{500}$-$Y_{X}$ relation used here has been calibrated
using X-ray hydrostatic mass estimates.  

Numerical simulations have suggested that hydrostatic mass estimates could be 
biased low by $\sim$10\% due to additional non-thermal pressure 
support of the gas in clusters \citep[e.g.,][]{lau09, men09}.  However, 
comparisons of hydrostatic and weak lensing mass estimates of low redshift 
($z < 0.3$) clusters show agreement at the $<$9\% level and suggest no 
significant bias \citep[e.g.,][]{zhang08,vikh09,zhang10}.
Additionally, the $M_{500}$-$Y_{X}$ relation assumes self-similar 
redshift scaling. The details of redshift-dependent departures 
from this scaling are not constrained observationally, although 
the results from simulations provide a 5\% upper limit for any 
evolution of the amplitude of this relation between $z=0$ and $0.6$.

Regardless of the exact reasons, if the SZ mass estimates are biased
low, there would be consequences for the cosmological constraints in V10.
It would generally favor a lower normalization of the SZ
significance-mass relation and hence a larger value of $\sigma_8$; however, 
these results should still be 
within the 68\% confidence region in V10 due to the priors imposed on the parameters defining the SZ
significance-mass relation which effectively include the 15\% systematic uncertainty 
for the masses.
The cosmological implications of including the X-ray mass
estimates with the V10 results will be the subject of a future paper.

\begin{deluxetable*}{llrrrr}
\tabletypesize{\scriptsize}
\tablecaption{Cluster masses from $M_{500}-Y_X$ and  $M_{500}-T_X$ relations \label{masstab}}
\tablewidth{0pt}
\tablehead{
\colhead{Name} &
\colhead{$z$} &
\colhead{$r_{500}$\footnotemark[a]} & 
\colhead{$M_{500,Y_X}$} &
\colhead{$M_{500,T_X}$} &
\colhead{$M_{500,SZ,V10}$\footnotemark[b]} \\
\colhead{} &
\colhead{} &
\colhead{$\mathrm{kpc}$} & 
\colhead{$10^{14} M_{\sun}$} & 
\colhead{$10^{14} M_{\sun}$} & 
\colhead{$10^{14} M_{\sun}$} \\
}
\tablecomments{Masses estimated from X-ray mass-proxy relations in equations \ref{eq500} and \ref{eqYgas}. SZ derived masses from V10 are shown for comparison.}
\startdata

J0000-5748 & 0.74 & $ 950 $ & $ 5.32 \pm 1.16 $ & $ 6.74 \pm 5.14 $ & $ 2.89 \pm 0.61 \pm 0.41 $ \\
J0509-5342 & 0.4626 & $ 1062 $ & $ 5.43 \pm 0.60 $ & $ 6.71 \pm 3.40 $ & $ 4.26 \pm 0.74 \pm 0.60 $ \\
J0516-5430 & 0.2952 & $ 1463 $ & $ 11.84 \pm 1.25 $ & $ 12.34 \pm 2.13 $ & $ 6.48 \pm 0.95 \pm 1.13 $ \\
J0528-5300 & 0.7648 & $ 775 $ & $ 2.97 \pm 0.89 $ & $ 3.05 \pm 4.00 $ & $ 2.83 \pm 0.60 \pm 0.38 $ \\
J0533-5005 & 0.8810 & $ 656 $ & $ 2.06 \pm 0.53 $ & $ 2.25 \pm 1.71 $ & $ 2.71 \pm 0.56 \pm 0.37 $ \\
J0546-5345 & 1.0665 & $ 840 $ & $ 5.33 \pm 0.62 $ & $ 5.25 \pm 1.13 $ & $ 3.25 \pm 0.51 \pm 0.44 $ \\
J0551-5709 & 0.4230 & $ 948 $ & $ 3.56 \pm 0.43 $ & $ 3.00 \pm 0.85 $ & $ 4.10 \pm 0.75 \pm 0.58 $ \\
J0559-5249 & 0.6112 & $ 1043 $ & $ 6.40 \pm 0.54 $ & $ 7.07 \pm 1.49 $ & $ 5.03 \pm 0.74 \pm 0.70 $ \\
J2331-5051 & 0.5707 & $ 972 $ & $ 4.70 \pm 0.51 $ & $ 4.91 \pm 1.36 $ & $ 4.63 \pm 0.73 \pm 0.66 $ \\
J2332-5358 & 0.32 & $ 1134 $ & $ 5.66 \pm 0.48 $ & $ 6.69 \pm 1.08 $ & $ 5.19 \pm 0.85 \pm 0.83 $ \\
J2337-5942 & 0.7814 & $ 1046 $ & $ 7.43 \pm 0.90 $ & $ 8.10 \pm 2.18 $ & $ 6.32 \pm 0.84 \pm 0.97 $ \\
J2341-5119 & 0.9983 & $ 847 $ & $ 5.06 \pm 0.66 $ & $ 5.20 \pm 1.69 $ & $ 4.05 \pm 0.58 \pm 0.63 $ \\
J2342-5411 & 1.08 & $ 647 $ & $ 2.47 \pm 0.32 $ & $ 2.39 \pm 0.63 $ & $ 2.65 \pm 0.50 \pm 0.37 $ \\
J2355-5056 & 0.35 & $ 1014 $ & $ 4.18 \pm 0.43 $ & $ 4.26 \pm 1.57 $ & $ 4.17 \pm 0.80 \pm 0.63 $ \\
J2359-5009 & 0.76 & $ 816 $ & $ 3.45 \pm 0.67 $ & $ 4.97 \pm 2.61 $ & $ 3.32 \pm 0.60 \pm 0.46 $ \\

\enddata
\footnotetext[a]{Estimated using the $M_{500}-Y_X$ relation (equation \ref{eqYgas}).}
\footnotetext[b]{Quoted masses from V10 include statistical and systematic 
uncertainties and have been scaled with our adopted $h$.}
\end{deluxetable*}

\section{SZ data analysis}
\label{sec:sze}

\subsection{Calculating $Y_{SZ}$}
\label{sec:calcsz}
The SPT observations, data reduction, and map-making used in this work are identical to V10, and are outlined 
in \S \ref{sec:spt}.  
The analysis differs from V10 at the point where the SPT maps are spatially filtered to identify the cluster 
candidates.   In our case, the cluster candidates have already been identified, and we instead 
want to calculate the SZ-inferred integrated Comptonization, $Y_{SZ}$, of each cluster.  We 
calculate $Y_{SZ}$ from the SPT 150~GHz maps by spatially filtering them using a filter 
motivated by the X-ray measurements of each cluster.  

In V10, cluster candidates are identified by spatially filtering 
the SPT maps with a matched 
filter \citep{haehnelt96, melin06}.  This is done in the Fourier domain by 
multiplying the map by 
\begin{equation}
\label{eqn:filter}
\psi(k_x,k_y) = \frac{B(k_x,k_y) S(|\vec{k}|)}{B(k_x,k_y)^2 N_{astro}(|\vec{k}|) + N_{noise}(k_x,k_y)}
\end{equation}
where $\psi$ is the matched filter, $B$ is the response of the
SPT instrument after timestream processing to signals on the sky,
$S$ is the assumed source template, and the noise power has been broken into
astrophysical ($N_{astro}$) and noise ($N_{noise}$) 
components.  The noise covariance matrix $N_{noise}$ includes contributions from atmospheric
and instrumental noise, while $N_{astro}$ includes power from primary and lensed
CMB fluctuations, an unresolved SZ background, and unresolved point sources.
In V10, the source template, $S$, was constructed from a projected spherical $\beta$-model. 

In this work, we instead use a SZ source template motivated from X-ray measurements of each 
cluster.  The SZ brightness is proportional to the line-of-sight integral of electron pressure, or 
density times temperature.   The profile is assumed to match the product of the best-fit 
gas density profile to the X-ray measurements of each cluster (equation \ref{n2model}), and the  
universal temperature profile of \citet{vikh06} (equation \ref{universalT}).  These profiles are multiplied 
together to give the radial pressure profile, and projected onto the sky by doing a line-of-sight 
integral through the cluster.  The radial pressure profile is truncated at $3 \times r_{500}$, 
where $r_{500}$ is estimated using equation \ref{eqYgas} and given in Table \ref{masstab}.  
In constructing the spatial filter, we only need to know the spatial shape of the source, so 
the SZ model is normalized to unity.  

For each of the fifteen clusters, we construct a spatial filter using equation \ref{eqn:filter} and 
a source model, as described above.  The spatially filtered SPT maps are a measure 
of the normalization of each source model at the cluster location.  Using the SPT maps alone to determine the 
cluster location would bias the $Y_{SZ}$ measurements high because of the freedom to maximize 
the SPT significance by position.  Therefore we use the X-ray measured 
position as a prior on the cluster location.  We define the uncertainty in the SPT normalization 
of the source model as the 
standard deviation of the spatially filtered map within a $90 \arcmin$ band in declination around the 
cluster.  The SPT maps are calibrated in units of $\mathrm{K}_{CMB}$, the equivalent CMB temperature 
fluctuation required to produce the same power fluctuation.  The SPT normalization 
is converted to Comptonization using the relation 
\begin{equation}
\label{eqn:filter2}
\Delta T = y \, T_{CMB} \left(x \frac{e^x +1}{e^x-1} - 4\right) [1 + \delta(x,T_e)],
\end{equation}
where $\Delta T$ is the measured decrement in units of $\mathrm{K}_{CMB}$, $y$ is the
Comptonization, $T_{CMB}$ is the CMB blackbody temperature of 2.725 K,
$x=h\nu/kT_{CMB}$, and $\delta(x,T_e)$ accounts for relativistic
corrections to the SZ spectrum \citep{itoh98, nozawa00}.
For the frequency dependent terms in equation \ref{eqn:filter2}, we calculate 
their SPT band averaged value, which would effectively correspond to an observing 
frequency, $\nu$, of 152.9 GHz.   The average SPT band
is measured from Fourier Transform Spectroscopy measurements of a sample
including more than 90\% of the SPT 150 GHz detectors. 

For the relativistic correction 
factor, $\delta(x,T_e)$, we assume an electron temperature of $T_{X,r500}/1.11$, where $T_{X,r500}$ 
is given in Table \ref{proxytable} and the factor of 1.11 is the average 
ratio of the X-ray spectroscopic and mass-weighted temperature measured in 
\citet{vikh06}.  The relativistic correction factor is only a weak function of temperature, between 
3 to 10 keV it varies by $\sim4$\%, and we expect this assumption to negligibly affect our results.  

The integrated Comptonization, $Y_{SZ,cyl}$, is calculated for each cluster
by integrating its source model over solid angle, $Y_{SZ,cyl} \propto \int y(\theta) d\Omega$, 
normalized to the best-fit SPT Comptonization, $y$.  
To more easily compare to the X-ray measurements, we convert our measurements to 
units of $M_\sun \mathrm{keV}$, and define $Y_{SZ,cyl}$ as, 

\begin{equation}
Y_{SZ,cyl} = \left(\frac{\mu_e m_p m_e c^2}{\sigma_T}\right) D_A^2 \int y(\theta) d\Omega, 
\end{equation}
where $D_A$ is the angular distance to the source, $\sigma_T$ is the Thomson cross-section, 
$m_e$ is the electron mass, $m_p$ is the proton mass, $c$ is the speed of light, 
$\mu_e$ is the mean molecular weight per free electron. 
In Table \ref{ysztable}, we give $Y_{SZ,cyl}$ integrated out to the angular radius corresponding 
to $r_{500}$ as determined from the X-ray measurements and given in Table \ref{masstab}. 
The uncertainty in $Y_{SZ}$ is calculated as the quadrature sum of the uncertainty in the SPT calibration, 
the measured SPT normalization of the source model, and an additional uncertainty in the 
assumed source model, which will be discussed in \S \ref{sec:model_uncertainty}.

\begin{deluxetable}{llrr}
\tabletypesize{\scriptsize}
\tablecaption{Measured cylindrical and spherical $Y_{SZ}$}
\tablewidth{0pt}
\tablehead{
\colhead{Name} &
\colhead{$z$} &
\colhead{$Y_{SZ,cyl,500}$} &
\colhead{$Y_{SZ,sph,500}$} \\
\colhead{\quad} &
\colhead{\quad} &
\colhead{$10^{14}$ $M_\sun$ keV} &
\colhead{$10^{14}$ $M_\sun$ keV} \\
}
\startdata
J0000-5748 & 0.74 & $3.0 \pm 0.6$ & $2.4 \pm 0.5$ \\
J0509-5342 & 0.4626 & $3.9 \pm 0.8$ & $3.1 \pm 0.6$ \\
J0516-5430 & 0.2952 & $15.3 \pm 2.4$ & $11.6 \pm 1.8$ \\
J0528-5300 & 0.7648 & $2.6 \pm 0.4$ & $2.1 \pm 0.4$ \\
J0533-5005 & 0.8810 & $1.8 \pm 0.5$ & $1.5 \pm 0.4$ \\
J0546-5345 & 1.0665 & $4.6 \pm 0.8$ & $3.5 \pm 0.6$ \\
J0551-5709 & 0.4230 & $3.7 \pm 0.8$ & $2.6 \pm 0.6$ \\
J0559-5249 & 0.6112 & $7.6 \pm 1.1$ & $5.8 \pm 0.9$ \\
J2331-5051 & 0.5707 & $5.0 \pm 0.8$ & $3.6 \pm 0.6$ \\
J2332-5358 & 0.32 & $5.6 \pm 1.0$ & $4.9 \pm 0.9$ \\
J2337-5942 & 0.7814 & $8.3 \pm 1.0$ & $7.1 \pm 0.9$ \\
J2341-5119 & 0.9983 & $4.9 \pm 0.7$ & $4.1 \pm 0.6$ \\
J2342-5411 & 1.08 & $2.3 \pm 0.5$ & $1.9 \pm 0.4$ \\
J2355-5056 & 0.35 & $2.6 \pm 0.6$ & $2.2 \pm 0.5$ \\
J2359-5009 & 0.76 & $2.1 \pm 0.6$ & $1.7 \pm 0.5$ \\
\enddata
\label{ysztable}
\end{deluxetable}

\subsection{Spherical Deprojection}
\label{sec:deprojection}

Both the X-ray and SZ observations are measuring a projected 
signal that is proportional to the integrated gas properties in a line 
of sight through the cluster.  This projection has different physical dependencies between X-ray and 
SZ observations that must be considered.  To ease this comparison 
we deproject each measurement so that they correspond to 
a physical observable that is proportional to a spherical volume integral of each signal.  

The $Y_X$ estimates are deprojected as described in \S \ref{imagingsec}.  For X-rays, the 
effect of the deprojection is smaller than for the SZ, because the X-ray flux is 
proportional to $n_e^2$ and only weakly dependent on temperature.  
This effect decreases the contribution to the X-ray signal from 
large radii where the density is lower.  Also, $Y_X$ will be proportional to a X-ray spectroscopic weighted 
temperature from gas between $0.15 r_{500}$ and $r_{500}$, while $Y_{SZ}$ is related to the 
gas mass weighted temperature.  We do not formally 
account for this difference in calculating either $Y_X$ or $Y_{SZ}$. However later in 
this section we will discuss this effect in interpreting their comparison.  

We calculate $Y_{SZ,sph}$ by spherically deprojecting the $Y_{SZ,cyl}$ measurements in Table \ref{ysztable}.  
The SZ signal is proportional to the line of sight integral of the 
electron density, $n_e$, times temperature, $T_e$.   For each cluster, we assume the density profile 
derived from the X-ray imaging analysis and assume a temperature profile of a universal form, 
\begin{equation}
T(r) = T_0 \frac{(x/0.045)^{1.9} + 0.45}{(x/0.045)^{1.9} + 1} \frac{1}{(1 + (x/0.6)^2)^{0.45}}, 
\label{universalT}
\end{equation}
where $x=r/r_{500}$ \citep[see][]{vikh06}.  
We then define $Y_{SZ,sph}$  as:
\begin{equation}
Y_{SZ,sph} = \frac{Y_{SZ,cyl}}{C} = Y_{SZ,cyl} \frac{\int_{sph} T(r)
n_e(r) dV}{\int_{cyl} T(r) n_e(r) dV},
\end{equation}
where $C$ is the ratio of the integrals of pressure in a cylindrical and
spherical volume through the cluster.  For these integrals we use an
integration radius of $r_{500}$ and truncate the density and temperature
radial models at $3 r_{500}$.   We note that varying the truncation radius
from between $3$-$5 r_{500}$ changes our measurement of $Y_{SZ}$ in Table
\ref{ysztable} by less than 1\%.  For the $r_{500}$ aperture, the median
and standard deviation of $C$ across the sample is $1.23 \pm 0.08$.  This
is consistent with the value of 1.203 that is calculated assuming the
universal pressure profile from \citet{arnaud09}.

As noted earlier in this section, even after the above spherical deprojection, $Y_X$ and $Y_{SZ,sph}$ are
not directly comparable because of their different weighting of the electron temperature.  $Y_X$ is 
proportional to the X-ray spectroscopic weighted temperature, $T_X$, while $Y_{SZ,sph}$ is proportional 
to the mass weighted temperature, $T_{mg}$.   The size of this difference has been estimated by several authors 
from X-ray measurements.  In \citet{vikh06}, they estimate $T_X/T_{mg} = 1.11$ for a sample of relaxed 
massive clusters with high angular resolution X-ray temperature measurements between 70 kpc and $r_{500}$.  This 
would imply that $Y_{SZ,sph}/Y_{X} = 1/1.11 \approx 0.90$.  In \citet{arnaud09}, a similar analysis 
was performed for a sample that included both relaxed and un-relaxed clusters, and it was estimated that 
$Y_{SZ,sph}/Y_{X} = 0.924 \pm 0.004$.  
For the measured density profiles and assumed temperature profiles for our
sample, we would expect an average ratio of $Y_{SZ,sph}/Y_{X} = 0.91 \pm 0.01$. 
To compute this expected ratio we have used the ``spectroscopic-like'' temperature 
\citep[e.g.,][]{mazzotta04}, given the density and temperature profiles used here. 

In \S \ref{sec:ycomparison} we compare the above ratios, predicted purely from X-ray 
observations, to the ratio including the $Y_{SZ,sph}$ calculated from the SZ data 
as described in \S \ref{sec:calcsz} and deprojected as described in this section.

\subsection{Model Uncertainty}
\label{sec:model_uncertainty}
 
The integrated Comptonization, $Y_{SZ}$, inferred from the SZ data depends on
the assumed spatial model through the construction of the spatial filter, and
the volume integral through the deprojection the SZ data.  For a typical cluster, 
the X-ray data constrains the cluster density profile with high 
signal-to-noise out to $\sim r_{500}/2$ with no information on 
the temperature profile.  Since a significant amount of the SZ signal is 
coming from larger radii than this, we would like to estimate
how much uncertainty our assumed profile is adding to the $Y_{SZ}$ estimates. 
To help do this, we calculate the $Y_{SZ}$ of each cluster assuming the universal pressure
profile of \citet{arnaud09}, variations of the temperature profile in 
equation \ref{universalT}, and an isothermal model.  

\citet{arnaud09} measure a universal pressure
profile from X-ray measurements for a representative set of local massive
clusters.  These measurements were deep enough to constrain the
cluster density and temperature profiles out to $r_{500}$ in each cluster.
They find that their sample is well fit by a universal pressure profile that is defined
only by $M_{500}$.  For each cluster in our sample, we re-calculate $Y_{SZ,sph}$ assuming the
universal pressure profile from \citet{arnaud09} and the X-ray measured 
values for $r_{500}$ and $M_{500}$ given in Tables \ref{ysztable} and \ref{masstab}, 
respectively.  Comparing these $Y_{SZ,sph}$ estimates to the values given in Table \ref{ysztable}, 
we find that this ratio has a mean and standard deviation of $1.01 \pm 0.15$ 
averaged over the sample.  There is one significant ($>20\%$) outlier, 
SPT-CL J0516-5430, whose $Y_{SZ,sph}$ deviates by a factor of 
0.59.  This cluster is a major merger, and the observed cluster profile is noticeably
poorly fit by the \citet{arnaud09} pressure profile, which does not capture the 
disturbed distribution of the central gas in this cluster. 
Leaving this cluster out of our comparison, the
average ratio becomes $1.03 \pm 0.10$.  Therefore we see no detectable bias 
when assuming the \citet{arnaud09} pressure profile, but these results suggest 
that there could be an additional $\sim 10\%$ uncertainty in our $Y_{SZ,sph}$ 
measurements from our assumed pressure profile.  

As an additional test, we vary the outer slope of the  \citet{vikh06} temperature profile 
in equation \ref{universalT}, letting the exponent on the term in the right hand side of 
the denominator vary between 0.16 to 0.80, from its starting value of 0.45.  This range of 
values matches the full range of effective slopes of the temperature profile at $r_{500}$ 
as measured in the sample of \citet{vikh06}.   Calculating new $Y_{SZ,sph}$ estimates 
assuming these new temperature profiles, we find they change the $Y_{SZ,sph}$ values 
given in Table \ref{ysztable} on average by a factor of  $1.09 \pm 0.06$ and $0.91 \pm 0.05$ 
for the exponent values of 0.16 and 0.80, respectively.  
While we consider such a significant systematic shift in the temperature profile unlikely 
across the whole sample, this level of variation in temperature profiles could 
contribute added statistical uncertainty in the $Y_{SZ,sph}$ estimates.  

As a limiting case, we also recalculate $Y_{SZ,sph}$ assuming that the 
cluster is isothermal and with a density profile corresponding to the 
best-fit profile to the X-ray measurements.   Comparing $Y_{SZ,sph}$ for the isothermal profile 
to the values given in Table \ref{ysztable}, we find that this ratio has a mean and 
standard deviation of $1.12 \pm 0.08$ averaged over the sample. This is  
certainly an extreme case because of the abundance of evidence for the temperature 
profile in clusters decreasing significantly by $r_{500}$ \citep[e.g.,][]{vikh06, arnaud09}.
Recent {\sl Suzaku}-observations of nearby clusters at large radii also indicate a drop 
of gas temperature by a factor $\sim 3$ at $r_{200}$ 
\citep[e.g.,][]{fujita08,george09,bautz09,reiprich09,hoshino10}.  We also note that the 
temperature profile with an exponent of 0.80 assumed in the previous paragraph would 
have a gas temperature that dropped by a factor of $\sim 5$ at $r_{200}$, so even 
this would seem like an extreme case.  

Overall, we are encouraged that the variation in $Y_{SZ,sph}$ is found to be
$\sim10\%$ when assuming a broad range of different temperature and pressure 
profiles found in the works of  \citet{vikh06} and \citet{arnaud09}.  Therefore we 
conservatively assume an additional 
$10\%$ statistical uncertainty in our $Y_{SZ,sph}$ estimates from any 
assumed model uncertainty, which we have added in 
quadrature to the uncertainties given in Table \ref{ysztable}.

\subsection{Point Source Contamination}
\label{sec:point_source}
Astrophysical point sources in the direction of the cluster can potentially fill 
in the SZ cluster decrement and bias the integrated Comptonization low.  There are 
generally two populations of sources that can contaminate the SZ signal: dusty or 
radio bright sources.  In V10, the potential contamination from both were discussed, and 
neither is expected to significantly bias the SZ measurements at 150 GHz
averaged over the sample.  We review some of those conclusions here.

In the Appendix, we discuss radio detections at 843 MHz by the Sydney University Molonglo 
Sky Survey (SUMSS) towards the clusters in this work.  The majority of 
radio sources in clusters have been found to have steep spectra with $\alpha < -0.5$ 
(where $S \propto \nu^{\alpha}$) \citep[e.g.,][]{coble07,lin09}.  For example, 
\citet{coble07} find a median spectral index of $-0.72$ between $1.4$ and $28.5$ GHz for 
radio sources towards a sample of massive clusters ranging from $0.14 < z < 1.0$.  
In V10, they noted that a typical cluster would suffer a decrease of 
$\Delta \xi = 1$ for a 2 mJy (5 mJy) source at 150 GHz. located at 0.5$\arcmin$ (1$\arcmin$) 
from the cluster center.  Assuming a spectral index of $-0.72$, a 2 mJy (5 mJy) source at 150 GHz
would be $\sim$83 (210) mJy at the SUMSS observing frequency of 843 MHz.  As
detailed in the Appendix, no source has been detected
above either threshold within 1$\arcmin$ of any cluster in this work.  
However, for any individual cluster radio source, extrapolating its 
flux from radio frequencies to the SPT observing frequency of 152.9 GHz is 
difficult because of the broad range and frequency dependence of the 
spectral indices that is typical for these sources \citep[e.g.,][]{coble07, lin09}.  In 
practice, extrapolating the flux of any SUMSS detected source to the 
SPT observing frequency will have factors of a few uncertainty without 
further observations spanning an intermediate range of frequencies.  With this 
caveat in mind, there appears to be no source strong enough to significantly bias the 
SZ flux estimate for any cluster in this work.  

Including every source detected within 0.5$\arcmin$ of each cluster center in our sample, 
the cumulative SUMSS flux is 224.5 mJy.  If we assume a spectral index of -0.72, this 
would imply an average flux of $\sim$0.35 mJy per cluster at 152.9 GHz.  This would 
correspond to an average decrease in the SZ flux of $\sim$2\%, assuming 
that these sources represent an overdensity to the background population.

The average radio source contamination in SZ surveys has 
been recently estimated in simulations of the 
microwave sky by \citet{sehgal10}.  These simulations were motivated by observations 
by \citet{lin09} measuring the radio source population characteristics in low redshift 
($z < 0.25$) X-ray detected clusters.  For the mass limit of this work, 
$M_{500}  \approx 3 \times 10^{14}  h^{-1} M_{\sun}$, these studies 
predict that in $\lesssim$1\% of clusters there would be radio source contamination 
large enough to effect the SZ flux measurement at the $>20\%$ level, with this 
result largely independent of redshift.   V10 estimates a similar
rate of contamination using a radio source count model \citep{dezotti05} and the 
measured over-abundance of radio sources near clusters \citep{coble07}.   Overall, 
the combination of the above results lead us to not expect any significant radio 
source contamination. 

Emission from sub-millimeter bright galaxies can also contaminate the SZ signal.  
These sources are typically dusty star forming galaxies that are very luminous but 
highly obscured, such that their luminosity is peaked at infrared (IR) wavelengths.  The number 
counts and fluxes of these sources have been characterized between 0.5 - 2.0 mm wavelength
by several experiments \citep[e.g.,][]{coppin06, devlin09, vieira09, austermann10}, and these measurements 
reasonably match 
simple analytic models describing their source population distribution \citep{lima10b}.  
These sources can largely be approximated as a random background that contributes 
additional signal in the SPT maps, and we have explicitly accounted for them in 
our matched filter in equation \ref{eqn:filter}.  However, this implementation did 
not account for any emission that could be correlated with clusters, or additional noise from 
the background objects being gravitationally lensed by the cluster.  The former was argued to be 
insignificant in V10 because: the IR overdensity towards clusters is expected to be $\ll$ 1 mJy 
at 150 GHz even out to highest redshift clusters in our sample, and observational 
measurements of the 100 $\mu$m flux towards local clusters ($z \sim 0.2$) are too low to 
significantly bias the SZ flux measurements even allowing for a strong redshift evolution
of these sources.  The latter has been characterized by \citet{lima10a} to increase the 
flux noise towards clusters by $\sim$60\%, which is at a level such that it would be 
$\lesssim$3\% of the SZ flux for the mass range of clusters in our sample.  We note
that because lensing is a surface brightness conserving process, this latter effect would 
not bias our results when averaged over a large sample of clusters.  

One noteworthy exception where point source contamination is evident 
is the cluster SPT-CL J2332-5358.   This cluster 
was the only candidate in V10 that was also coincident with a point source detected 
in the SPT 220 GHz maps at $>5\sigma$.  
Similar 220 GHz detected point sources have a source flux, $S$, that scales 
with frequency as $S \propto \nu^{\alpha}$ with 
$\alpha = 3.3 \pm 0.7$ between 150 and 220 GHz  \citep{vieira09}.  Extrapolating 
its measured 220 GHz source flux would predict a flux of $11.0 \pm 2.7$ mJy 
at 150 GHz.  To the 150 GHz SPT maps, we subtract a point source of this 
brightness, after convolving it with the 150 GHz SPT transfer function.  We then 
repeat the same cluster extraction method as in V10 to calculate a new cluster 
position and SPT-significance, $\xi$, which is given in Table \ref{obstable}.  
This point source corrected 150 GHz map is used when calculating 
the $Y_{SZ}$ given in Table \ref{ysztable}, and the added uncertainty in 
$Y_{SZ}$ from the uncertain point source spectral index is added in quadrature 
with the other uncertainties outlined in \S \ref{sec:calcsz}.

\section{Scaling relations}
\label{sec:scaling}

Here we discuss the X-ray and SZ scaling relations for the sample. 
We consider the integrated SZ flux $Y_{SZ}$ for both cylindrical and spherical 
volumes of integration. The integrated cylindrical estimate is deprojected to a spherical 
estimate as described in \S \ref{sec:deprojection}.

To measure the slope and normalization of the scaling relations 
we perform linear regression in the presence of intrinsic scatter. 
We use the Bayesian method of \citet{kel07} and maximize the probability 
of a linear model in $\log$-$\log$ space, accounting for measurement 
uncertainties in both axes. 
For the $Y_{SZ}$ scaling relations below we fit the relations with another 
method, accounting for the $\xi$ selection. We investigate this fitting method 
using simulated cluster samples as detailed in \S \ref{sec:ycomparison}.
We have assumed self-similar $E(z)$ scaling for these scaling relations unless 
otherwise stated. 
The best fit parameters for these relations are listed in Table \ref{scalingtab}.

\subsection{$M_{g}$ - $T_X$, $L_X$ - $T_X$ and $L_X$-$M_{500}$}
The scaling between the ICM gas mass ($M_{g}$) and the spectroscopic 
temperature ($T_X$) is an important test of the properties of this 
SZ-selected sample. 
We want to know if the SZ selection biases the observables in any way 
or if they are consistent with those from X-ray selected samples. 
The values of $T_X$ and $M_{g}$ within $r_{500}$ used here 
were derived as described in \S \ref{specsec} and \S \ref{imagingsec} and 
are given in Table \ref{proxytable}. 
This relation can provide clues about the high-$z$ behavior 
of the $M$-$T_X$ and $M$-$M_{g}$ scaling relations (equations \ref{eq500} and 
\ref{eqfgas}). 
To accurately constrain the evolution of these 
relations, however, either hydrostatic masses or masses obtained via 
an independent measurement are needed for high-$z$ clusters. 

\begin{figure*}
\begin{center}
\begin{tabular}{cc}
\includegraphics[width=3in]{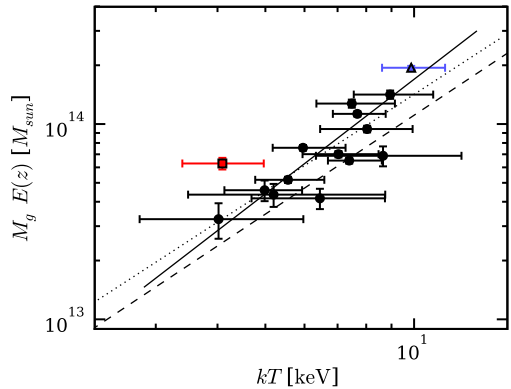} \\ 
\end{tabular}
\caption{$M_{g}-T_X$ relation with self-similar $E(z)$-scaling. 
The data from the SPT sample are shown as circles with error bars and the solid line 
shows the best fit scaling relation. For comparison we also show the best fit relations using 
the low-$z$ (dashed line) and high-$z$ (dotted line) samples from \citet{vikh09}. These 
data were fitted using the same method described here. 
SPT-CL J0551-5709 is shown as a red square and SPT-CL J0516-5430 is shown with a blue triangle.  
\label{xrayscaling2}}
\end{center}
\end{figure*}

We fit the relation 
\begin{equation}
E(z) M_{g} = 10^A \left(\frac{T_X}{5~\mathrm{keV}}\right)^B,
\end{equation}
to the data as described above, accounting for intrinsic scatter. 
Figure \ref{xrayscaling2} shows the $M_{g}-T_X$ relation 
with the best fit power law (solid line). 

To compare our results with previous studies using X-ray selected 
samples we use the gas mass and temperature data for the low-$z$ (median $z\approx0.05$) and 
high-$z$ (median $z\approx 0.48$) as measured by \citet{vikh09}. We choose these samples since the 
data analysis is analogous to ours. 
We fit the scaling relation to these samples using the same method. 
For our SPT-selected sample (median $z=0.74$) we find $A = 13.65 \pm 0.10$ and $B=1.95 \pm 0.66$ 
(see Table \ref{scalingtab}). 
When fitting the low-$z$ sample, we find $A = 13.54 \pm 0.02$ and $B = 1.66 \pm 0.08$, 
at 5 keV.  
For the high-$z$ sample from \citet{vikh09}, 
the best fit parameters are  $A=13.66 \pm 0.03$ and $B = 1.64 \pm 0.21$ in 
even better agreement with our findings. 
This offset between low-$z$ and high-$z$ samples indicate deviations from self-similarity. 
The slope appears steep in our relation, although consistent with both the low- and high-$z$ 
X-ray selected samples as well as the self-similar slope, $B = 1.5$.
A positive trend of the gas mass fraction with mass has been observed previously 
\citep[e.g.,][]{mohr99, vikh06} and is likely what is causing the steeper than self-
similar slope of the relation. 

Numerical simulations \citep[e.g.][]{kra05} also show an increasing
gas mass fraction with redshift, presumably due to a different distribution 
of mass progenitors at high redshift.
The SPT-selected cluster
sample is consistent with this scenario.  The normalization agrees well
with that of the high-$z$ sample of \citet{vikh09} while 
still consistent with the low-$z$ sample at the $1 \sigma$ level.  
We cannot rule out that this behavior is partly related to 
deviations from self-similarity in the $M$-$T_{X}$ relation at high 
redshift.  
However, the results are consistent with previous findings from X-ray 
selected cluster samples.

\begin{figure*}
\begin{center}
\begin{tabular}{cc}
\includegraphics[width=3in]{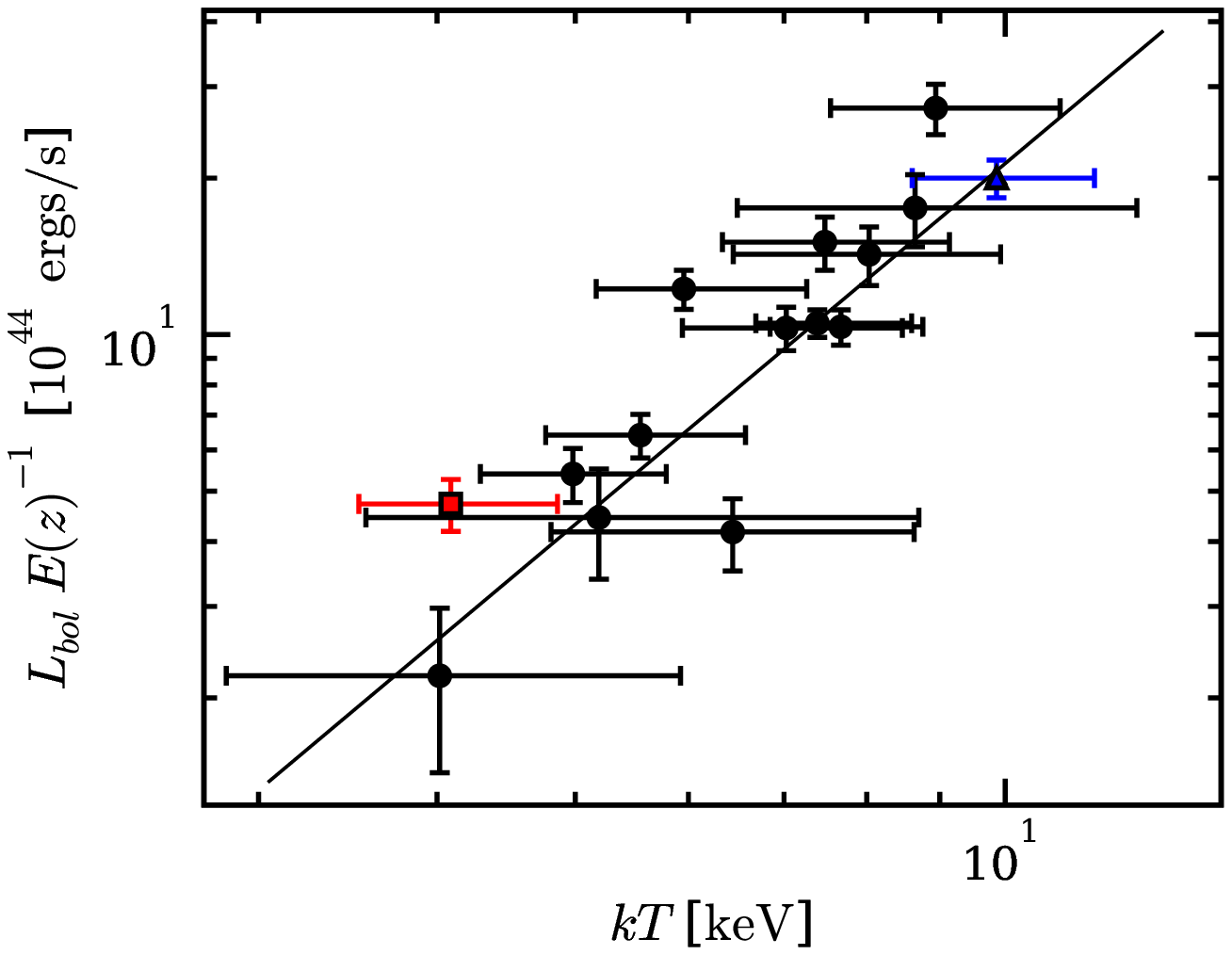} & 
\includegraphics[width=3in]{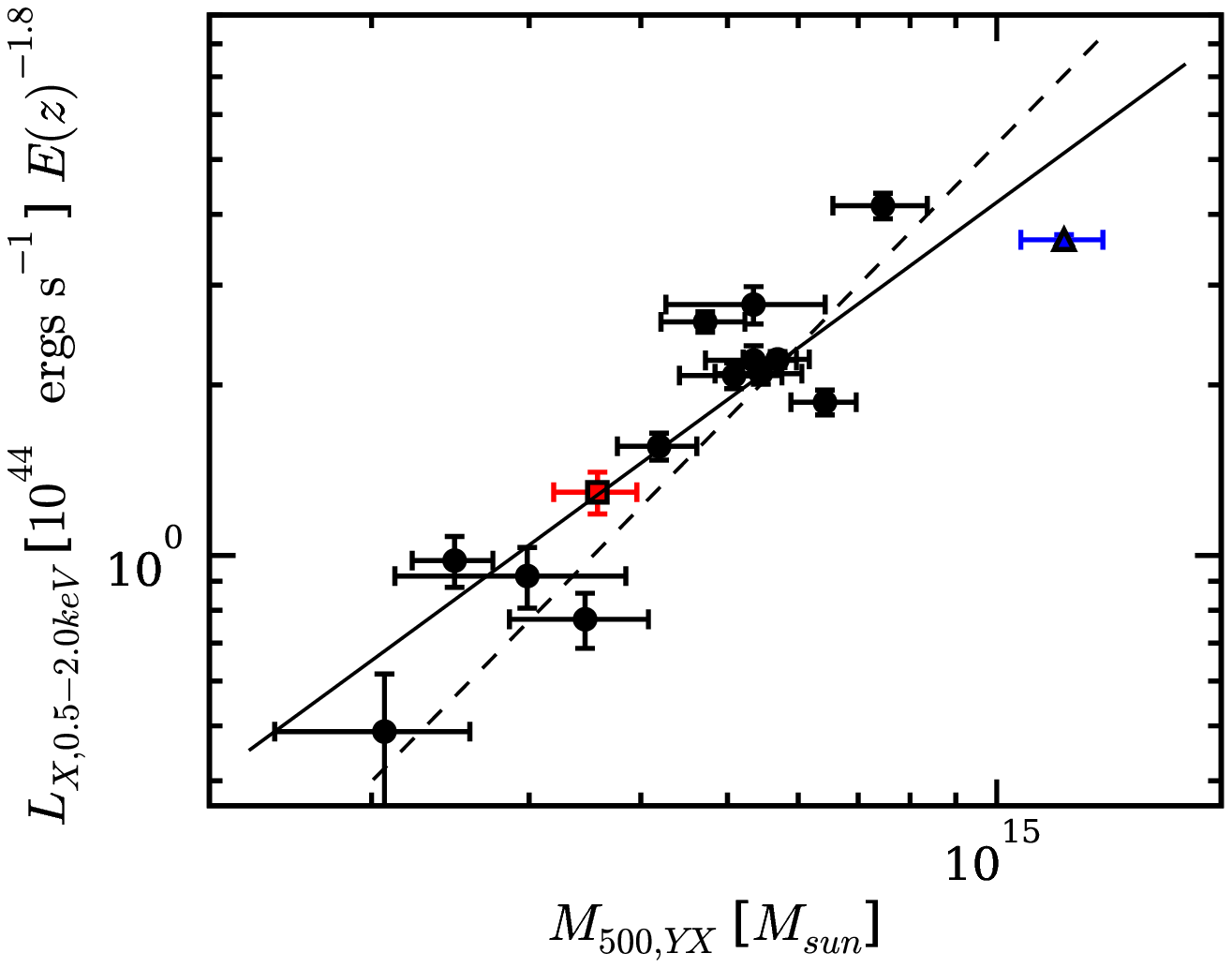} \\
\end{tabular}
\caption{$L_{bol}-T_X$ (left) and $L_{X}-M_{500,Y_X}$ (right) relations for the sample.  For the $L_{bol}-T_X$ relation we have applied self-similar $E(z)$ scaling. For the $L_{X}-M_{500,Y_X}$ relation we have applied $E(z)^{1.85}$-scaling to enable comparison with \citet{vikh09} shown with a dashed line. Solid lines represent the best fit relation. SPT-CL J0551-5709 is shown as a red square and SPT-CL J0516-5430 is shown with a blue triangle.  
\label{xrayscaling}}
\end{center}
\end{figure*}

We also investigate the luminosity-temperature relation \citep[e.g.,][]{edge91,mar98} for the sample where 
we compare the luminosity in the $0.5$-$2.0$ keV band within $r_{500}$, including the core, $L_X$, to the 
temperature estimate within $r_{500}$, $T_X$,  with the core excised.
We fit the relation 
\begin{equation}
E(z)^{-1} L_{X} = 10^A \left(\frac{T_X}{5~\mathrm{keV}}\right)^B,
\end{equation}
and find a shallow slope, $B=1.92 \pm 0.60$ (see Table \ref{scalingtab}), marginally inconsistent with most 
previous works which typically find a slope close to $3$ \citep[e.g.][]{mantz09}. 
For completeness, we also estimate the bolometric luminosity by extrapolating 
the spectrum for all frequencies using the core excised temperature. This introduces 
additional uncertainties on the luminosity due to the uncertainty in 
the temperature estimate. 
For the $L_{bol}$-$T_X$ relation, we find a steeper slope of $B=2.31 \pm 0.85$, 
consistent with previous work. The best fit relation is shown in Figure \ref{xrayscaling} 
(left). 

We investigate the luminosity-mass relation \citep[e.g.,][]{reip02} and adopt the best-fit 
$E(z)$-scaling found by \citet{vikh09} to fit for the relation 
\begin{equation}
E(z)^{-1.85} L_{X} = 10^A \left(\frac{M_{500,Y_X}}{10^{15}~M_{\sun}}\right)^B.
\end{equation}
The best fit relation is shown in Figure \ref{xrayscaling} (solid line) compared 
to the best fit in \citet{vikh09} (dashed line).
We find a slope of $B=1.16\pm 0.20$, shallow compared to the $B=1.61 \pm 0.14$ 
found by \citet{vikh09}, while the normalization agrees very well near 
the median mass of the sample at $5 \times 10^{14}~M_{\sun}$. 
From the relation in \citet{vikh09} we expect a normalization of $A = 44.25$ at 
$5 \times 10^{14}~M_{\sun}$, which is in good agreement with our measured value of
 $A = 44.28 \pm 0.07$.
The slope of the fit is mostly driven by SPT-CL 0516-5430 (blue triangle) which 
has a high derived mass for its measured luminosity, possibly caused by the 
observed merging activity in the cluster. Excluding this object from the fit 
changes the slope to $B=1.45\pm0.29$, consistent with \citet{vikh09}. 

In summary, we find a good agreement between the X-ray observables measured for this 
SZ-selected sample when comparing our results to X-ray selected samples where similar 
analysis methods were used.

\subsection{$Y_{SZ}$ scaling relations}
\label{sec:ycomparison}

\subsubsection{Fitting method}
\label{sec:fitting}
The selection of clusters above a fixed S/N threshold can bias the 
SZ-X-ray scaling relations. 
The SZ flux, $Y_{SZ}$, is correlated with the SZ S/N on which the sample 
is selected and ignoring this effect will lead to bias in the $Y_{SZ}$-scaling relation 
parameters. 
The steepness of the cluster mass function makes this bias more of a problem 
since the number of clusters in a given mass bin grows exponentially 
towards lower masses. This causes more clusters to be near the selection threshold 
where the effect is most prominent. 
The measurement uncertainty in $Y_{SZ}$ and the intrinsic scatter in the scaling 
relation cause clusters with low mass (or $Y_{X}$) to scatter over the selection 
threshold.  
This is visible as a tail towards low mass (or $Y_{X}$) near the $Y_{SZ}$-selection threshold in the scaling 
relation plots. If ignored, this bias will lead to a biased slope and 
a higher normalization near the threshold in the fitted scaling relation parameters. 
Below we describe our method to account for the SZ selection in our scaling
relation fits.

We estimate how the selection cut in $\xi$ translates to a selection in 
$Y_{SZ,sph}$ using simulated SPT observations.  We repeat 
the procedure used in V10 to estimate the SZ selection function and 
SZ-significance mass relation, where observations of simulated SZ maps 
are added to the dominant sources of astrophysical confusion and 
instrumental noise, mock observed, and processed through the SPT analysis 
pipeline.  In contrast from V10, we keep track of the predicted value of 
$Y_{SZ,sph}$ for each cluster, which is estimated using the 
$Y_{sph}-M_{500}$ scaling relation from \citet{arnaud09} and the simulated 
cluster's mass, $M_{500}$.  Also, we measure $\xi$ for each 
simulated cluster in the same way as we do for the SPT observations, where 
we record the maximum significance S/N over different spatial filters and 
in a single realization of the astrophysical confusion and noise.  

With these results we find a best-fit relation 

\begin{equation}
\xi=5.90~\left(\frac{Y_{SZ}}{2 \times 10^{14} M_{\odot} \mathrm{keV}}\right)^{0.64} ~ E(z)^{0.16}
\end{equation}
with intrinsic scatter in $\log Y_{SZ}$ of 0.10.
Comparing this result with the measured $\xi$ and $Y_{SZ}$ for our sample we find 
that the simulation-based relation is offset slightly high in $\xi$. 
When fitting the normalization of the above relation with the observed data we find 
a best fit value of $5.56\pm0.31$, about 6\% lower than the simulations suggest.
The reason for this offset may be related to systematic uncertainties in the 
simulation of SPT observations. 
We adopt this best fit normalization from the observed data, and use the above relation 
to construct a selection function in $\log Y_{SZ}$. 
We estimate the selection probability as an error function in $\log Y_{SZ}$ 
at this threshold with the width set by the intrinsic scatter in this relation. 
To account for the uncertainty in the simulations we use include a 10\% 
gaussian uncertainty on the normalization and a 20\% gaussian uncertainty on the intrinsic 
scatter, which we marginalize over for our results.

The $Y_{SZ}$ scaling relation fit is performed as follows. 
First, we calculate the probability of measuring a certain $Y_{SZ}$ for a given 
$M$ or $Y_X$ and a given scaling relation with log-normal scatter. This is then convolved 
with the measurement uncertainty in $Y_{SZ}$. 
The $Y_{SZ}$ selection cut is applied by multiplying the probability of measuring 
$Y_{SZ}$ with an error function in $\log Y_{SZ}$ as described above.
The likelihood is then calculated as the product of probabilities for all clusters 
and is maximized to obtain the scaling relation parameters.

The uncertainties on the calculated $Y_{SZ}$ are correlated with the measurement error in 
$Y_{X}$ and $M_{500,Y_X}$. The $r_{500}$ radius used for integration of $Y_{SZ}$ is estimated 
from $M_{500,Y_X}$ and any measurement scatter in $Y_{X}$ will contribute to scatter 
in $Y_{SZ}$ of similar magnitude. 
This will lead to scatter along the scaling relation and for this reason we 
ignore these effects. We fit for the intrinsic scatter in the $Y_{SZ}$-$M_{500,Y_X}$ and 
$Y_{SZ}$-$Y_{X}$ relations while noting that the scatter in $Y_{SZ}$ and $Y_{X}$ with 
the true cluster mass will also be partially correlated.

To test our fitting method for potential biases we run it on mock samples drawn from 
a fiducial mass function. We perform this test both in the limit of a sample with 1000 clusters 
and for many samples of 15 clusters. To estimate a $Y_{SZ}$ for the mock samples we assume 
a $Y_{SZ}$-$M$ relation with log-normal scatter and convolve this with a linear measurement 
scatter of $0.4 \times 10^{14}~M_{\sun}$. We assume that we know the relation between $\xi$ and 
$Y_{SZ}$ and its scatter to reproduce the sample selection function. 
In the limit of a large sample of clusters, we find no measurable bias in 
either the fitted slope or normalization. For 1000 generated mock samples of 15 clusters 
we find that the scatter in normalization and slope between the samples and the 
average measurement error on the parameters over the samples are consistent. 
The normalization is reproduced 
to within 0.1\% of the input value while we find that the slope is biased low by 
2.0\%. The average measurement error on the slope from the 15 cluster samples is 
13.5\% so this bias is of little significance.

\subsubsection{$Y_{SZ}$-$Y_{X}$ relation}
The relationship between $Y_{SZ}$ and $Y_{X}$ is determined by details of the 
gas and temperature distribution in the cluster.  It is effectively measuring a 
relationship between the mass weighted and X-ray spectroscopic weighted temperature.  
In \citet{arnaud09}, X-ray measurements of both relaxed and unrelaxed 
clusters were used to approximate a $Y_{SZ}$-$Y_{X}$ relation, for a spherical 
integration to $r_{500}$.  Their work found a relationship with a 
normalization of $0.924\pm0.004$, implying a lower mass weighted temperature.  
This result is consistent with previous X-ray measurements \citep[e.g.,][]{vikh06}, and 
our expectation of $0.91\pm0.01$ for the measured gas density profiles and the assumed temperature profile 
in this work.  
Hydrodynamical simulations predict a similar ratio, but
also find that X-ray measurements could overestimate the integrated
pressure at some level \citep[e.g.,][]{nag07}.  This would imply that the
X-ray predictions of the normalization of the $Y_{SZ}$-$Y_{X}$ relation
could be biased high; for example, the results of \citet{nag07} suggest a
bias at the few percent level.

Recently there has been some evidence that the SZ signal from 
clusters could be lower than this expectation from comparison of WMAP SZ
observations of X-ray selected clusters \citep{kom10}, however similar analyses 
have led to different conclusions \citep{melin10}.  
By fitting a 
normalization of the $Y_{SZ}$-$Y_{X}$ relation to the clusters in this work, we can 
test for a similar offset, which could be indicative of differences in the real 
gas profiles from those we've assumed or some other systematic bias in 
either the SZ or X-ray measurement. 

\begin{figure*}[htb!]
\begin{center}
\begin{tabular}{cc}
\includegraphics[width=3in]{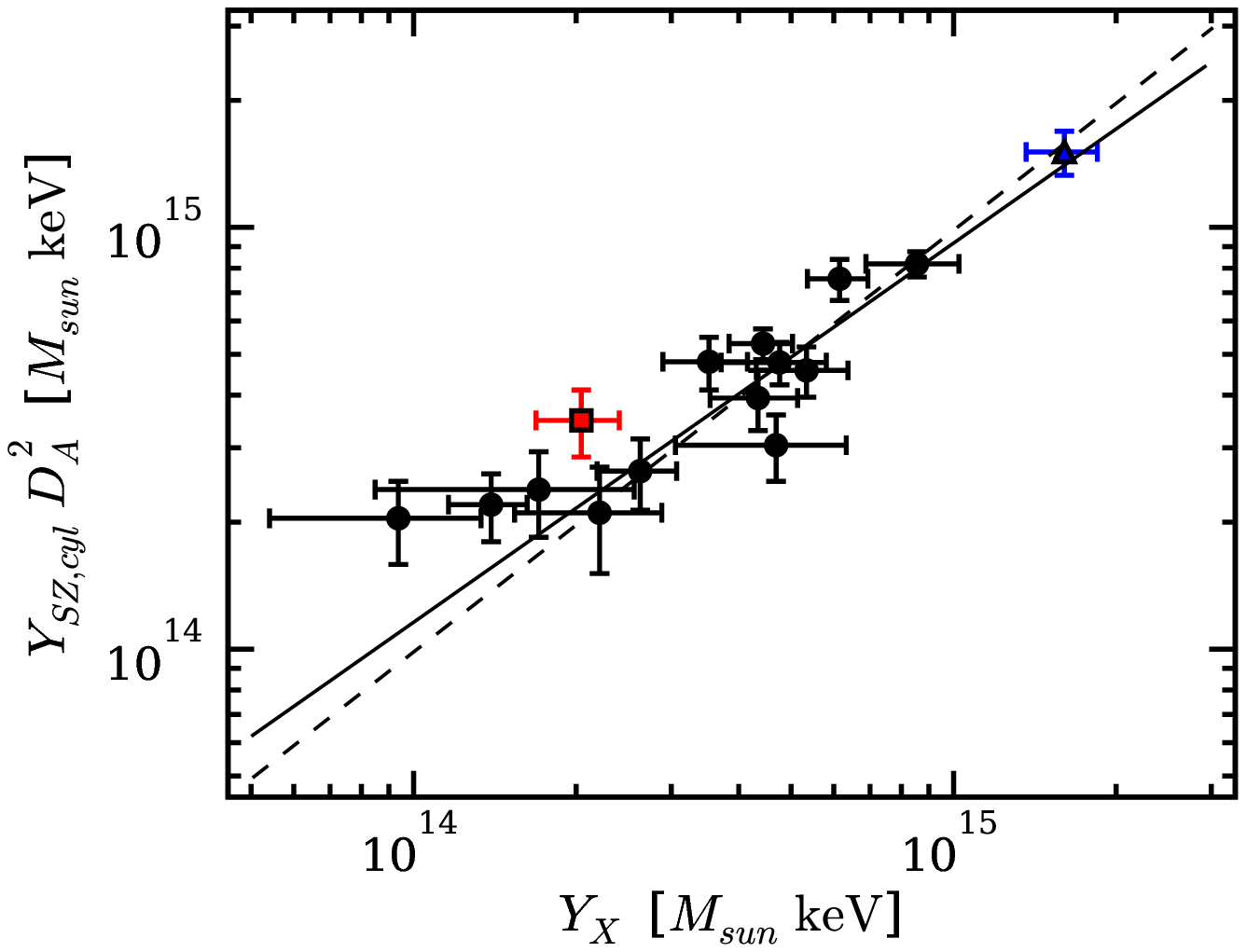} & 
\includegraphics[width=3in]{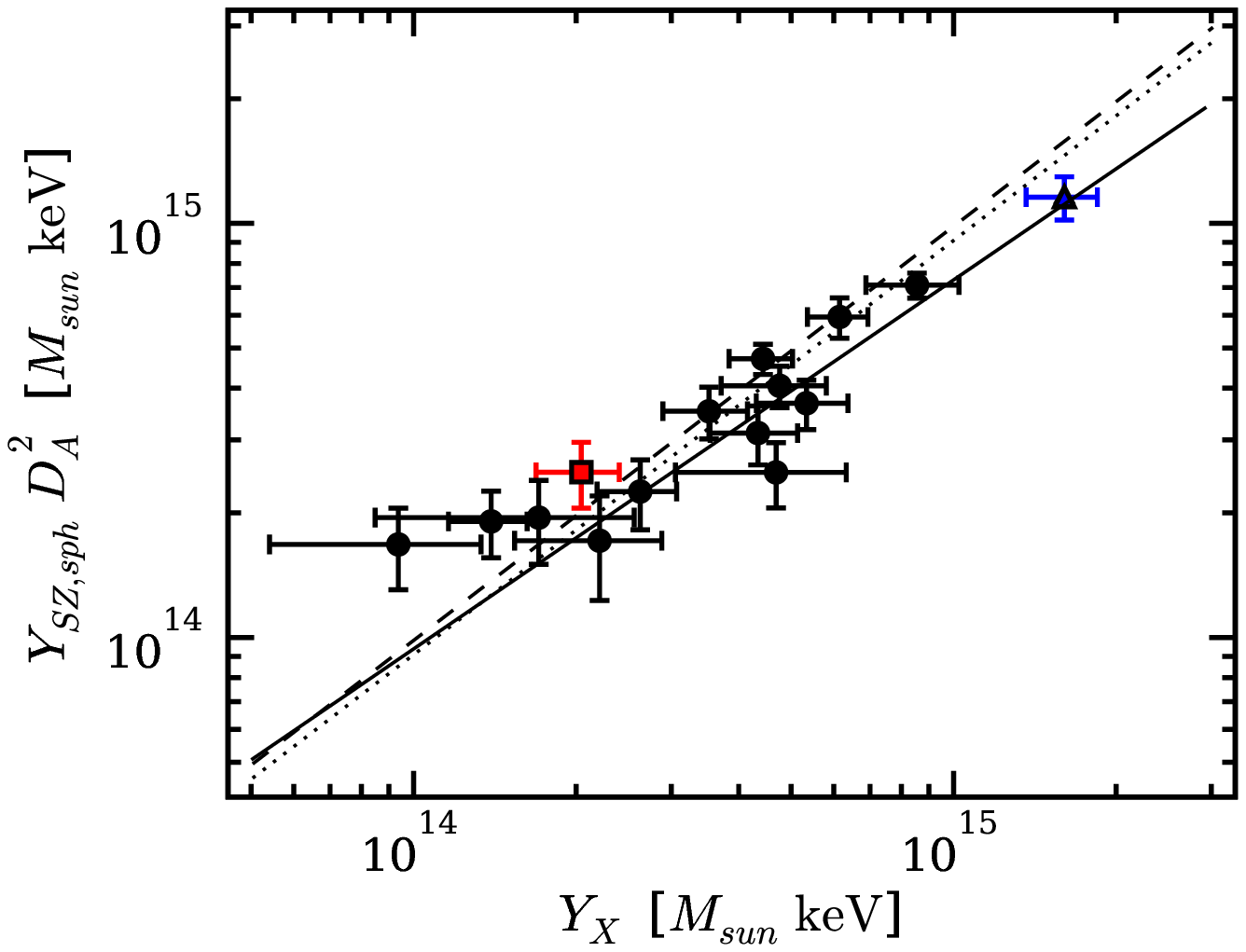} \\
\end{tabular}
\caption{The $Y_{SZ}-Y_X$ relation for the sample. The plots show uncorrected $Y_{SZ,cyl}$ estimates (left) and the deprojected $Y_{SZ,sph}$ estimates assuming a universal temperature profile (right). The solid line shows the best fit relation. The dashed lines represent equality and the dotted line is the best fit relation from \citet{arnaud09}. SPT-CL J0551-5709 is shown as a red square and SPT-CL J0516-5430 is shown with a blue triangle.  
\label{YszYx}}
\end{center}
\end{figure*}

Using the method described in \S \ref{sec:fitting}, we fit a scaling relation between 
$Y_{SZ}$ and $Y_X$ of the form
\begin{equation}
\label{eqn:ysz_v_yx}
Y_{SZ} = 10^A \left(\frac{Y_X}{3 \times 10^{14}~M_{\sun}~\mathrm{keV}}\right)^B.
\end{equation}
In Figure \ref{YszYx}, we show the $Y_{SZ}$-$Y_{X}$ relation for both cylindrical (left) and spherical (right)  
$Y_{SZ}$, denoted by $Y_{SZ,cyl}$ and $Y_{SZ,sph}$ respectively.  In Table \ref{scalingtab}, we give 
fits to equation \ref{eqn:ysz_v_yx} given several different assumptions, which are described and 
discussed further below. 

For both cylindrical and spherical $Y_{SZ}$ we find a slope consistent with unity. 
The dashed line in the plots is the $Y_{SZ} = Y_{X}$ relation and the dotted line (right) shows 
the expected $Y_{SZ,sph}/Y_{X}$ ratio of 0.924 found in the work of \citet{arnaud09}. 
We also fit the scaling relations with the slope fixed to 1 (see Table \ref{scalingtab}). 
The normalization of our $Y_{SZ,sph}$-$Y_{X}$ fit implies an average ratio of $0.82 \pm 0.07$, 
consistent with the prediction of \citet{arnaud09} at the $1.5 \sigma$ level.  The most 
significant outlier from the $Y_{SZ,sph}$-$Y_{X}$ relation, SPT-CL J0000-5748, also has 
the lowest  $Y_{SZ,sph}/Y_{X}$ ratio in our sample. If we exclude this cluster from the 
fit, we find an even better agreement with a normalization of $0.87 \pm 0.06$.
The marginally lower normalization found in this relation when compared to the 
X-ray prediction is expected at some level from hydrodynamical simulations which 
predict that $Y_{SZ}$ could be biased high when estimated using X-ray data. 

As an additional test, we fit the $Y_{SZ}$-$Y_{X}$ relation including a factor of
$E(z)^C$, and fit for $C$ while keeping the slope fixed to $1$. 
We find $C = -0.36\pm0.58$, consistent with no evolution. 
Further investigating this, we divide the sample in two redshift bins. 
The low-$z$ bin consists of the 7 clusters at $z<0.7$ while the high-$z$ bin 
consists of the 8 clusters at $z>0.7$. Fixing the slope to $1$, we find 
a ratio of $0.88\pm0.12$ for the low-$z$ sample, and a ratio of of $0.72\pm0.14$ 
for the high-$z$ sample.  If we exclude the most significant outlier from the 
high-$z$ sample, SPT-CL J0000-5748, we find a ratio of $0.83 \pm 0.09$.

As discussed at the beginning of this section, the $Y_{SZ,sph}/Y_{X}$ ratio is expected to be 
lower than $1$ due to the different weighting of the temperature in the two estimates. 
As explained in \S \ref{sec:model_uncertainty}, our measurement of $Y_{SZ}$ 
is sensitive to the assumed temperature profile, whereas the $Y_{X}$ measurement is not. 
This causes the measured $Y_{SZ,sph}/Y_{X}$ ratio to change under different 
assumptions about the temperature profile.  
However, the expected ratio of 
the gas-mass weighted and the spectroscopic weighted temperatures 
changes by a similar amount as the temperature profile changes.  The overall 
effect makes the measured ratio  
to be fairly independent to our assumptions of the shape of the temperature 
profile.  This is particularly true for 
variations in the slope of the temperature profile at large radii that we 
tested in \S \ref{sec:model_uncertainty}.

Overall we find a normalization consistent with our expectations given 
the gas profiles in this work, and believe this result to be largely 
independent of our assumed temperature profile.
We find no significant evidence for a redshift dependent evolution in the
normalization of the $Y_{SZ}$-$Y_{X}$ relation, but note that the high-$z$
sub-sample does favor a $\sim$1$\sigma$ lower normalization.  This could be
marginal evidence for redshift evolution, however further X-ray and
SZ observations will be needed to make any statistically significant
statements.

\subsubsection{$Y_{SZ}$-$M_{500,Y_X}$ relation}
Finally, we investigate the relation between the SZ flux, $Y_{SZ}$, and the X-ray derived mass, $M_{500,Y_X}$.  
This is not an independent result from the previous section because the X-ray masses are calculated directly 
from the $Y_X$ measurements and the $M_{500,Y_X}-Y_X$ relation in \citet{vikh09}.  However, it is useful in 
understanding the SZ mass calibration, and can be compared to previous measurements of this relation.  

\begin{figure*}[htb!]
\begin{center}
\begin{tabular}{c}
\includegraphics[width=3in]{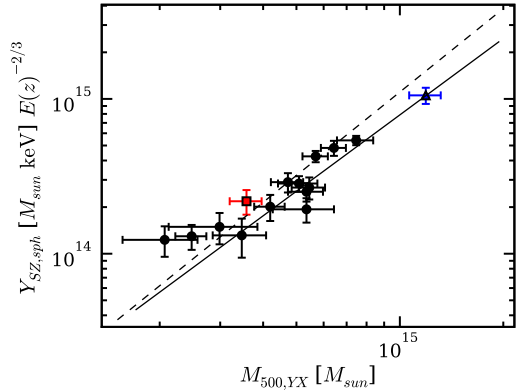} \\ 
\end{tabular}
\caption{The $Y_{SZ}-M_{500}$ relation for the sample. The plots show the $Y_{SZ,sph}$ estimates, deprojected assuming a 
universal temperature profile. 
Masses are estimated from the $M_{500}$-$Y_X$ relation.
The dashed line shows the best fit relation from \citet{arnaud09}. SPT-CL J0551-5709 is shown as a red square and SPT-CL J0516-5430 is shown with a blue triangle. 
\label{YszM}}
\end{center}
\end{figure*}

Using the method of \S \ref{sec:fitting}, we fit a scaling relation between $Y_{SZ}$ and $M_{500,Y_X}$ of the form
\begin{equation}
\label{eqn:ysz_v_m500}
Y_{SZ} = 10^A \left(\frac{M_{500,Y_X}}{3 \times 10^{14}~M_{\sun}}\right)^B E(z)^{2/3}.
\end{equation}
In Figure \ref{YszM} we show the $Y_{SZ}-M_{500,Y_X}$ relation, and in Table \ref{scalingtab} we give
the best-fit parameters to equation \ref{eqn:ysz_v_m500}.    The slope of this relation is found to 
be $1.67 \pm 0.29$ for $Y_{SZ,sph}$-$M_{500,Y_X}$ with intrinsic scatter of $0.09 \pm 0.05$ (see Table \ref{scalingtab}).  
This slope is consistent with the self-similar expectation of $5/3$ and previous measurements 
\citep[e.g.,][]{bonamente08, arnaud09}.  
In \citet{arnaud09}, a slope of $1.790 \pm 0.015$ is measured for their predicted 
$Y_{SZ,sph}$ using X-ray observables.

We also fit the relation keeping the slope fixed at the value expected from X-ray studies, $1.79$, 
and note that the normalization of the $Y_{SZ}$-$M$ relation is lower than that of 
\citet{arnaud09}. 
The \citet{arnaud09} results imply a normalization $A = 14.115 \pm 0.003$, using our adopted $h$ while we 
find $A = 14.03 \pm 0.04$, around a $2 \sigma$ offset. 
The best fit relation is shown in Figure \ref{YszM} (solid line) 
with the \citet{arnaud09} relation shown (dashed line) for comparison.

It should be noted that part of the offset is due to differences in the 
mass estimates in our work and in the work of \citet{arnaud09}.
The normalization in the $M-Y_X$ relation (equation \ref{eqYgas}), 
used here to estimate cluster masses, is different from the normalization of the 
$M-Y_X$ relation used in \citet{arnaud09} to derive the 
$Y_{SZ}$-$M_{500,Y_X}$ relation in that work. 
For our adopted $h$, we find a mass, $M_{500,Y_X}=4.83\pm0.17 \times 10^{14} ~ M_{\odot}$, at 
$Y_X = 3\times10^{14} ~ M_{\odot}~\mathrm{keV}$ from equation \ref{eqYgas}. 
If we instead use the $M-Y_X$ relation from \citet{arnaud09}, we find 
$M_{500,Y_X}=4.64\pm0.12 \times 10^{14} ~ M_{\odot}$, at the same $Y_X$.
To account for this scaling relation offset we shift the normalization 
of the \citet{arnaud09} $Y_{SZ}$-$M_{500,Y_X}$ relation down by a factor 
$(4.83/4.64)^{1.79}=1.074$, resulting in $A = 14.084\pm0.003$.  This is reduces the offset to 
our measurement, $A = 14.03 \pm 0.04$, to a similar level to what we 
found for the $Y_{SZ}$-$Y_{X}$ relation in the previous section. 

It is important to note that the intrinsic scatter in this scaling relation does not directly reflect the low
scatter relationship between $Y_{SZ}$ and the gravitational mass. The mass is derived directly from $Y_{X}$ and therefore
includes the scatter in the $Y_{X}$-mass relationship.  Independent measurements of the gravitational mass, e.g.,
through weak lensing or hydrostatic masses from X-ray data, are necessary to diagnose scatter in the $Y_{SZ}$-$M$ 
relation.

\begin{deluxetable*}{llllll}
\tabletypesize{\scriptsize}
\tablecaption{X-ray and SZ scaling relations $\log (Y)=A + B \log(X/X_{pivot}) + C \log(E(z))$. \label{scalingtab}}
\tablewidth{0pt}
\tablehead{
\colhead{Relation} &
\colhead{$A$} &
\colhead{$B$} & 
\colhead{$C$} & 
\colhead{$\sigma_y$} & 
\colhead{$X_{pivot}$} \\
}
\tablecomments{Self-similar $E(z)$-scaling has been assumed here except for the $L_{X}$-$M_{Y,500}$ relation, where the best fit evolution from \citep{vikh09} is adopted. $\sigma_y$ represents the intrinsic scatter in $\log~Y$.}
\startdata
$L_X$-$T_X$   & $44.16 \pm 0.09$  & $1.92 \pm 0.60$ & $1$ & $0.11 \pm 0.06$ & $5~\mathrm{keV}$ \\
$L_{bol}$-$T_X$ & $44.64 \pm 0.14$  & $2.31 \pm 0.85$ & $1$ & $0.17 \pm 0.08$ & $5~\mathrm{keV}$ \\
$L_{X}$-$M_{Y,500}$ & $44.63 \pm 0.07 $ & $ 1.16 \pm 0.20 $ & $1.85$ & $ 0.11 \pm 0.04 $ & $10^{15}~M_{\sun}$ \\
$M_g$-$T_X$   & $13.65 \pm 0.10$  & $1.95 \pm 0.66$ & $-1$ & $0.11 \pm 0.06$ &  $5~\mathrm{keV}$ \\

$Y_{SZ,sph}$-$Y_X$ & $ 14.40 \pm 0.08 $ & $ 0.96 \pm 0.18 $ & $0$ &  $ 0.09 \pm 0.04 $ & $3 \times 10^{14}~M_{\sun}~\mathrm{keV}$ \\
$Y_{SZ,sph}$-$Y_X$ & $ 14.39 \pm 0.04 $ & $ 1             $ & $0$ &  $ 0.09 \pm 0.04 $ & $3 \times 10^{14}~M_{\sun}~\mathrm{keV}$ \\
$Y_{SZ,sph}$-$Y_X$ & $ 14.44 \pm 0.08 $ & $ 1             $ & $-0.36\pm0.58$ &  $ 0.09 \pm 0.04 $ & $3 \times 10^{14}~M_{\sun}~\mathrm{keV}$ \\

$Y_{SZ,cyl}$-$Y_X$ & $ 14.50 \pm 0.05 $ & $ 0.90 \pm 0.17 $ & $0$ &  $ 0.07 \pm 0.05 $ & $3 \times 10^{14}~M_{\sun}~\mathrm{keV}$ \\
$Y_{SZ,cyl}$-$Y_X$ & $ 14.49 \pm 0.04 $ & $ 1             $ & $0$ &  $ 0.09 \pm 0.04 $ & $3 \times 10^{14}~M_{\sun}~\mathrm{keV}$ \\
$Y_{SZ,cyl}$-$Y_X$ & $ 14.54 \pm 0.09 $ & $ 1             $ & $-0.35\pm0.56$ &  $ 0.09 \pm 0.04 $ & $3 \times 10^{14}~M_{\sun}~\mathrm{keV}$ \\

$Y_{SZ,sph}$-$M_{Y,500}$ & $ 14.06 \pm 0.10 $ & $ 1.67 \pm 0.29 $ & $2/3$ & $ 0.09 \pm 0.05 $ & $3 \times 10^{14}~M_{\sun}$ \\
$Y_{SZ,sph}$-$M_{Y,500}$ & $ 14.03 \pm 0.04 $ & $ 1.79          $ & $2/3$ & $ 0.09 \pm 0.04 $ & $3 \times 10^{14}~M_{\sun}$ \\

\enddata
\end{deluxetable*}

\section{Conclusions}
We have presented the results from X-ray observations of a subset 
of 15 clusters from the first SZ selected 
cluster catalog from the South Pole Telescope cluster survey.  We report 
the X-ray properties of this sample, including measurements of $M_g$, $T_X$, and $Y_X$, 
and have used $T_X$ and $Y_X$ to estimate the total masses of the clusters.  
We find generally good agreement between the expected X-ray properties of this sample 
and those expected from scaling relations.
However, we find some indication of deviations from self-similar evolution 
of the $M_{g}$-$T$ relation compared to other local cluster samples 
at a level consistent with the explanation of an evolving gas mass fraction 
in high redshift clusters.  

Using the X-ray measured cluster positions and gas profiles, 
we have re-analyzed the SZ measurements to calculate each cluster's integrated 
Compton-$y$ parameter, $Y_{SZ}$.  We further use the X-ray measured gas profiles to 
deproject the SZ measurements so that they correspond to a spherical 
integrated Comptonization, $Y_{SZ,sph}$, that is more directly comparable 
to the X-ray measurements. 

We have calculated scaling relations between $Y_{SZ,sph}$ and the X-ray measured
quantities $Y_X$ and $M_{500,Y_X}$.  We fit the
$Y_{SZ,sph}$-$Y_{X}$ relation, and find a slope consistent with unity,
$0.96 \pm 0.18$. Fixing this slope to 1, we re-fit the relation and find a
normalization that implies a ratio of $Y_{SZ,sph}/Y_{X} = 0.82 \pm 0.07$.
This normalization effectively corresponds to the ratio between the mass
weighted and X-ray spectroscopic weighted temperature.  For the
spherically symmetric density and temperature profiles assumed in this work, we would
have expected a ratio of $0.91 \pm 0.01$, consistent with predictions from other
X-ray studies of clusters \citep[see e.g.][]{vikh06, arnaud09}.  
We therefore find 
a normalization of the $Y_{SZ,sph}$-$Y_{X}$ relation that is  
marginally consistent with, although lower than this prediction. This indicates that the 
SZ and X-ray measured pressure largely agree.  We find, however, that the lower 
normalization is more pronounced in the $z>0.7$ high-$z$ sub-sample, with 
a ratio of $0.72\pm0.14$.  Further 
X-ray and SZ observations of high redshift clusters are needed to determine if 
this is a real effect.

Using the $Y_X$ measurement as a proxy for the total cluster
mass with a relation calibrated in \citet{vikh09}, we find similar
results when we fit a $Y_{SZ,sph}-M_{500,Y_X}$ relation.  We find a slope
consistent with the self-similar expectation of $Y_{SZ} \propto M^{5/3}$
and a normalization consistent with the predictions from X-ray
measurements by \citet{arnaud09}.  We have compared the $Y_X$ inferred 
total mass to the SZ significance-inferred 
total mass from \citet{van10}.   Considering only the clusters used in their 
cosmological analysis, we find an average ratio 
of $M_{500,SZ}$ to $M_{500,Y_X}$ of $0.89 \pm 0.06$, which is within
the $\sim15\%$ systematic uncertainty on the simulation-based mass estimates in \citet{van10}.  

This work is encouraging for future studies of SZ-selected clusters.  It
suggests that SZ mass estimates are consistent with X-ray based mass
estimates used in other cluster cosmology studies.  This result is
important for the use of SZ-selected cluster samples to constrain
cosmology, and demonstrates that the X-ray measurements can play a valuable
role in calibrating SZ surveys.  This work also highlights the potential
power of SZ surveys to study cluster evolution due to the broad redshift
range of SZ-selected samples.  
The SPT has now surveyed 
an additional $\sim$900 deg$^2$ to a similar depth to the $\sim$178 deg$^2$
used in this work, and expects to cover over 2000 deg$^2$ by December 2011.  
These measurements will produce a catalog of hundreds of massive 
SZ-discovered clusters that extend to high redshift.  Joint X-ray and SZ 
measurements of these clusters promise to place interesting constraints on cluster
formation and gas physics for the most massive young clusters in
the universe.

\section*{Acknowledgments}
The South Pole Telescope is supported by the National Science
Foundation through grants ANT-0638937 and ANT-0130612.  Partial
support is also provided by the NSF Physics Frontier Center grant
PHY-0114422 to the Kavli Institute of Cosmological Physics at the
University of Chicago, the Kavli Foundation and the Gordon and Betty
Moore Foundation.

K.\ Andersson is supported in part by NASA through SAO Award Number 
2834-MIT-SAO-4018 issued by the Chandra X-ray Observatory Center, 
which is operated by the Smithsonian Astrophysical Observatory 
for and on behalf of NASA under contract NAS8-03060.

The McGill group acknowledges funding from
the National Sciences and Engineering Research Council of Canada, the
Quebec Fonds de recherche sur la nature et les technologies and the
Canadian Institute for Advanced Research. The following individuals acknowledge additional support:
B.\ Stalder from the Brinson Foundation,
B.\ Benson from a KICP Fellowship,
R.\ Foley from a Clay Fellowship,
D.\ Marrone from Hubble Fellowship grant HF-51259.01-A,
N.W.\ Halverson acknowledges support from an Alfred P. Sloan Research
Fellowship.
M.\ Brodwin from the Keck Foundation, and A.T. Lee from
the Miller Institute for Basic Research in Science, University of California Berkeley.

Facilities: Chandra (ACIS), XMM-Newton (EPIC), Blanco (MOSAIC II), Magellan:Baade (IMACS), Magellan:Clay (LDSS2),

\clearpage

\clearpage
\appendix

\section{Notes on individual objects}
\label{sec:indiv}
\begin{figure*}[htb!]
\begin{center}
\begin{tabular}{cc}
\includegraphics[width=3.2in]{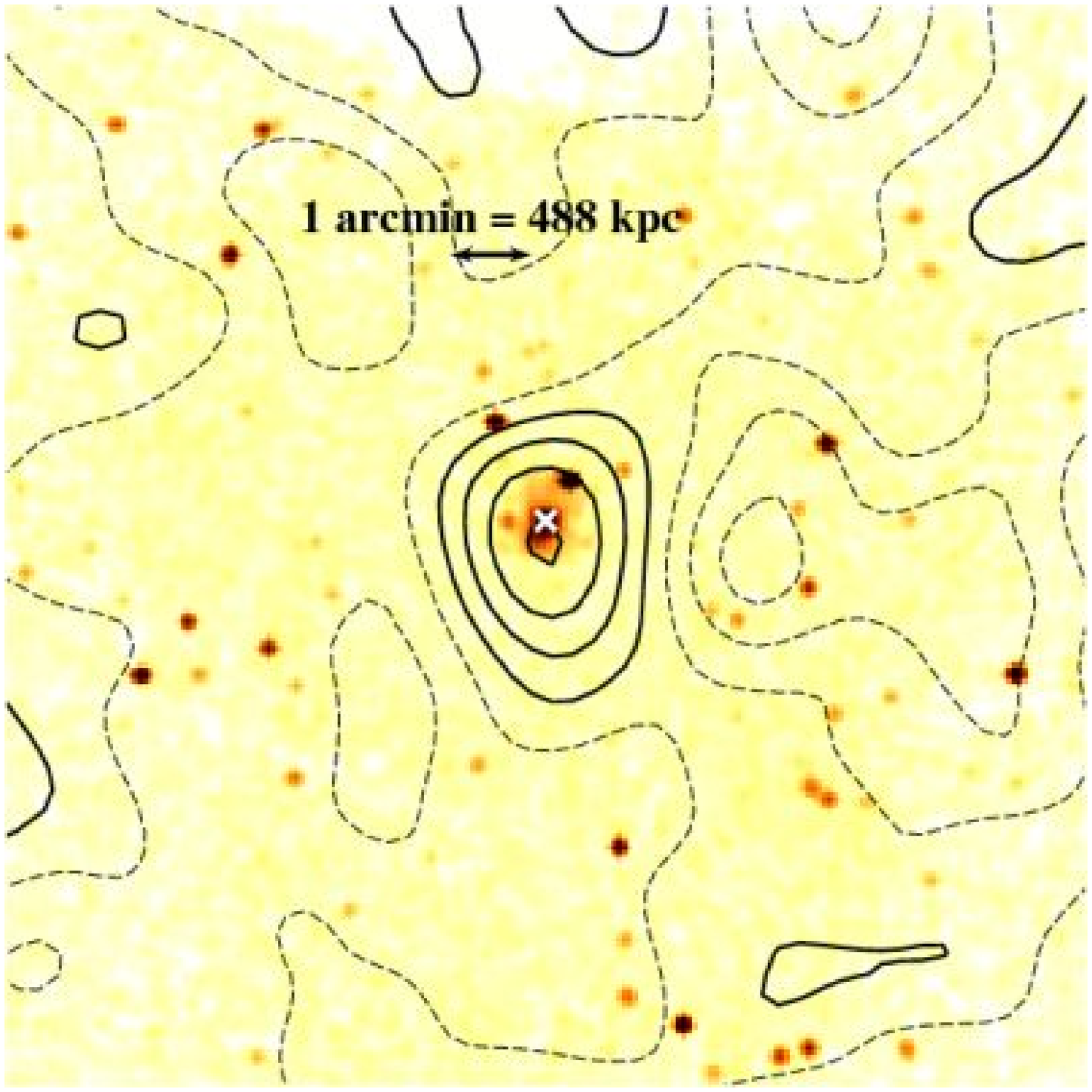} &  
\includegraphics[width=3.2in]{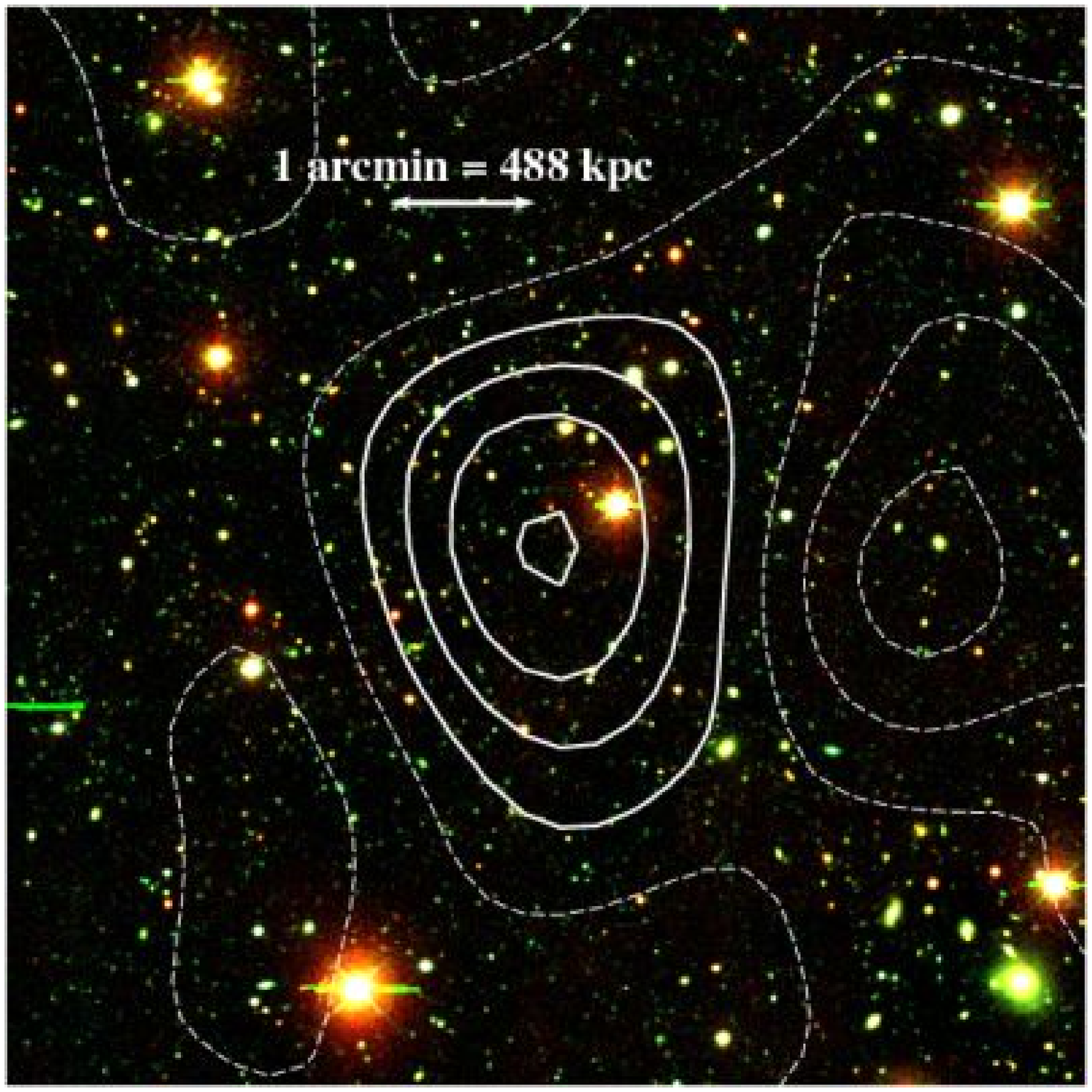} \\
\end{tabular}
\caption{SPT-CL J2342-5411, z=1.08 \label{2342app}}
\end{center}
\end{figure*}

In this section we discuss the X-ray emission for each of the objects in the sample and 
note any extended substructures or other features in the X-ray images. 
Nearby radio sources, as listed in the SUMSS survey 
are quoted with fluxes at 843 MHz, and we also discuss any particularly bright X-ray point-sources.
The X-ray images ($\sim 15\arcmin \times 15\arcmin$) in the $0.5$-$2.0$ keV band are shown in Figures 
\ref{0000app}-\ref{2359app} with optical $grz$-images from 
the Magellan Baade 6.5-m telescope \citep[see][]{high10}.
For SPT-CL J0546-5345 and SPT-CL J2342-5411 we instead show $grz$-images from the BCS survey since 
Magellan images were not available in all filters for these objects. 
Both X-ray and optical images are overlaid with SZ significance, $\xi$,
contours from V10.  The contours are from spatially filtered SPT maps
where the clusters appear as positive significance detections, and where
the spatial filtering often causes noticeable negative significance
ringing around the cluster.  The contour levels are spaced at 1.5$\sigma$
increments starting from zero with dashed contours for $\xi \leq 0$ and
solid contours for $\xi > 0$.
The position of the BCG (Stalder et al., in prep.) is marked with a white cross in the X-ray 
images. Large circles and arrows show the locations of interesting X-ray features, small circles 
correspond to locations of SUMSS sources mentioned in the text. 
North is up and east is left.

\subsection*{SPT-CL J0000-5748}
\label{0000}
\begin{figure*}[htb!]
\begin{center}
\begin{tabular}{cc}
\includegraphics[width=3.2in]{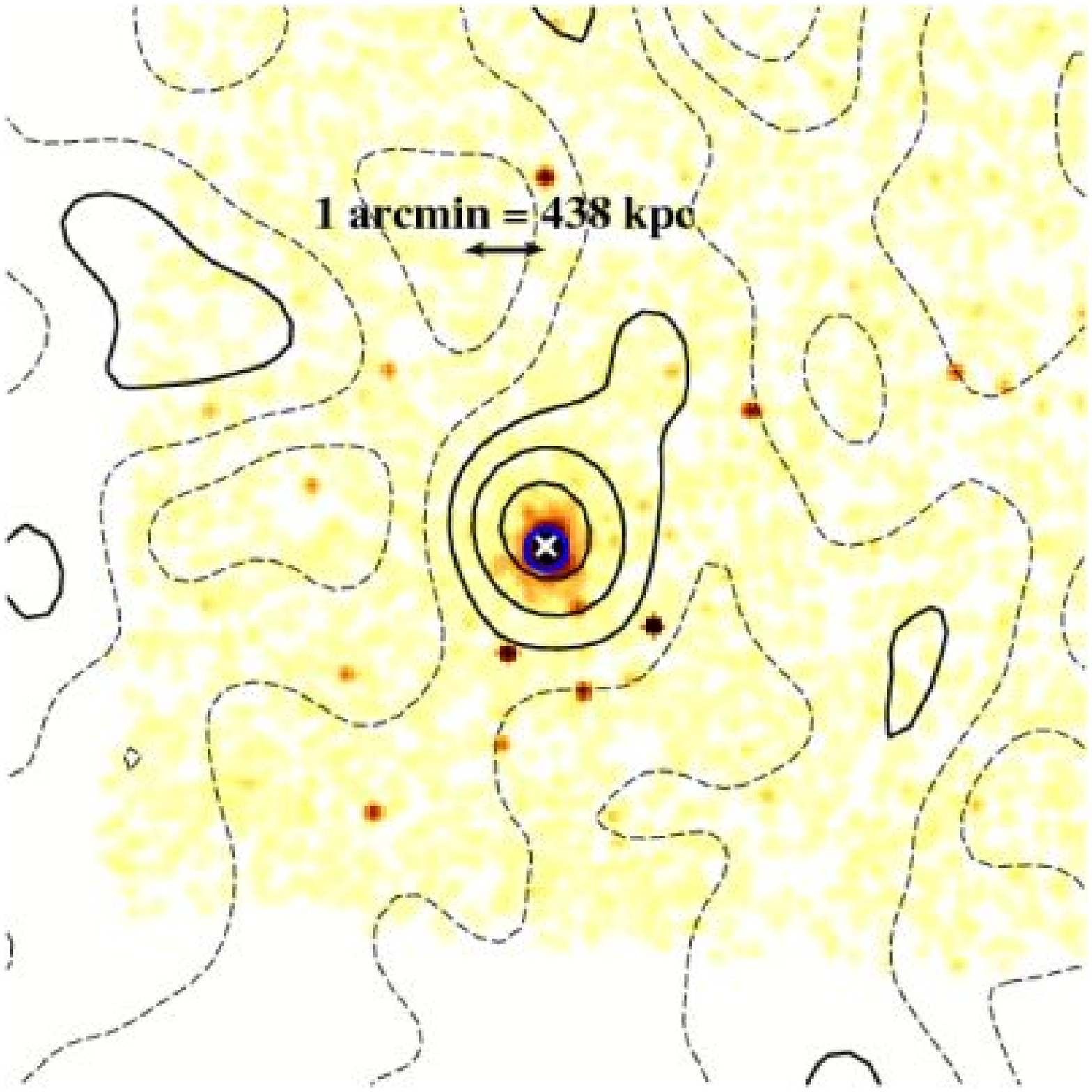} & 
\includegraphics[width=3.2in]{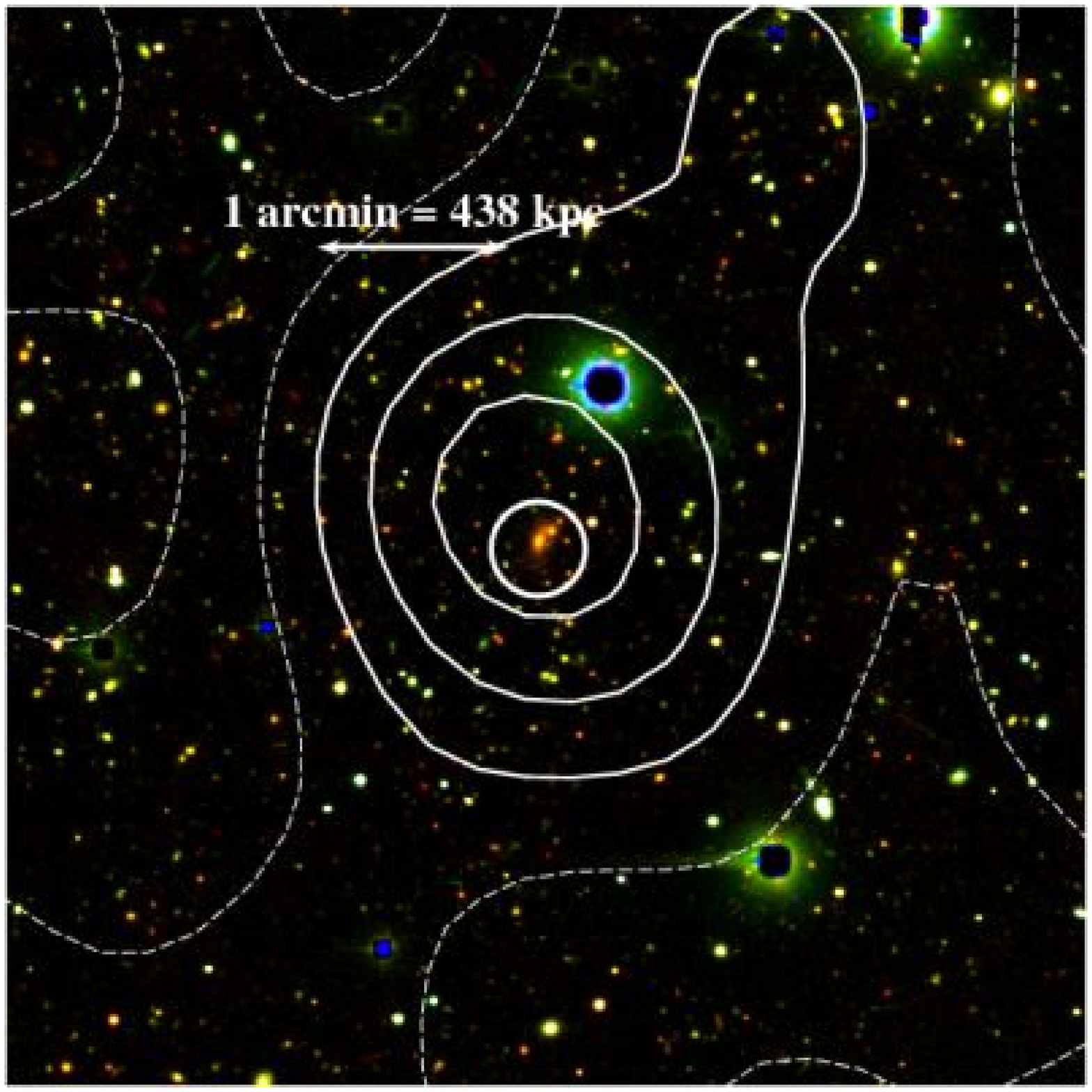} \\
\end{tabular}
\caption{SPT-CL J0000-5748, z=0.74 \label{0000app}}
\end{center}
\end{figure*}

This cluster is the highest redshift ($z=0.74$) cluster in the sample with a sharp central 
brightness peak (see Figure \ref{0000app}). The X-ray peak is coincident with the BCG and is 
also associated with a SUMSS radio source with a flux of $40.4\pm 1.5$ mJy indicative of 
a central AGN. The central brightness peak is broader than the {\sl Chandra} PSF and we do not 
associate the central X-ray emission with AGN emission.  

\subsection*{SPT-CL J0509-5342}
\label{0509}
\begin{figure*}[htb!]
\begin{center}
\begin{tabular}{cc}
\includegraphics[width=3.2in]{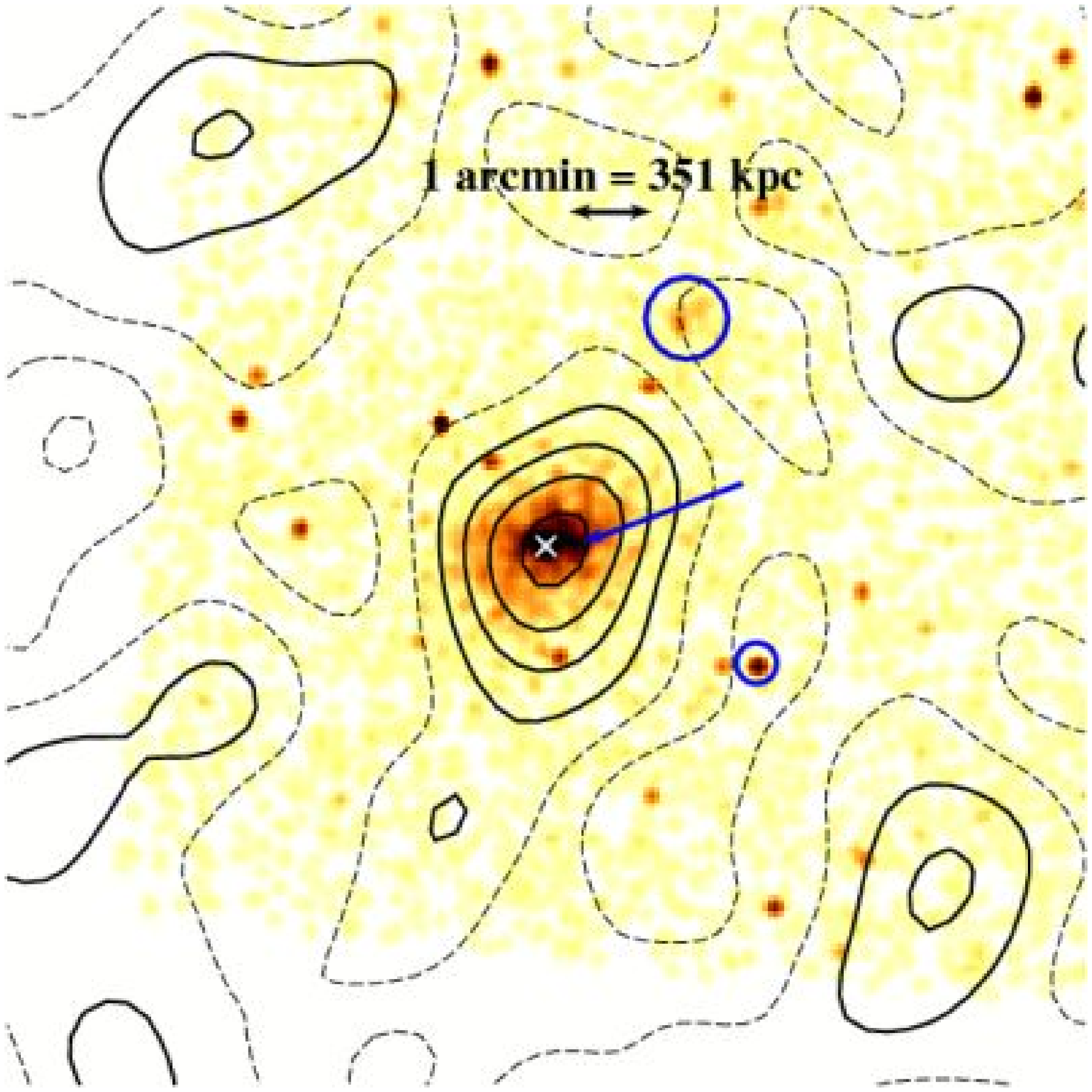} & 
\includegraphics[width=3.2in]{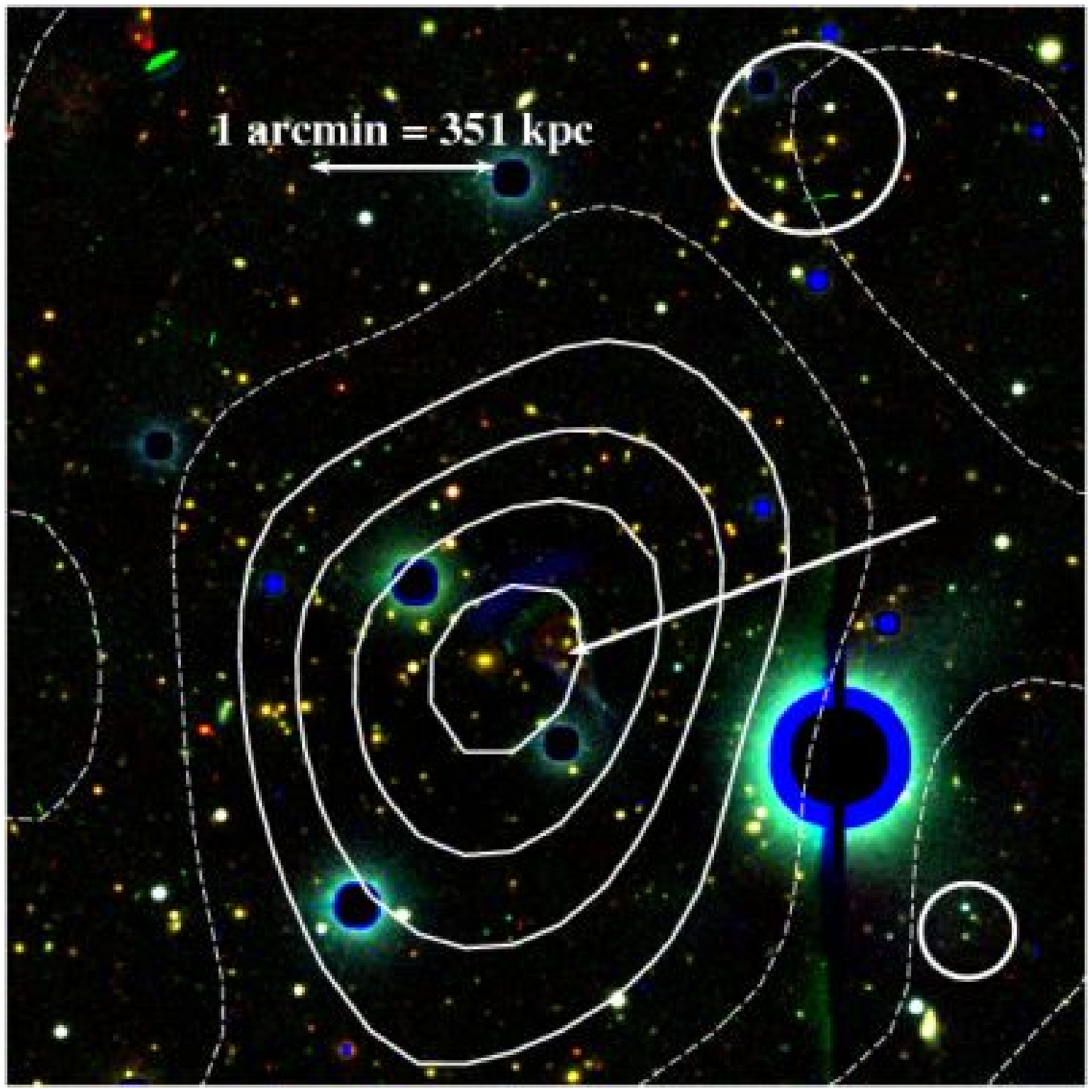} \\
\end{tabular}
\caption{SPT-CL J0509-5342, z=0.4626 \label{0509app}}
\end{center}
\end{figure*}

There is a clear double-peak in the central X-ray emission of this cluster (see Figure 
\ref{0509app}). The main peak is associated with the central BCG. The secondary 
peak is located $\sim 20 \arcsec$ west of the main component (arrow) and the X-ray brightness 
ratio is $1/4$. The secondary peak is also associated with two elliptical galaxies. The peak of the 
SZ flux is located between the two components.

The main component is the most centrally compact core in this sample which indicates that 
a possible merger is in an early stage.
A bright SUMSS radio source with a flux of $125.7$ mJy is located $2.8 \arcmin$ WSW of the cluster 
and is also coincident with an X-ray point source. 
This is not likely associated with the cluster.
A possible third cluster component shows up as an extended X-ray source $3.5 \arcmin$ NNW of the cluster 
(large circle) and is coincident with two elliptical galaxies in the optical image. 

\subsection*{SPT-CL J0516-5430}
\label{0516}
\begin{figure*}[htb!]
\begin{center}
\begin{tabular}{cc}
\includegraphics[width=3.2in]{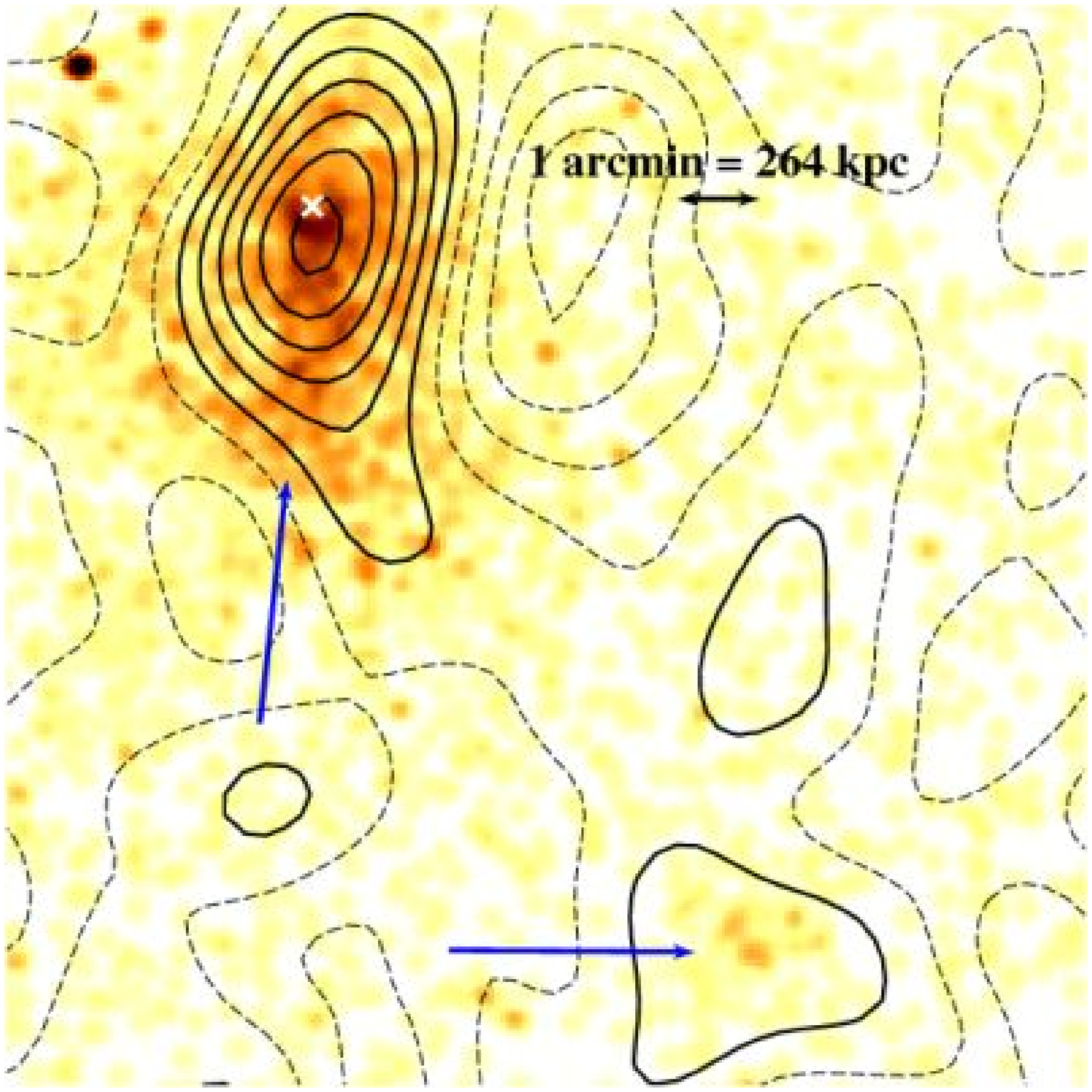} & 
\includegraphics[width=3.2in]{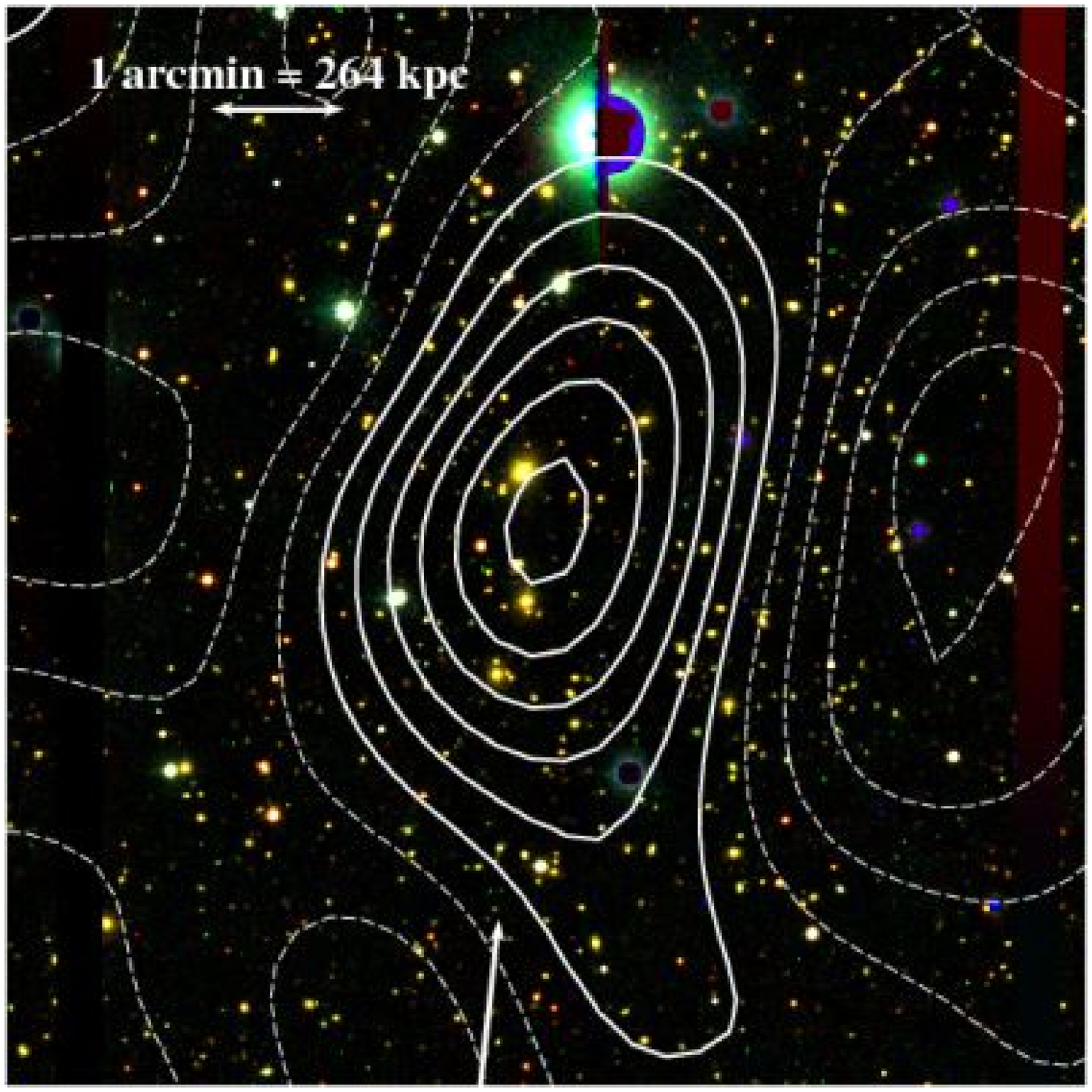} \\
\end{tabular}
\caption{SPT-CL J0516-5430, z=0.2952 \label{0516app}}
\end{center}
\end{figure*}

This apparently merging cluster is very elongated along the N-S direction and also features a string of galaxies 
with the same alignment (see arrow, Figure \ref{0516app}). The BCG is also offset by $30 \arcsec$ from 
both the X-ray and SZ peaks. 
There is a second extended X-ray component $9.5 \arcmin$ SW of the main cluster (SW arrow). This 
possible subcluster also shows up in the SZ map as a S/N$>2$ detection.  The subcluster is 
outside of the field of the Magellan observation and no galaxy appears to be associated with 
this component in a Digitized Sky Survey image.

\subsection*{SPT-CL J0528-5300}
\label{0528}
\begin{figure*}[htb!]
\begin{center}
\begin{tabular}{cc}
\includegraphics[width=3.2in]{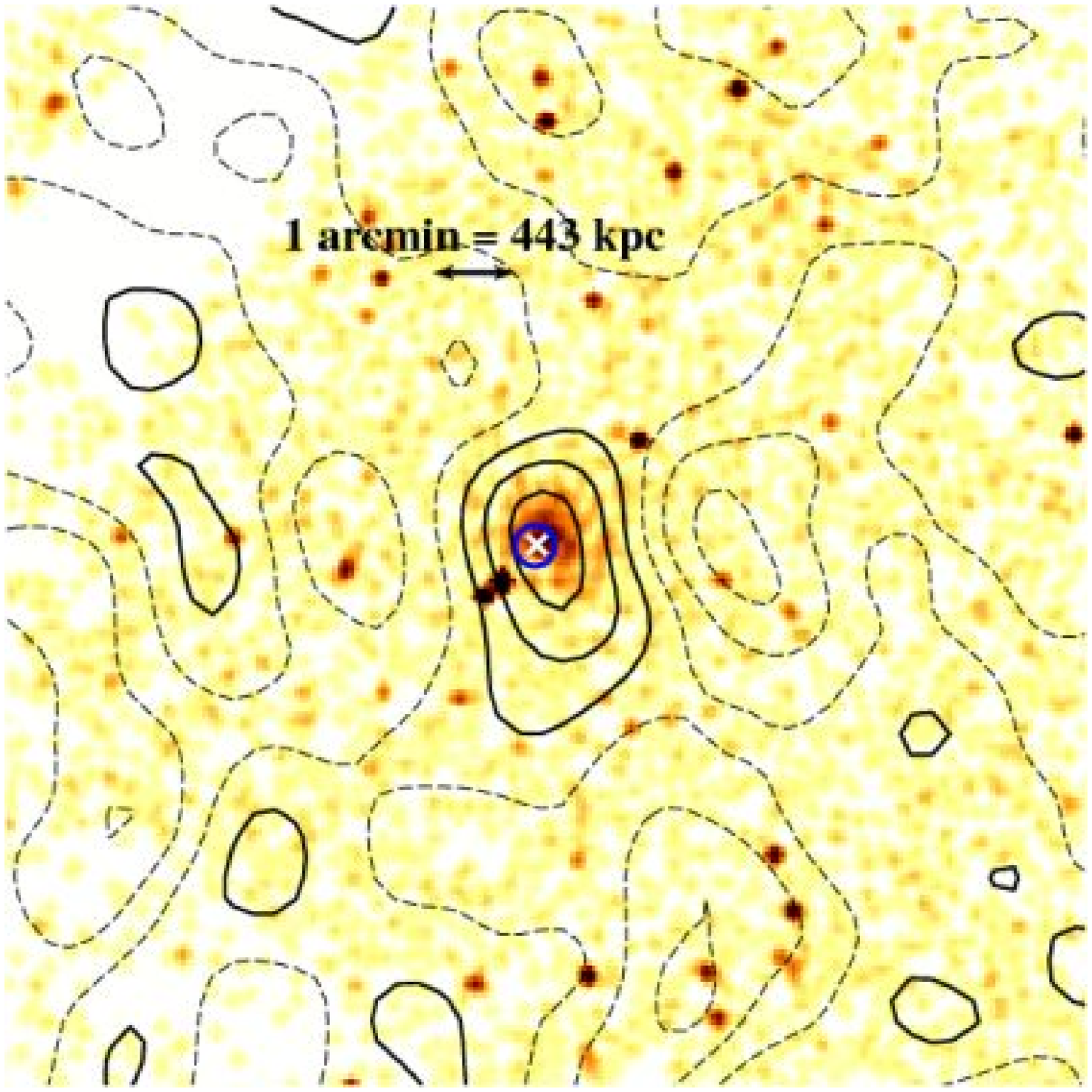} & 
\includegraphics[width=3.2in]{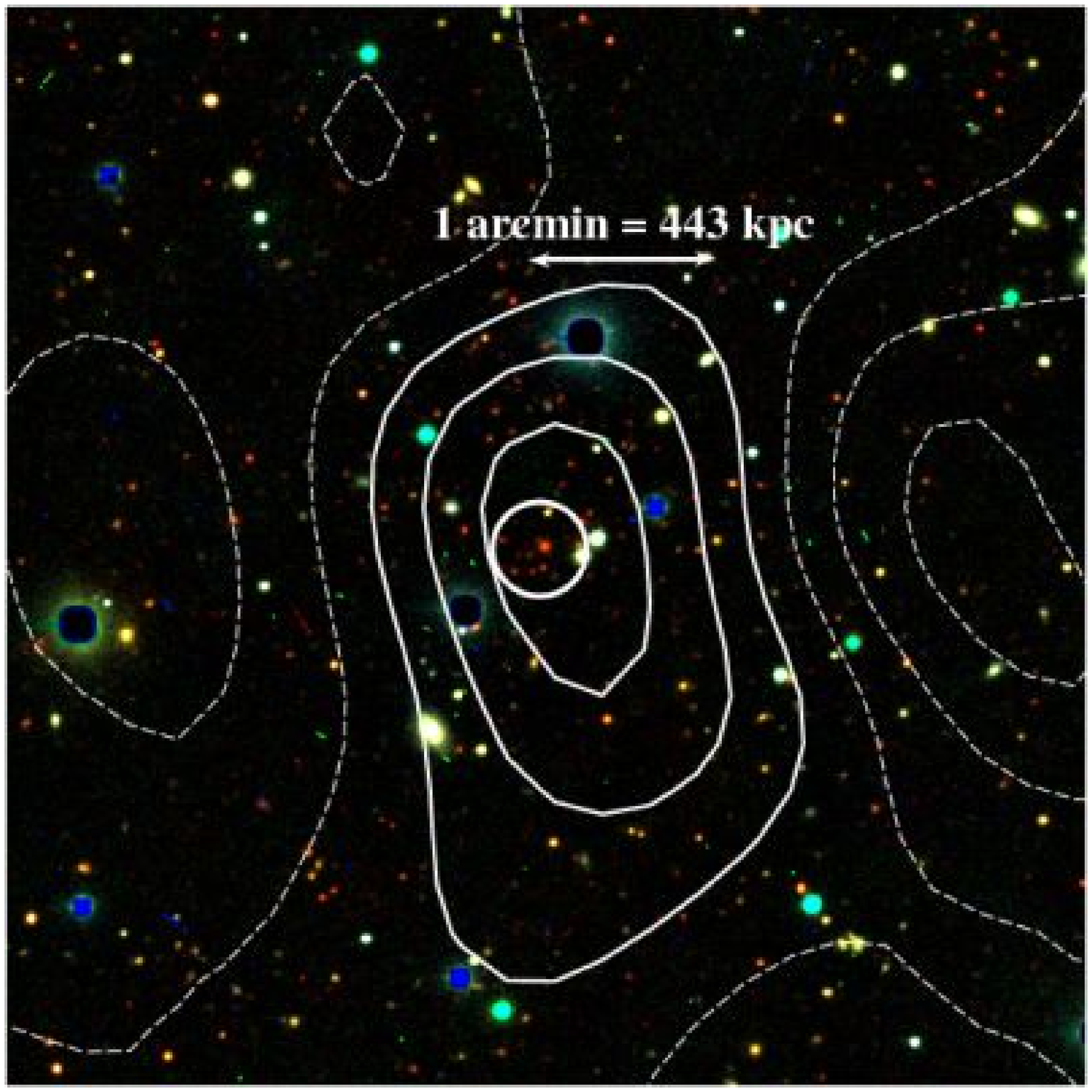} \\
\end{tabular}
\caption{SPT-CL J0528-5300, z=0.7648 \label{0528app}}
\end{center}
\end{figure*}

This is a faint cluster ($L_X(0.5-2.0\mathrm{kev}) = 1.9\pm0.2~10^{44}$ erg$~$s$^{-1}$) without much 
structure and the BCG is offset by $15 \arcsec$ from the 
X-ray peak (Figure \ref{0528app}). The BCG is coincident with a SUMSS radio source with a flux of 
$61.2\pm 2.0$ mJy likely associated with a central AGN. 

\subsection*{SPT-CL J0533-5005}
\label{0533}
\begin{figure*}[htb!]
\begin{center}
\begin{tabular}{cc}
\includegraphics[width=3.2in]{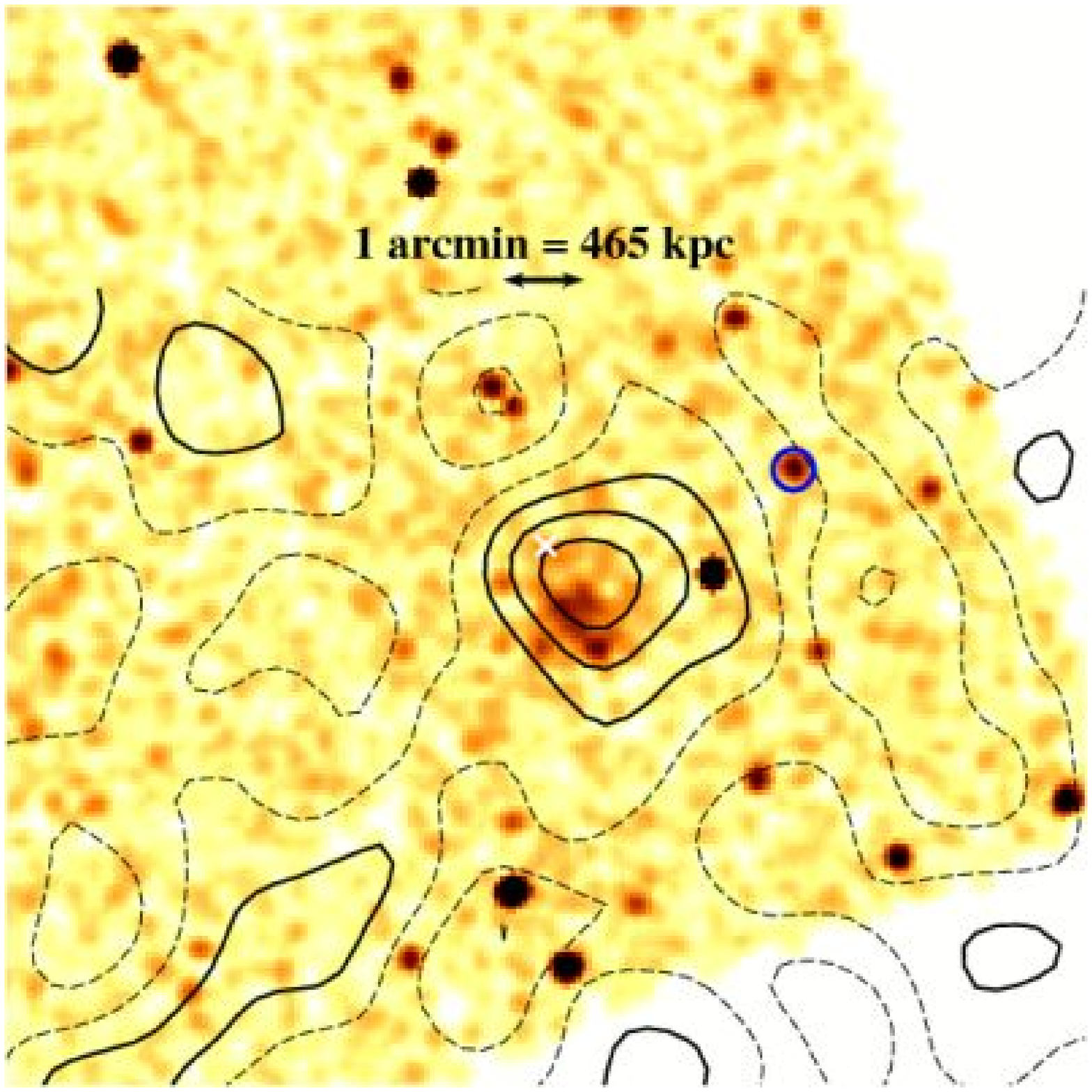} & 
\includegraphics[width=3.2in]{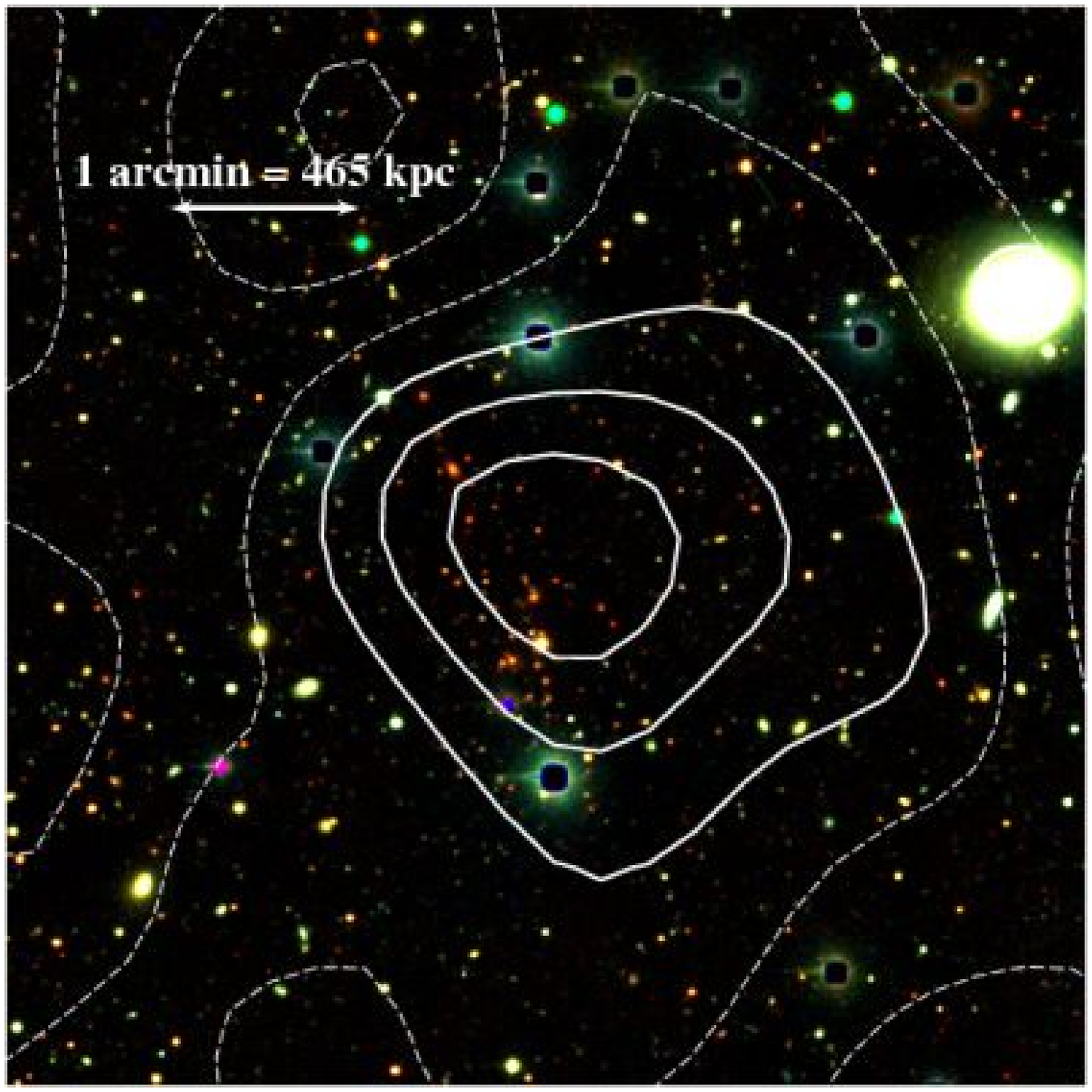} \\
\end{tabular}
\caption{SPT-CL J0533-5005, z=0.8810 \label{0534app}}
\end{center}
\end{figure*}

This is another faint cluster ($L_X(0.5-2.0\mathrm{kev}) = 1.2\pm0.3~10^{44}$ erg$~$s$^{-1}$) with 
no well defined core. The BCG is offset $45 \arcsec$ from the central X-ray emission indicating 
an unrelaxed state (Figure \ref{0534app}). 
The SZ and X-ray peaks are also offset by $30\arcsec$. 
A local galaxy at $z=0.0147$ is located $3 \arcmin$ NW of the cluster (circle) but is not likely to 
affect X-ray or SZ measurements.

\subsection*{SPT-CL J0546-5345}
\label{0547}
\begin{figure*}[htb!]
\begin{center}
\begin{tabular}{cc}
\includegraphics[width=3.2in]{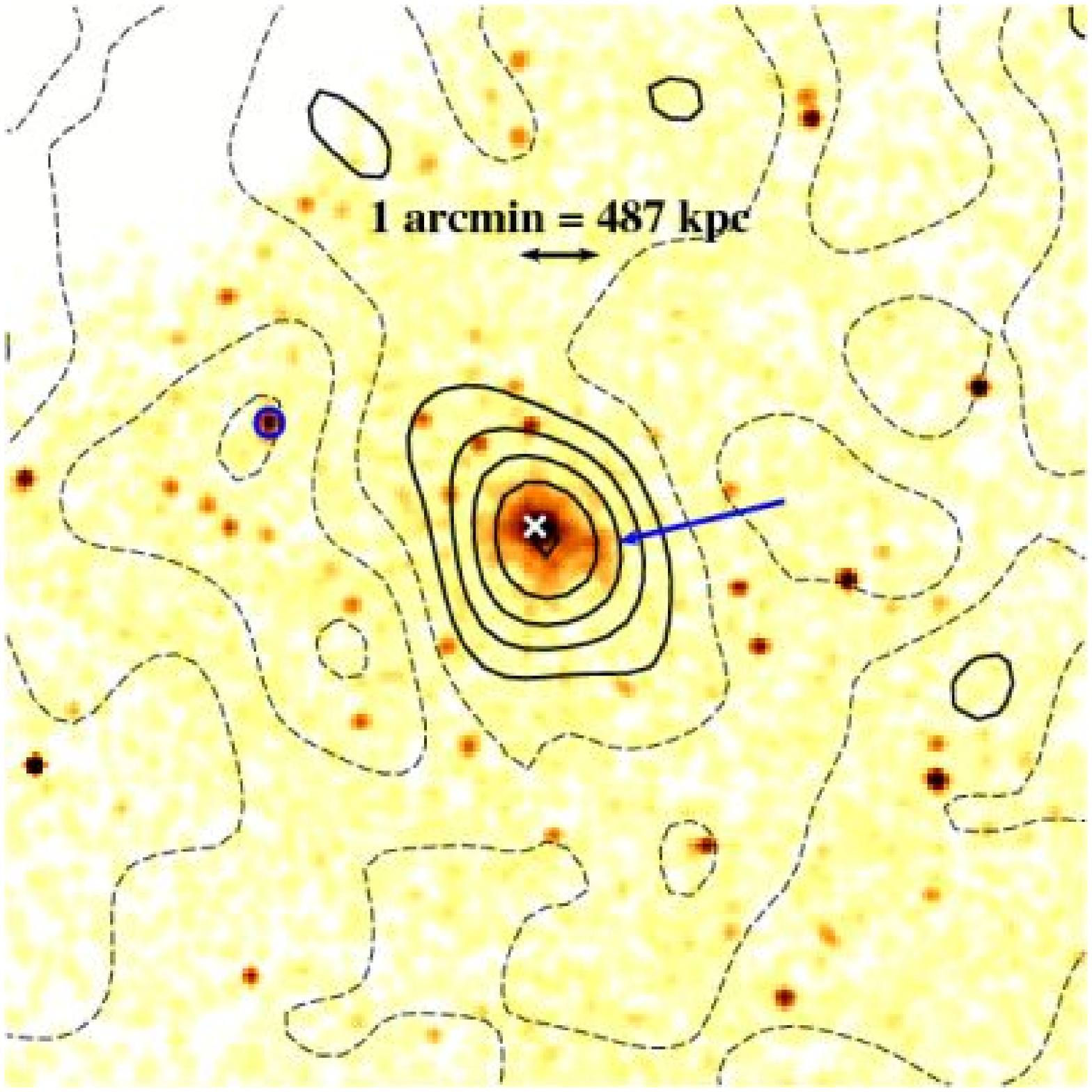} & 
\includegraphics[width=3.2in]{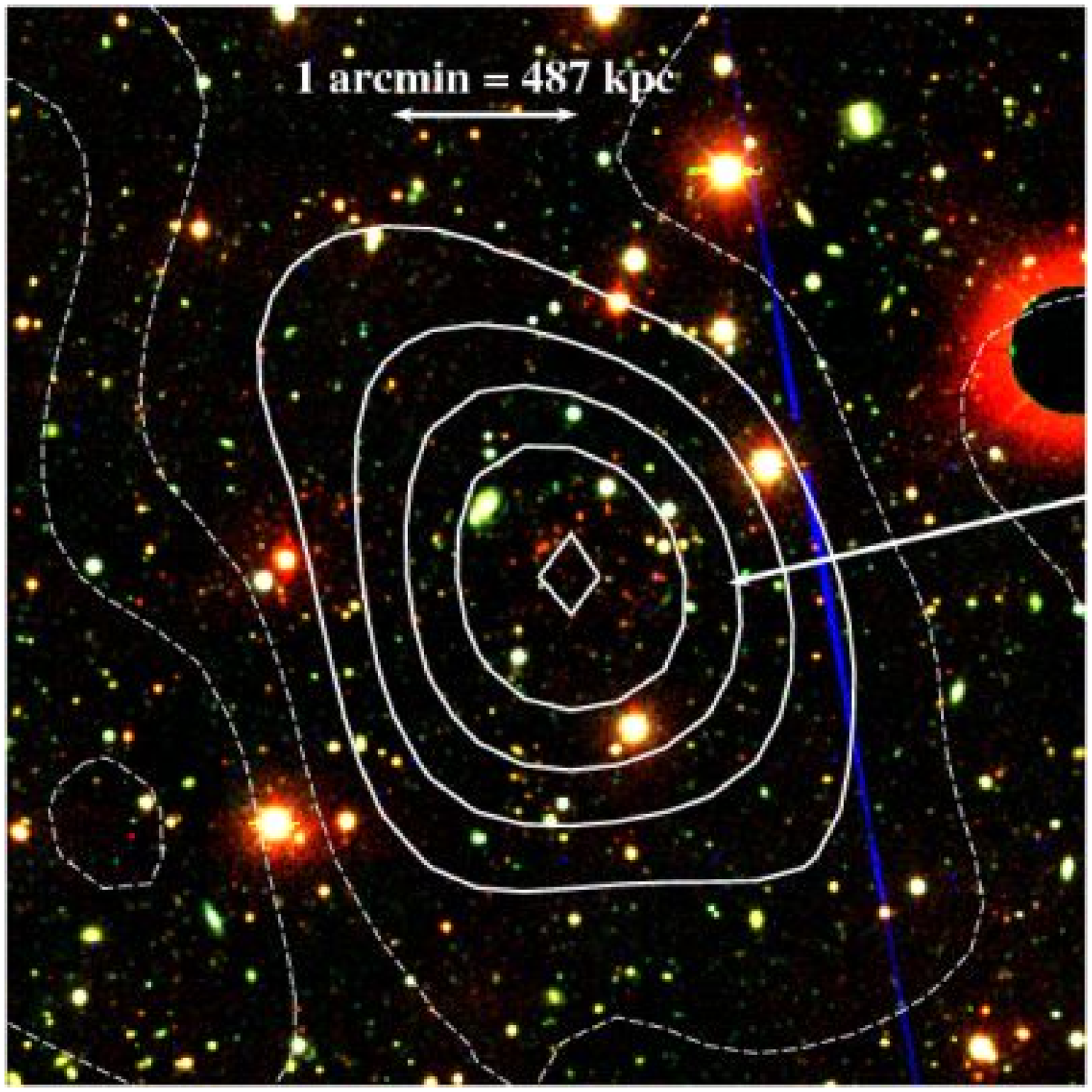} \\

\end{tabular}
\caption{SPT-CL J0546-5345, z=1.0665 \label{0547app}}
\end{center}
\end{figure*}

The X-ray image in Figure \ref{0547app} shows a substructure extending $\sim 1\arcmin$ 
SW from the main cluster (arrow) suggesting that a minor merger may be taking place. 
This elongation is aligned with an apparent extension of the SZ signal further 
supporting this scenario. 
One of the X-ray point sources, $6\arcmin$ NW, is associated with a SUMSS radio source 
with a radio flux of $19.1 \pm 0.9$ mJy. 

\subsection*{SPT-CL J0551-5709}
\label{0552}
\begin{figure*}[htb!]
\begin{center}
\begin{tabular}{cc}
\includegraphics[width=3.2in]{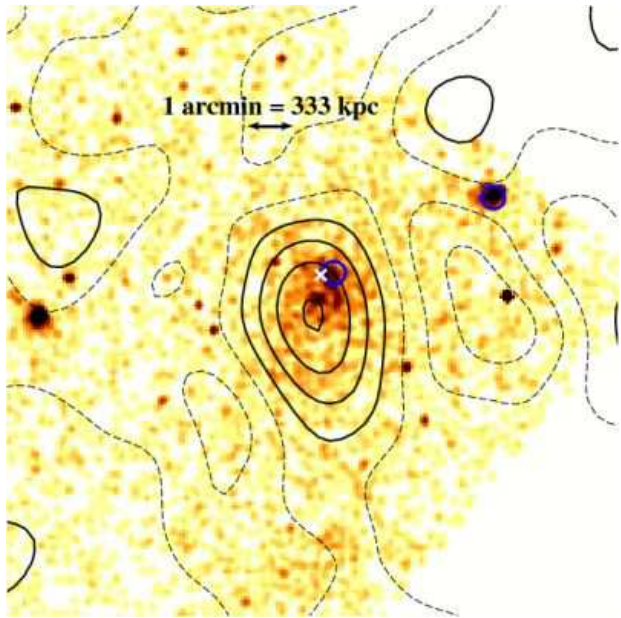} & 
\includegraphics[width=3.2in]{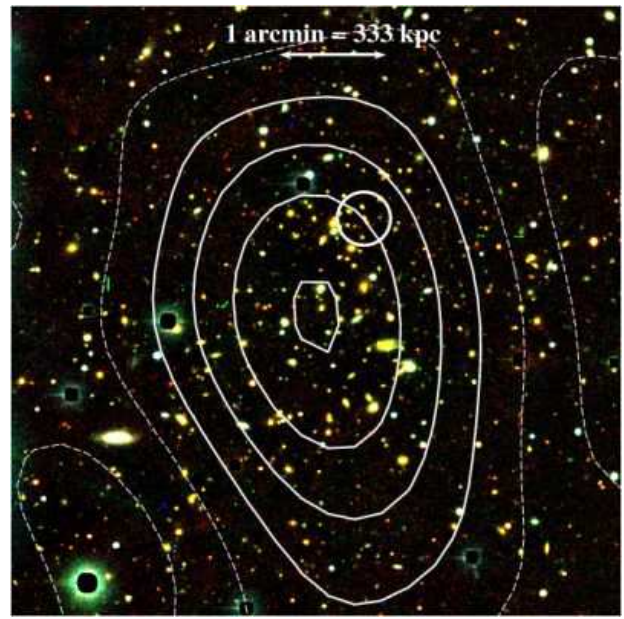} \\
\end{tabular}
\caption{SPT-CL J0551-5709, z=0.4230 \label{0551app}}
\end{center}
\end{figure*}

This apparently merging cluster has a disturbed X-ray morphology and the X-ray and SZ peaks are offset 
by $30\arcsec$. A SUMSS radio source with flux of $22.7 \pm 1.6$ mJy is located $15 \arcsec$ 
east of the BCG in a region with an overdensity of cluster galaxies (Figure \ref{0551app}). 
Another radio source is located $5\arcmin$ NE of the cluster and is associated with an 
X-ray source and a local galaxy.

The X-ray morphology and temperature structure of this object suggest that 
it is projected on top of a low redshift structure. Extended emission can be seen 
both towards the S and NW, covering a large area beyond $r_{500}$. 
The temperature decreases from $4.4~$keV to $3.3~$keV when increasing the 
aperture radius from $0.5~r_{500}$ to $r_{500}$, implying that this extended emission 
has very low temperature.
The emission is likely associated with Abell S0552, located at this position. 
\citet{high10} find a strong red sequence at $z=0.09$ at the cluster location. 
The X-ray emission associated with this local cluster will likely bias any 
measurement of the temperature due to the contribution of the low-$z$ flux at large 
radii. We attempt to correct for this bias as described in \S \ref{sec:yxresults}. 

\subsection*{SPT-CL J0559-5249}
\label{0559}
\begin{figure*}[htb!]
\begin{center}
\begin{tabular}{cc}
\includegraphics[width=3.2in]{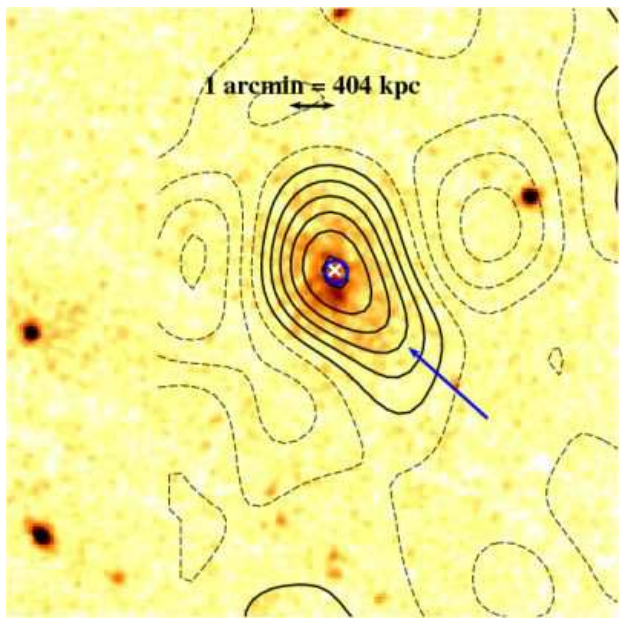} & 
\includegraphics[width=3.2in]{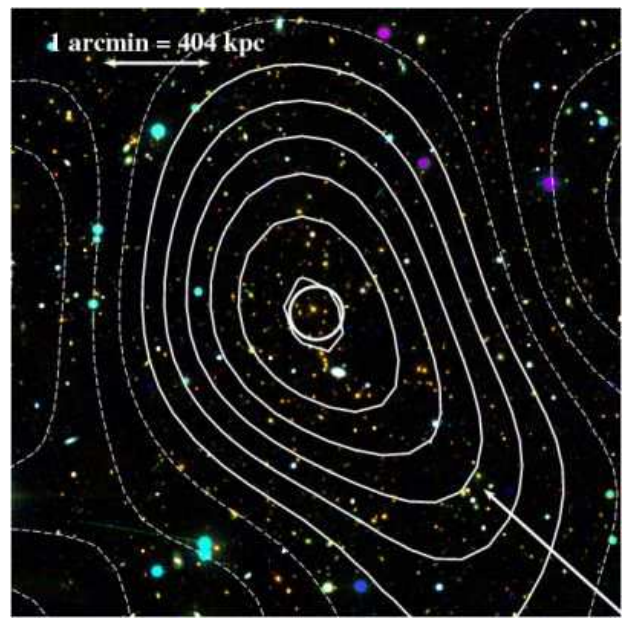} \\
\end{tabular}
\caption{SPT-CL J0559-5249, z=0.6112 \label{0559app}}
\end{center}
\end{figure*}

The peak of the X-ray emission is offset to the south by $\sim 25\arcsec$ with respect to the 
BCG location and SZ peak in this cluster (Figure \ref{0559app}). The X-ray emission 
south of the BCG is associated with a group of galaxies at this location. 
The BCG is coincident with a $50.1 \pm 2.7$ mJy SUMSS radio source. 
On larger scales, the X-ray emission and SZ signal are extended to the SW, clearly 
visible $2.5 \arcmin$ from the cluster core (arrow in Figure \ref{0559app}). 
This cluster is clearly in a merging state.

\subsection*{SPT-CL J2331-5051}
\begin{figure*}[htb!]
\begin{center}
\begin{tabular}{cc}
\includegraphics[width=3.2in]{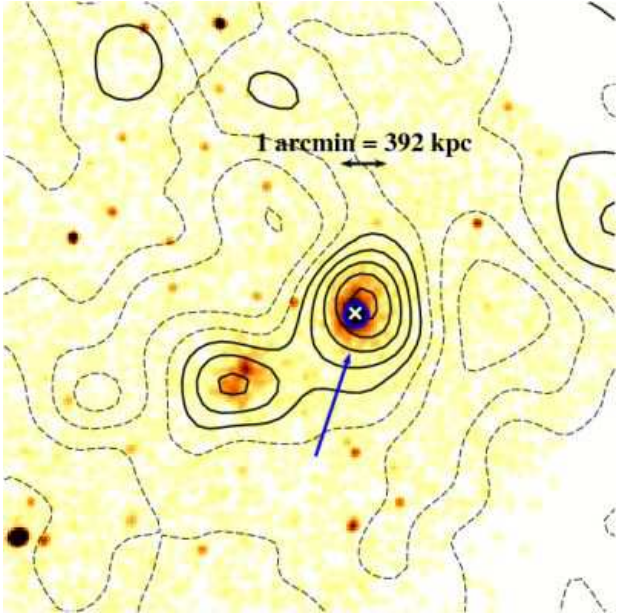} & 
\includegraphics[width=3.2in]{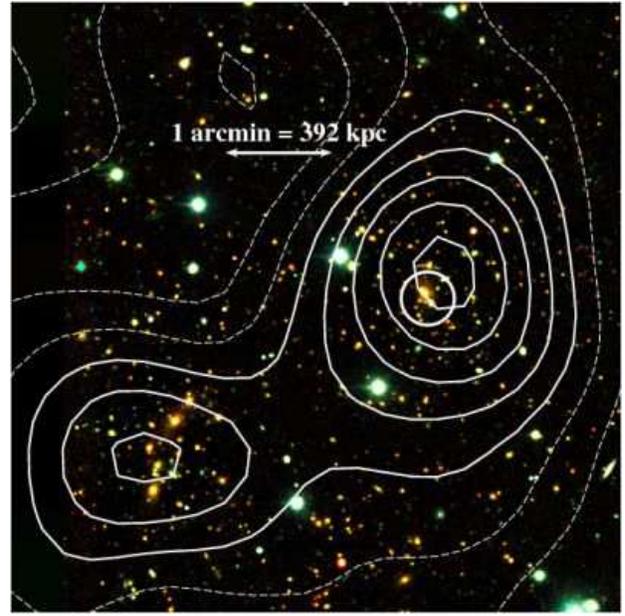} \\
\end{tabular}
\caption{SPT-CL J2331-5051, z=0.5707 \label{2331app}}
\end{center}
\end{figure*}

As noted in V10, this cluster appears to be a pair of merging clusters with 
the smaller component located $2.5 \arcmin$ SE of the main cluster. 
A spectral analysis of the smaller component indicates that it has a 
luminosity of $2 \times 10^{43}$ erg s$^{-1}$, roughly 5\% of the main cluster.
As can be seen in Figure \ref{2331app}, the location of the X-ray 
emission and SZ decrement at the subcluster show very good agreement. 
This subcluster was also confirmed \citep[see][]{high10} to be at the same redshift 
as the main cluster. 
Additionally, this cluster has a well defined X-ray cavity $18 \arcsec$ 
S of the X-ray peak (arrow). This cluster also hosts a cool core, 
coincident with the BCG and a $13.3 \pm 1.0$ mJy SUMSS radio source.

\subsection*{SPT-CL J2332-5358}
\begin{figure*}[htb!]
\begin{center}
\begin{tabular}{cc}
\includegraphics[width=3.2in]{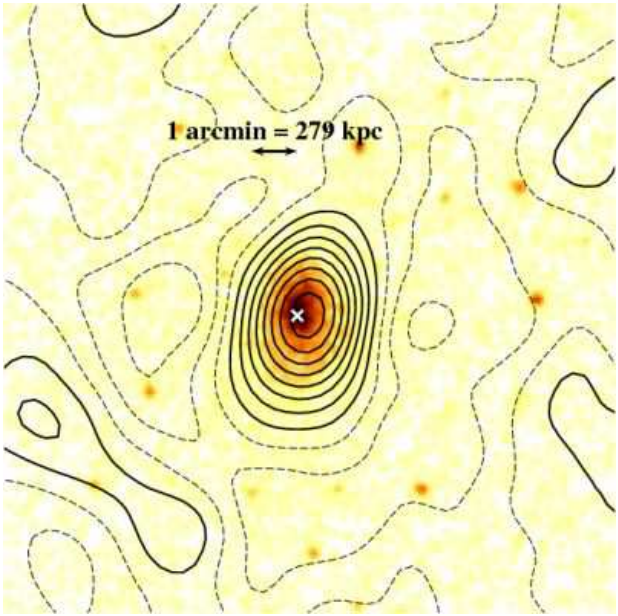} & 
\includegraphics[width=3.2in]{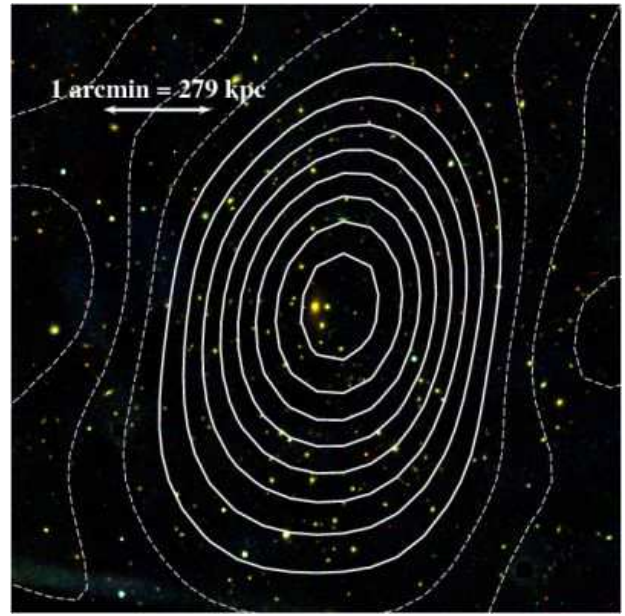} \\
\end{tabular}
\caption{SPT-CL J2332-5358, z=0.32 \label{2332app}}
\end{center}
\end{figure*}

This nearby ($z=0.32$) regular cluster shows no sign of significant structure in our 
{\sl XMM-Newton} observation with 7 ks of good data. 
The location of the BCG, the X-ray peak and the peak of the SZ-signal 
show good agreement and the cluster is likely relaxed (see Figure \ref{2332app}).
This cluster is also the only candidate in V10 with a point source detected 
in the SPT 220 GHz maps at $>5\sigma$. 
We correct the SZ flux and position of the cluster by subtracting this 
source from the 150 GHz assuming a dust spectral index as described in 
\S \ref{sec:point_source}. 

\subsection*{SPT-CL J2337-5942}
\begin{figure*}[htb!]
\begin{center}
\begin{tabular}{cc}
\includegraphics[width=3.2in]{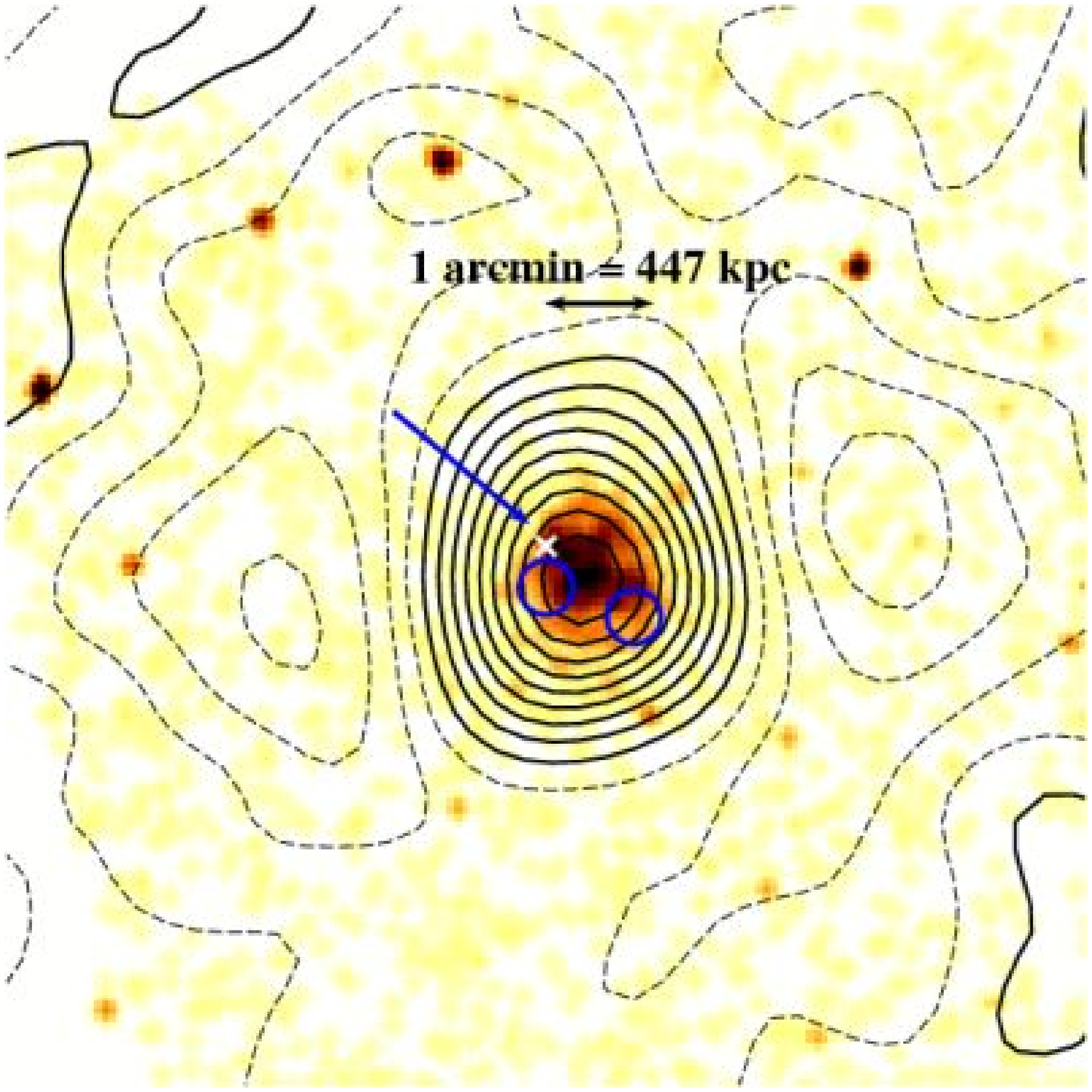} & 
\includegraphics[width=3.2in]{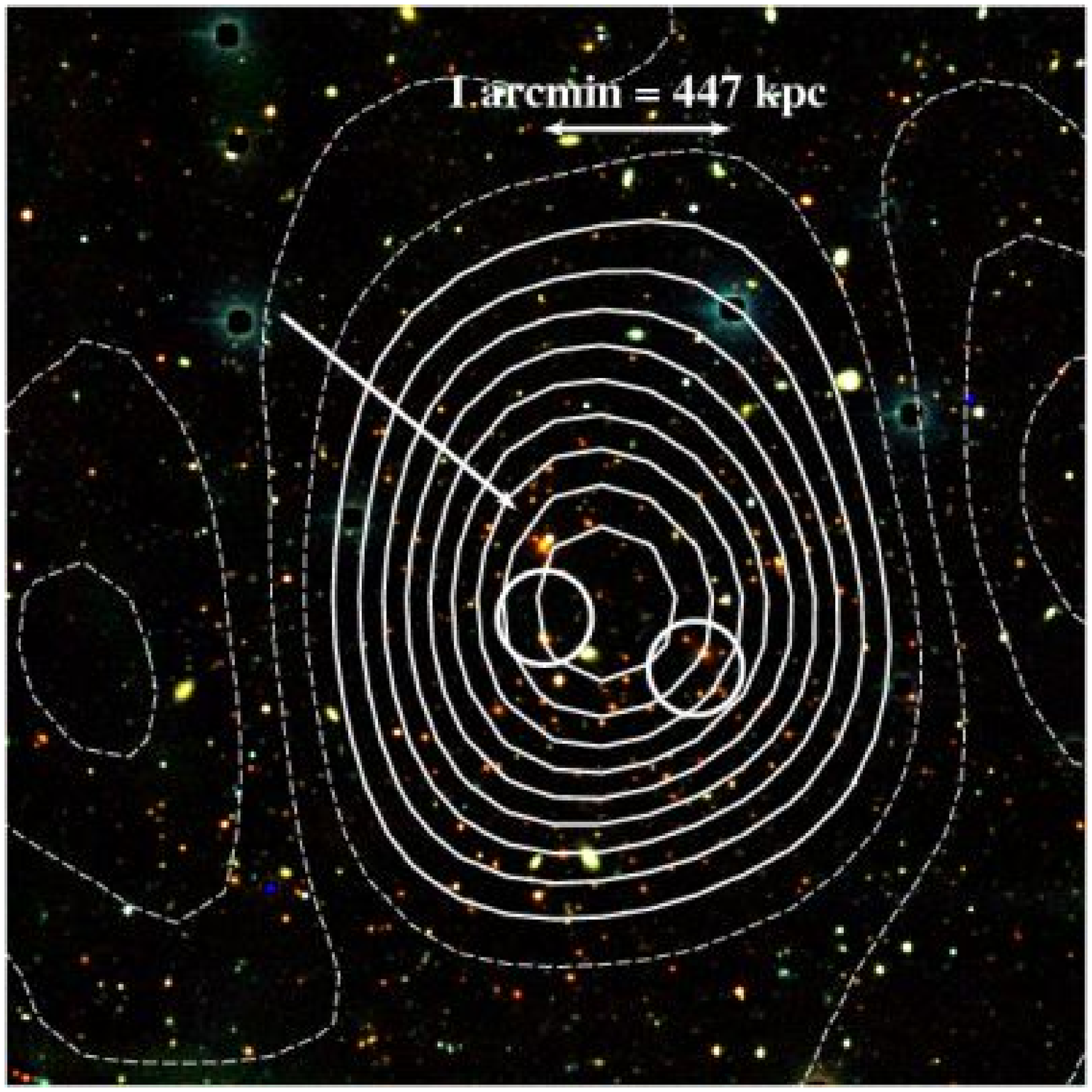} \\
\end{tabular}
\caption{SPT-CL J2337-5942, z=0.7814 \label{2337app}}
\end{center}
\end{figure*}

This cluster shows a sharp X-ray surface brightness discontinuity NE of the cluster 
center (arrow in Figure \ref{2337app}). Interestingly, the BCG is also located in this region, 
providing tentative evidence that this is a compact gas core moving through the ambient cluster 
medium, driving a shock front. 
The cluster shows an overall irregular morphology without a clear peak in the 
X-ray emission. This is the cluster with the highest SZ signal in the sample 
and is the second most massive with $M_{500,Y_X} = 7.43\pm0.80 \times 10^{14} M_{\odot}$.
There are two SUMSS radio sources with $7.1 \pm 1.0$ mJy and $7.8 \pm 1.0$ mJy fluxes 
$33 \arcsec$ and $23 \arcsec$ respectively from the cluster center.

\subsection*{SPT-CL J2341-5119}
\begin{figure*}[htb!]
\begin{center}
\begin{tabular}{cc}
\includegraphics[width=3.2in]{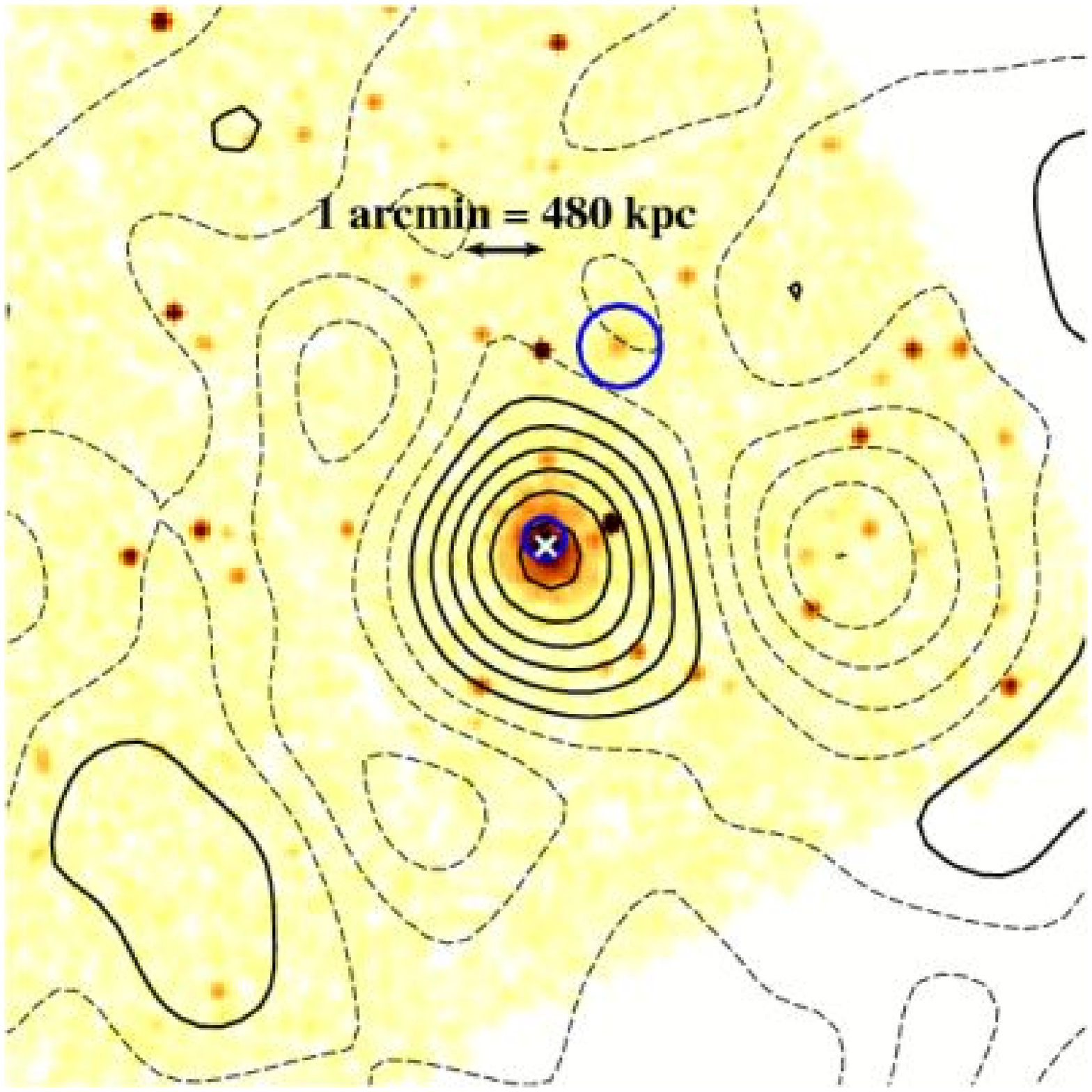} & 
\includegraphics[width=3.2in]{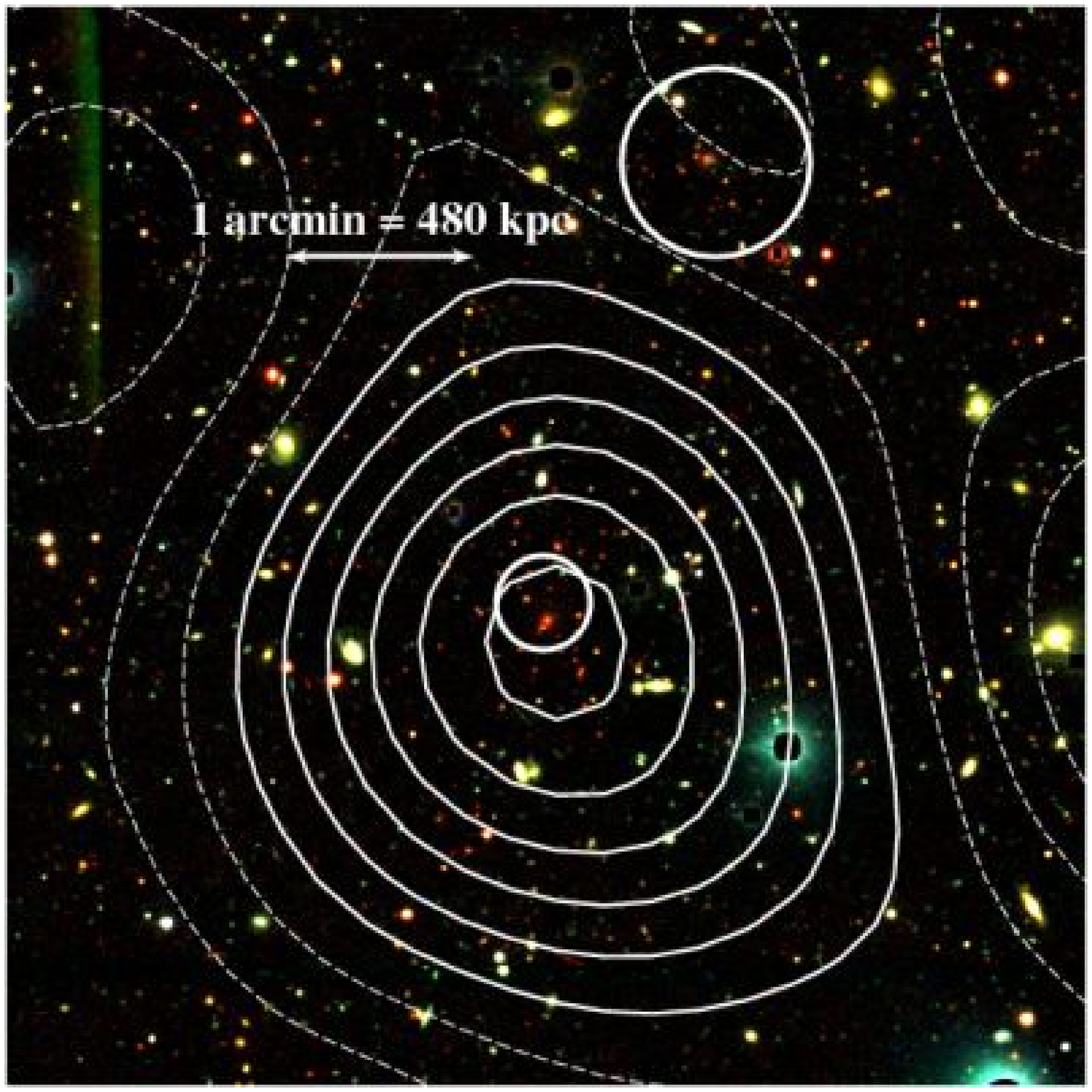} \\
\end{tabular}
\caption{SPT-CL J2341-5119, z=0.9983 \label{2341app}}
\end{center}
\end{figure*}

In this cluster there are no signs of significant structure within the cluster 
in the X-ray data (see Figure \ref{2341app}). 
However, $2.5 \arcmin$ NNW of the cluster center (large circle) there is 
a faint extended X-ray source at the location of a bright galaxy indicating that 
this is a satellite galaxy group about to merge with the main cluster. 
The BCG shows good correspondence with the X-ray peak and the main cluster appears 
relaxed. There is a $7.9 \pm 1.1$ mJy SUMSS radio source $6 \arcsec$ N of the BCG 
which could be associated with a central AGN.

\subsection*{SPT-CL J2342-5411}
This high redshift ($z=1.08$) cluster shows no signs of significant merger activity and 
is likely relaxed. However there is a possible secondary component or tail to the SW, 
$15 \arcsec$ from the core.

\subsection*{SPT-CL J2343-5521}
\label{2343nondet}
In a 70 ks {\sl Chandra} observation, no X-ray source was found at the location of 
this SPT detection.
This field was also imaged with BCS and Magellan with no optical counterpart found.  
The optical data suggested that this cluster would have to be at $z \gtrsim 1.2$ 
to not find an optical counterpart in either observation \citep{high10}.  
The long {\sl Chandra} observation allows us to put strict upper limits on the 
luminosity. Assuming the cluster is at $z=1$, the $5 \sigma$ upper limit 
is $L_X (0.5-2.0 \mathrm{keV}) < 3 \times 10^{42} \mathrm{ergs}~\mathrm{s}^{-1}$. 
The same limit assuming $z=2$ is $10^{43}$ $\mathrm{ergs}~\mathrm{s}^{-1}$. 
Additionally, no cluster counterparts could be identified in a 3.6 $\mu$m Spitzer/IRAC observation.
The X-ray, IR, and optical data all strongly support that this is a 
single band false detection by SPT.

\subsection*{SPT-CL J2355-5056}
\begin{figure*}[htb!]
\begin{center}
\begin{tabular}{cc}
\includegraphics[width=3.2in]{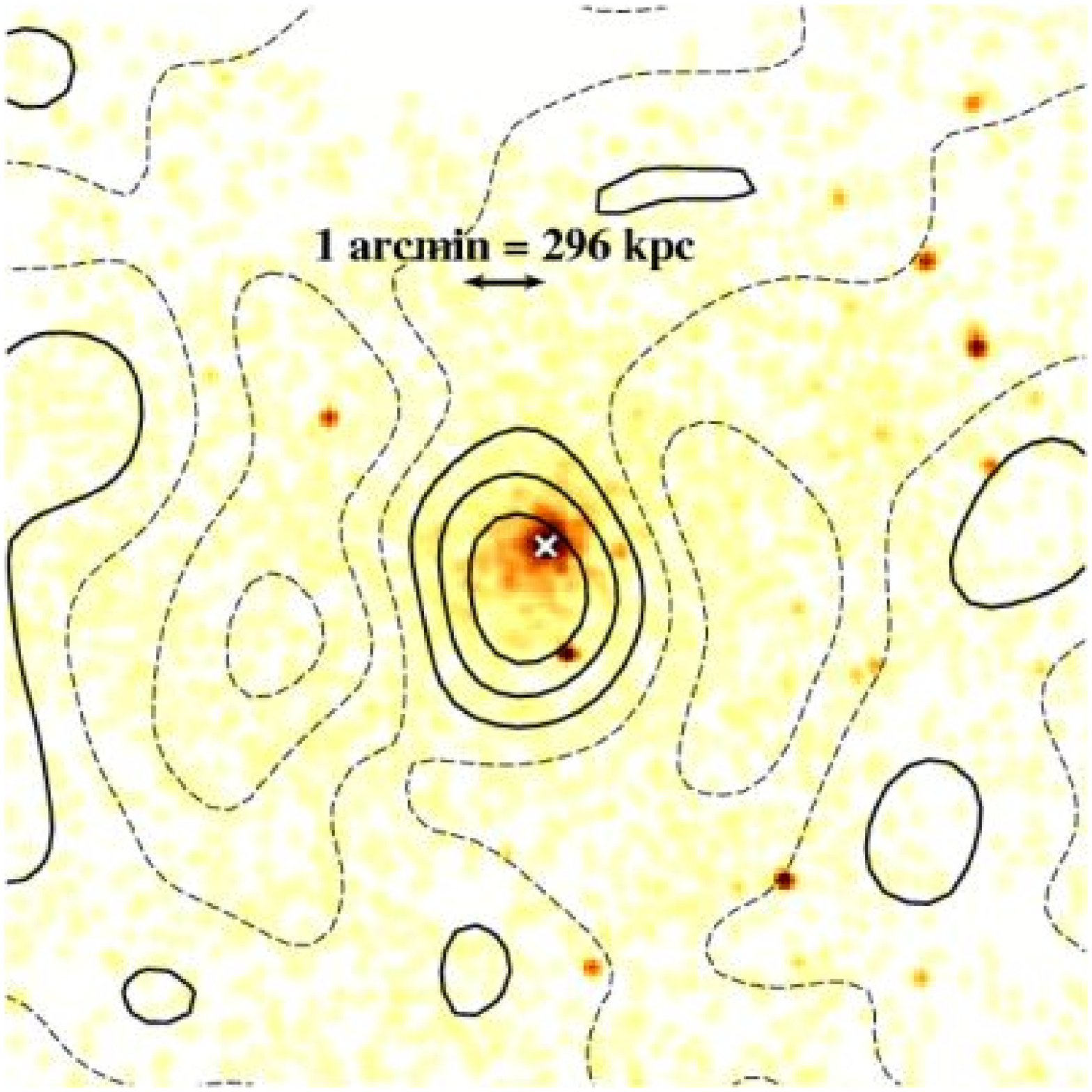} & 
\includegraphics[width=3.2in]{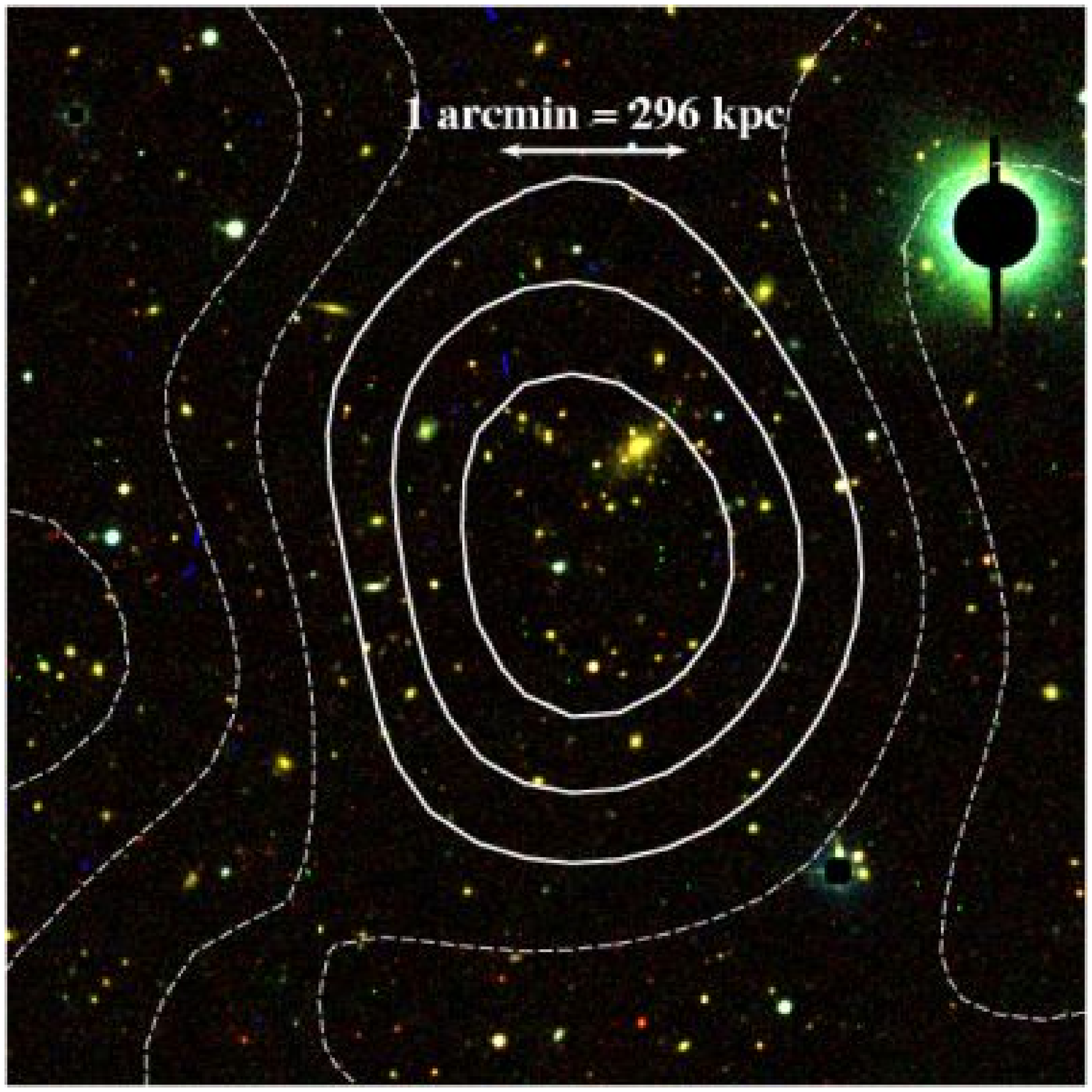} \\
\end{tabular}
\caption{SPT-CL J2355-5056, z=0.35 \label{2355app}}
\end{center}
\end{figure*}

This cluster shows good agreement between the position of the BCG and the X-ray peak and 
shows no significant structure in the X-ray image (Figure \ref{2355app}). 
The SZ peak is offset $\sim 35\arcsec$ to the south of the BCG while there is no 
evidence for extended X-ray emission in this direction. 
The X-ray emission is peaked in the center, likely due to a cool core.

\subsection*{SPT-CL J2359-5009}
\label{2359}
\begin{figure*}[htb!]
\begin{center}
\begin{tabular}{cc}
\includegraphics[width=3.2in]{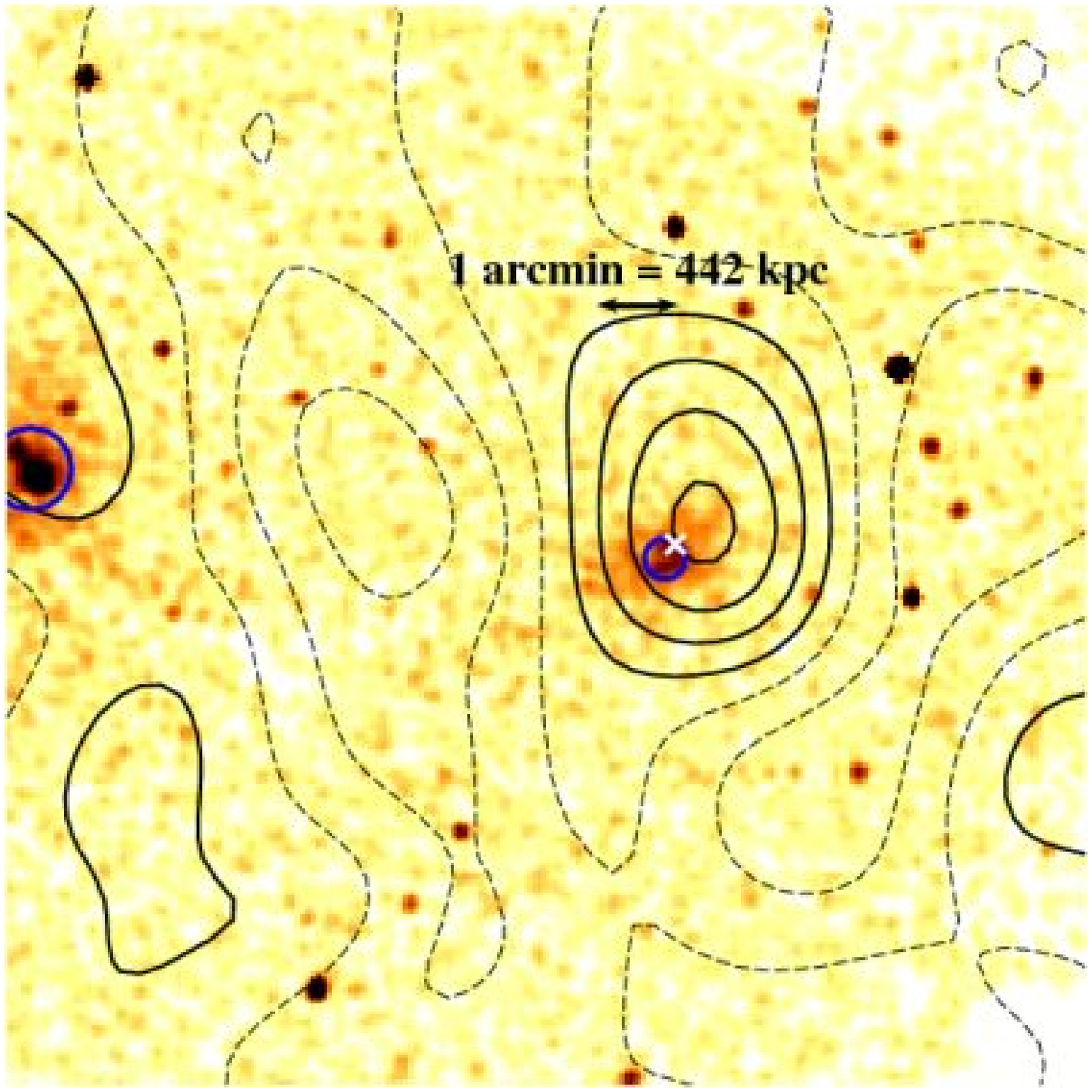} & 
\includegraphics[width=3.2in]{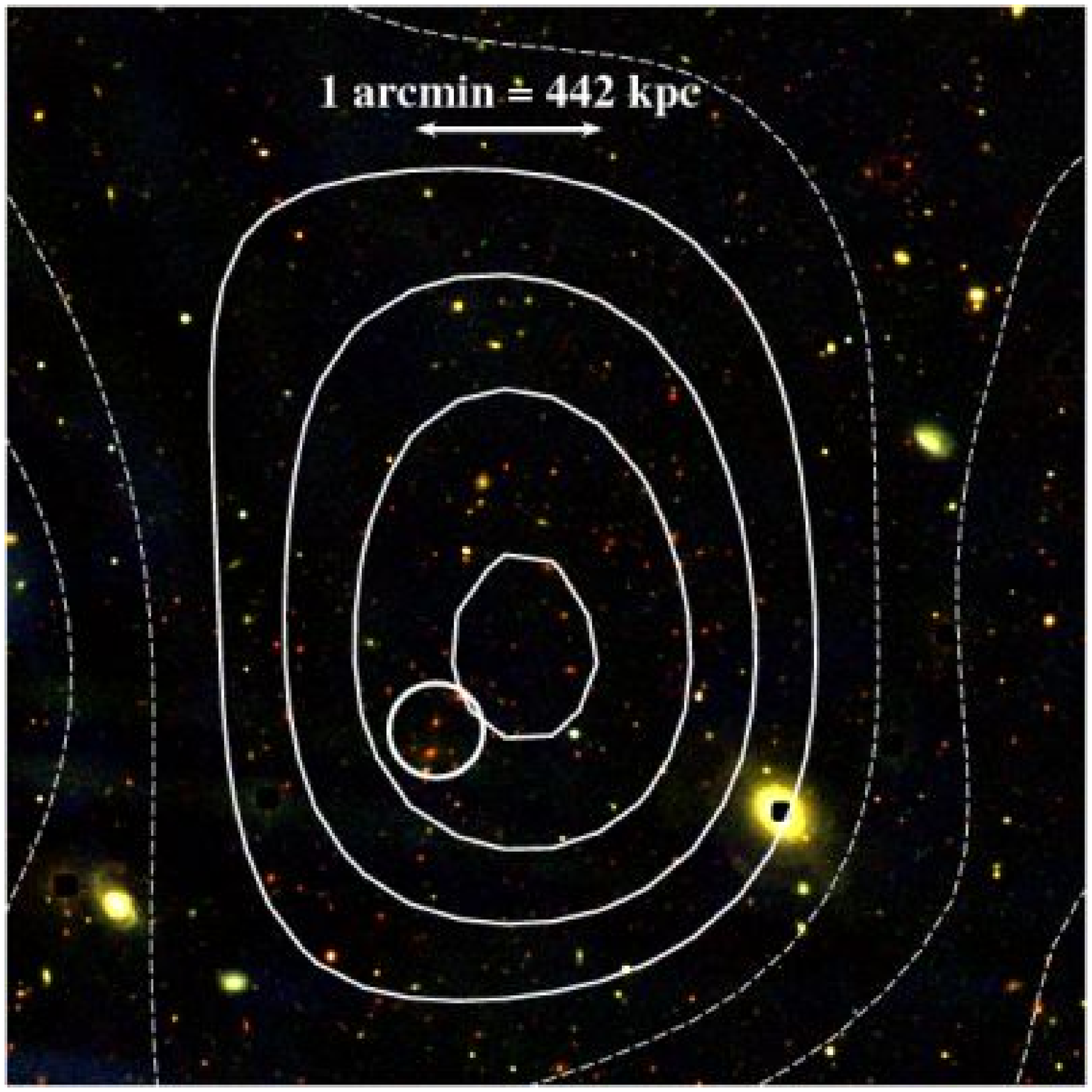} \\

\end{tabular}
\caption{SPT-CL J2359-5009, z=0.76 \label{2359app}}
\end{center}
\end{figure*}

The X-ray image in Figure \ref{2359app} reveals a $45\arcsec$ offset between 
the location of the SPT SZ detection and the X-ray peak. The BCG position 
is offset from the SZ peak by $33\arcsec$. 
Located east of the cluster is a local ($z=0.029$) galaxy pair (circle, $7.5\arcmin$ E) 
which is clearly seen in both the X-ray and optical images. 
Another local galaxy ($z=0.047$) is associated with a faint X-ray source ($1.8\arcmin$ W), 
this is excluded from our X-ray analysis.
A very bright point source is also visible in the X-ray image ($3.5\arcmin$ NW), 
likely associated with an AGN. 
We do not find a counterpart to this source in the NASA/IPAC extragalactic database (NED) and 
estimate a flux of $1.6 \times 10^{-13}$ erg s$^{-1}$ cm$^{-2}$ in the $0.5$-$2.0$ keV band.
A SUMSS radio source is located $15\arcsec$ SE of the BCG with a flux 
of $21\pm1.1$ mJy. 
The X-ray morphology is elongated along the east-west direction, possibly indicating 
merging activity. This is also supported by the $\sim 70$ kpc offset between the 
X-ray peak and the BCG position.

\end{document}